\documentclass[10pt, aps, showpacs, preprintnumbers, prd, nofootinbib, preprint]{revtex4}
\usepackage{feynmf}
\usepackage{amsmath}
\usepackage{graphicx}
\usepackage{mathrsfs}
\usepackage{stmaryrd}
\usepackage{amsfonts}
\usepackage{wrapfig}
\usepackage{subfig}
\usepackage{float}
\usepackage{bbm}
\usepackage{hyperref}
\usepackage{cleveref}

\begin{document}
\def\sech{\mathop{\rm{sech}}\nolimits}
\def\arcsinh{\mathop{\rm{arcsinh}}\nolimits}

\title{A Clash-of-Symmetries Mechanism from Intersecting Domain-Wall Branes}
\author{Benjamin D. Callen}\email{bdcallen@student.unimelb.edu.au}
\affiliation{ARC Centre of Excellence for Particle Physics at the Terascale, School of Physics, The University of Melbourne, Victoria 3010, Australia}
\author{Raymond R. Volkas}\email{raymondv@unimelb.edu.au}
\affiliation{ARC Centre of Excellence for Particle Physics at the Terascale, School of Physics, The University of Melbourne, Victoria 3010, Australia}

\begin{abstract}
 We present a new Clash-of-Symmetries mechanism in the context of an intersecting domain-wall brane model in 5+1-dimensional Minkowskian spacetime recently proposed by the authors. This new application of the Dvali-Shifman idea is designed for localizing gauge fields on a domain-wall intersection and we employ it by adding a gauge group $G$ and giving the scalar fields which form lump-like profiles gauge charges. These fields in turn break $G$ to two different subgroups $H_{1}$ and $H_{2}$ on each domain wall, and the gauge fields of these subgroups are taken to be localized to the respective walls by the confinement dynamics of $G$. There is then a further breaking on the domain-wall intersection to $H_{1}\cap{}H_{2}$ and gauge fields of this overlap group can then be localized to the intersection if they belong inside non-Abelian subgroups of both $H_{1}$ and $H_{2}$ which are spontaneously broken on the intersection and confining in the 4+1D bulks of the respective domain-wall branes. This mechanism has some similarities to the Clash-of-Symmetries mechanism on a single domain wall except that in this case $H_{1}$ and $H_{2}$ need not be isomorphic. We then give some interesting examples of the mechanism in an $SU(7)$ gauge theory, several of which result in the localization of the Standard Model gauge group.  
\end{abstract}

\maketitle

\newpage

\section{Introduction}
\label{sec:introduction}

Braneworld models, in which we live on a 3+1-dimensional brane or subspace embedded in a higher dimensional space, have been a popular application of extra dimensions to solving particle physics problems for many years \cite{antoniadisnewdimattev, exotickkmodels, pregeometryakama, gibbonswiltshire, originalbranepaper, newdimatmillimeter, randallsundrum1, randallsundrum2}. Branes are frequently used to trap Standard Model fields in models with large extra dimensions \cite{originalbranepaper, newdimatmillimeter} and in models where the extra dimensions are warped with the geometry of anti-de Sitter space \cite{randallsundrum1, randallsundrum2}. Braneworld models have been useful for resolving the hierarchy problem \cite{originalbranepaper, randallsundrum1} and also the fermion mass hierarchy problem and other flavor problems  \cite{rs1plusa4csaki, mirabellischmaltz2000, muchenmahanthappayutprimemodel}. The model proposed by Arkani-Hamed, Dimopoulos and Dvali \cite{originalbranepaper} can be extended to an 
arbitrary number of dimensions and in particular the Randall-Sundrum type 2 model (RS2) can naturally be extended beyond five dimensions \cite{arkanihamedinfinitenewdimensions}.

 Domain-wall brane models are extra-dimensional models of the universe in which branes are generated dynamically from field theory rather than being fundamental objects placed in the theory by hand like D-branes. The idea of our 3+1D universe being trapped to the world volume of a domain wall was first put forward by Rubakov and Shaposhnikov\cite{rubshapdwbranes}. Using a dynamically generated object as a prototype brane also has the appeal that all dimensions are treated on an equal footing, so that translational invariance is broken spontaneously rather than explicitly. If we take the brane to be dynamically constructed rather than fundamental, then it also follows that the Standard Model fields must be \emph{dynamically localized} to the brane rather than placed on the brane by hand. This means that we need a mechanism or several mechanisms to trap scalars, fermions, gauge bosons and gravitons on the domain wall. Since domain-walls are generated by scalar fields, fermions and scalars are very easy to trap since we can couple them to the background scalar fields through Yukawa and quartic interactions respectively. In the case of fermions interacting with a single domain wall, it is always the case that a single chiral zero mode is localized to the wall when a 5D fermion is coupled to the wall \cite{jackiwrebbi, modetower, rsgravitydaviesgeorge2007}, which is important since chirality must be reproduced if we are to localize the Standard Model on the wall. For scalars, when quartic interactions are introduced, there will always exist a lowest energy mode localized to the wall and its squared mass is dependent on the parameters of its interaction with the wall and its bulk mass: potentially the squared mass of the lowest energy mode can be tachyonic and this means we can localize a realistic Higgs sector \cite{modetower, rsgravitydaviesgeorge2007}. In both the fermionic and scalar cases, there exists a tower of massive localized Kaluza-Klein modes after the zero mode and lowest energy mode, respectively. After coupling the domain wall to gravity, one can show that gravitons and thus gravity are localized to the domain wall \cite{rsgravitydaviesgeorge2007}, and that the warped geometry in the presence of this domain wall closely resembles that of the RS2 model \cite{randallsundrum2}.
 
  On the other hand, gauge bosons are notoriously difficult to dynamically localize to a domain wall. They cannot be localized to the wall in a similar manner as fermions and scalars by introducing some cubic or quartic coupling between the gauge boson fields and the scalar field generating the domain wall since this will mean gauge invariance and gauge charge universality will be lost \cite{dubrub}. Instead, the only mechanism for trapping gauge bosons without destroying gauge invariance or gauge charge universality that is known to be plausible is the Dvali-Shifman mechanism \cite{dsmech}. Under this mechanism, a non-Abelian gauge group $G$ is spontaneously broken to a subgroup $H$ in the interior of the domain wall by an additional scalar field and the bulk where $G$ is unbroken is taken to be in confinement phase. If we take the 't Hooft-Mandelstam picture of confinement being the result of the formation of a magnetic monopole condensate, then it follows that the bulk in this case will behave as a dual superconductor. This makes it pretty obvious what happens on the domain wall in the case that $G=SU(2)$ and $H=U(1)$: for a test charge placed on the wall, the electric field lines of $H=U(1)$ are free to propagate but they will be repelled by the dual Meissner effect from the bulk. Similarly, if we place the test charge in the bulk, the electric field lines will still be repelled from the bulk and they will form a flux string which will then diverge out onto the wall and behave as if the test charge was in fact on the wall. Under the Dvali-Shifman conjecture, this generalizes to the case where $G$ is a larger non-Abelian gauge group and $H$ is non-Abelian. An alternative way to view this mechanism is in terms of the mass gaps that appear in confining, non-Abelian theories. In the bulk, all the gauge bosons of $H$ must exist in a $G$-glueball state which has a mass of order the mass gap naturally arising in the confining field theory in the bulk. However, on the brane, the same bosons are either massless photons or they exist in glueballs if they are non-Abelian. It then follows that if the mass gap in the bulk is much larger than any of the mass gaps of the non-Abelian factors of $H$, there will be a energy cost for a $H$ boson localized on the wall to propagate into the bulk. It is important to note that the Dvali-Shifman mechanism remains a conjecture and that in the single wall case it relies on 5D confinement.
  
   Assuming that the Dvali-Shifman mechanism works in higher dimensions, we have all the key ingredients to construct a realistic domain-wall brane model. In Ref.~\cite{firstpaper}, a realistic model was constructed where $G=SU(5)$ and $H=SU(3)\times{}SU(2)\times{}U(1)$. This model has some interesting phenomenology since the different SM fermions and scalars are naturally split, in a way analogous to the mechanism for separating fermions first given by Arkani-Hamed and Schmaltz \cite{splitfermions} and to the `families as neighbours' idea of Dvali and Shifman \cite{familiesasneighboursined}. It was further shown that the fermion mass hierarchy problem as well as quark mixing could be explained naturally in the model \cite{su5branemassfittingpaper} and that by adding the discrete flavor symmetry $A_{4}$ that appropriate lepton mixing could be generated as well \cite{su5a4braneworldcallen}. The domain-wall brane framework has also been extended to higher gauge groups such as $SO(10)$ \cite{jayneso10paper} and $E_{6}$ \cite{e6domainwallpaper}. 
   
    The $E_{6}$ model in Ref.~\cite{e6domainwallpaper} is particularly interesting as it is based on a generalization of the Dvali-Shifman mechanism called the Clash-of-Symmetries (CoS) mechanism \cite{o10kinks, abeliankinkscos, clashofsymmetries, e6domainwallpaper, pogosianvachaspaticos, vachaspaticos2, pogosianvachaspaticos3}. The condition of the original Dvali-Shifman mechanism where $G$ was unbroken is not a necessary one: one just has to ensure that the subgroup preserved on the wall is contained by a larger non-Abelian subgroup of $G$ which is in confinement phase in the bulk. It was realized that a smaller subgroup on the wall could still be localized if $G$ was broken respectively to isomorphic but \emph{differently embedded} subgroups $H$ and $H'$ on each side of the domain wall. In the interior of the wall, the symmetry respected is the intersection of these subgroups $H\cap{}H'$ and some of the factors of this final subgroup will be localized to the domain wall provided they are proper subgroups of non-Abelian factors of both $H$ and $H'$ which are confining in the respective halves of the bulk. In proposing the CoS mechanism, we have many tools in our framework in which to extend domain-wall brane models to larger gauge groups. 
    
     Apart from extending the gauge group, we can also consider extending the dimensionality of the bulk spacetime. In doing this one might naturally consider how to construct a model based on domain walls in 5+1D spacetime. Given that domain walls by themselves are codimension-1 defects, it is necessary to introduce two domain walls. There are two options for dimensionally reducing 5+1D to 3+1D with two domain walls. One is to construct what is called a domain ribbon in which a first domain wall is generated by one scalar field and then a second scalar field gets localized to the first domain wall with a tachyonic mass so that it in turn forms the second domain wall in the world volume of the first\cite{morrisnestedwalls, britobazeiadomainribbon}: a wall within a wall. A second idea is to set up domain walls which intersect. Some early attempts at this second option are given in \cite{hksigmamodelintersectingwalls, troitskyvoloshinintersectingdw, exactdomainwalljunction}.
     
   In a previous paper \cite{intersectingdwpaper}, the authors proposed a model in 5+1D based on the discrete group $\mathbb{Z}_2\times{}\mathbb{Z}_{2}$ with four real scalar fields in which two of the scalar fields generate intersecting domain walls and the other two attain lump-like profiles parallel to each of the walls. It was found that there existed a small, special region of parameter space generating analytic solutions. It was also shown that fermions and scalars could be localized to the domain wall intersection, with the couplings to the lumps shifting the profiles away from the center. To construct a realistic model with a Standard Model localized to the domain-wall intersection then requires that we introduce mechanisms for the localization of gravity and the localization of gauge bosons. This paper focuses on the latter. 
 
  Just as there is more freedom in constructing braneworlds based on solitons such as domain walls in 5+1D and higher, there is clearly also more freedom in how we localize gauge fields from the Dvali-Shifman mechanism assuming that 5+1D non-Abelian theories have a confinement phase. Although we are unaware of any work which attempts to prove that a confinement phase exists in 5+1D non-Abelian gauge theories, we are encouraged by lattice gauge simulations which have shown that there exist confining phases in 4+1D SU(2) \cite{5dconfinement} and SU(5) \cite{damiengeorgephd} Yang-Mills gauge theories. The simplest scenario one could think of in both intersecting and nested wall scenarios is a simple codimension-2 generalization of the standard Dvali-Shifman picture on a single wall where a scalar field attains a tachyonic mass in the center of the defect or intersection region and breaks a non-Abelian gauge group $G$ to a subgroup $H$ with the entire 5+1D bulk around the core of the defect in confinement phase. With domain ribbons, one could imagine a nested Dvali-Shifman scenario where we use scalar fields to break $G$ to a subgroup $H$ on the first wall with another scalar field localized to the first domain wall breaking a non-Abelian factor of $H$ to yet a smaller gauge group on the core of the domain ribbon. 
  
   This paper focuses on an application of the Dvali-Shifman mechanism suited for intersecting domain walls and which is the natural one to consider in the context of the model proposed in Ref.~\cite{intersectingdwpaper}, namely that of what we call an intersecting Clash-of-Symmetries mechanism. Here, we utilize the two scalar fields which attain one-dimensional lump-like profiles parallel to each domain wall by giving them gauge charges so that they break $G$ to two subgroups $H_{1}$ and $H_{2}$ on the respective domain walls. Here, the 5+1D bulk away from both domain walls is assumed to be in confinement phase so that $H_{1}$ and $H_{2}$ are localized to the respective walls by the standard Dvali-Shifman mechanism. On the intersection of these walls, there in general is a further symmetry breaking to the overlap of these subgroups $H_{1}\cap{}H_{2}$. We in turn assume that the non-Abelian factors of $H_{1}$ and $H_{2}$ are in confinement phase in the 4+1D bulk of the respective domain walls outside the intersection. This means that non-Abelian factors of $H_{1}\cap{}H_{2}$ are localized by Dvali-Shifman dynamics if they are proper subgroups of both $H_{1}$ and $H_{2}$. Further, Abelian factors of $H_{1}\cap{}H_{2}$ are localized if their generators can be written completely in terms of generators belonging to the non-Abelian factors of both $H_{1}$ and $H_{2}$. Given that the scalar fields generating lumps need not be in the same representation or have the same symmetry breaking pattern, in this version of the CoS mechanism we need not have $H_{1}$ and $H_{2}$ isomorphic. In general, the clashing groups $H_{1}$ and $H_{2}$ are determined by the 4+1D energy densities (or brane tensions) of the two perpendicular kink-lump pairs which can be calculated in terms of the kink-lump solutions that we set as the boundary conditions at infinity around the plane spanned by the two extra dimensions. Given that the 4+1D energy density is degenerate for single kink-lump solutions which break $G$ to different embeddings of the same subgroup, it is then the minimization of the 3+1D junction tension or energy density which arises due to interactions between the perpendicular kink-lump solutions which determines the exact form of the resultant $H_{1}\cap{}H_{2}$ on the intersection. 
   
    After laying out the details of the intersecting CoS mechanism, we give several toy models based on the gauge group $SU(7)$. It turns out that it is possible to localize a Standard Model gauge group\footnote{Let us note that by saying that a gauge group is localized, we mean that all the gauge bosons associated with that gauge group are localized.} under this mechanism with $G=SU(7)$. The first three examples we give are all with the fields attaining lump-like profiles in the adjoint representation. The first example is one where the lump fields attain vacua such that $H_{1}=SU(5)\times{}SU(2)\times{}U(1)$ and $H_{2}=SU(4)\times{}SU(3)\times{}U(1)$. We show that a particular intersecting CoS solution yields a localized Standard Model gauge group with the hypercharge generator proportional to $\rm{diag}(-2/3, -2/3, -2/3, +1, +1, -2, +2)$. This arrangement does have some problems since a single kink-lump solution breaking $SU(7)$ to $H_{1}=SU(5)\times{}SU(2)\times{}U(1)$ is not the most stable one for the interaction potential between the two scalar fields involved in this kink-lump pair, but we give some suggestions about how to overcome this, including adding a cubic invariant for the lump field and accepting metastability or alternatively extending the model to a sextic potential. We also find that we can embed the SM fermions in the anomaly-free combination $\overline{7}+\overline{7}+\overline{7}+21$ and we outline how to embed the electroweak Higgs doublet and the additional Higgs fields required to break the semi-delocalized $U(1)$ groups that we get in addition to the SM. 
    
     The second example we give is one in which $H_{1}$ and $H_{2}$ are differently embedded subgroups isomorphic to $SU(4)\times{}SU(3)\times{}U(1)$. This can also yield a localized Standard Model gauge group but this time with a hypercharge generator which acts on the fundamental as $\rm{diag}(-2/3, -2/3, -2/3, -1, -1, +2, +2)$. This seems like it might not work due to the highly unusual form of this hypercharge generator but it actually turns out that the SM fermions can still be embedded into $SU(7)$ multiplets with the correct quantum numbers, this time in the anomaly-free combination $7+\overline{21}+35$. This model has the advantage over the previous one in that the energetics of the single kink-lump solutions used as the boundary conditions can be assured in a model of the form given in Ref.~\cite{intersectingdwpaper} without resorting to a sextic potential or other additional physics.

    The third example we give for adjoint scalars is one in which we show that this form of the Clash-of-Symmetries mechanism can also be used to implement the approach taken in Ref.~\cite{e6domainwallpaper} by localizing a grand unification group to the domain wall. Here, we have $H_{1}$ and $H_{2}$ as differently embedded subgroups isomorphic to $SU(6)\times{}U(1)$, yielding an $SU(5)$ gauge group which is fully localized to the intersection along with some semi-delocalized $U(1)$ gauge groups which must be broken. 
    
    The last example we give is one in which we have one of the lump-forming fields in the $21$ representation and the other in the $35$ representation. The $21$ can naturally break $SU(7)$ to $H_{1} = SU(5)\times{}SU(2)$ and the $35$ can naturally induce a breaking to $H_{2} = SU(4)\times{}SU(3)$. This is the most elegant example we give in the paper since we attain the same Standard Model gauge group as we get in the first example with adjoint scalars with the generators corresponding to the semi-delocalized $U(1)$ generators that we got previously already broken naturally. Furthermore, we can choose parameters such that the desired solution is the most energetically favorable one. 
    
     In Sec.~\ref{sec:dsmechandcosmechsinglewall}, we give a review of the Dvali-Shifman and Clash-of-Symmetries mechanisms, which includes giving the conditions necessary for localization of both Abelian and non-Abelian gauge fields under the Clash-of-Symmetries mechanism. In Sec.~\ref{sec:intersectingdwsolution}, we review the intersecting kink-lump solution given in Ref.~\cite{intersectingdwpaper}. In Sec.~\ref{sec:intersectingcos}, we outline the proposal for the intersecting Clash-of-Symmetries mechanism, again outlining the necessary conditions for localization which are similar to those for the original CoS mechanism. In Sec.~\ref{sec:su7cos}, we give all four of the examples we have discussed applying this mechanism in the case that $G=SU(7)$. Section \ref{sec:conclusion} is our conclusion.

\section{The Dvali-Shifman and Clash-of-Symmetries mechanisms in the single domain-wall scenario}
\label{sec:dsmechandcosmechsinglewall}

 To employ the Dvali-Shifman mechanism \cite{dsmech} that we discussed in the introduction in 4+1D spacetime, we need to introduce a singlet scalar field $\eta$ which generates the domain wall along with an additional gauge-charged scalar field $\chi$ which condenses in the interior of the domain wall, breaking $G$ to $H$. As a simple example, consider $G=SU(2)$ and $H=U(1)$ and an $SU(2)\times{}\mathbb{Z}_{2}$-invariant scalar field theory with $\chi$ charged under the adjoint representation $3$ of $SU(2)$. Under the discrete symmetry $\mathbb{Z}_{2}$, $\eta\rightarrow{}-\eta$ and $\chi\rightarrow{}-\chi$. The scalar potential of this theory may be written as
\begin{equation}
\begin{aligned}
\label{eq:dvalishifmanmodel}
V(\eta, \chi) &= \frac{1}{4}\lambda_{\eta}(\eta^2-v^2)^2+\lambda_{\eta\chi}(\eta^2-v^2)\rm{Tr}[\chi^2]+\mu^2_{\chi}\rm{Tr}[\chi^2] \\
              &+\lambda_{\chi}\rm{Tr}[\chi^2]^2.
\end{aligned}
\end{equation}

 We want to generate a kink-lump solution. To do this $\chi$ must go to zero at spatial infinity while $\eta$ interpolates between nonzero vacua from negative infinity to positive infinity along a direction $y$. In the interior of the wall, where $\eta$ is zero, we then want $\chi$ to attain a tachyonic mass. Since $\chi$ is an adjoint scalar field, some component of it proportional to some linear combination of the $SU(2)$ generators must condense and all other components vanish. Without loss of generality, we choose the component proportional to the Pauli matrix $\sigma_{3}$, $\chi_{1}$, to condense, and we normalize our generators to $\rm{Tr}[T_{i}T_{j}]=1/2\delta_{ij}$. To generate a stable kink-lump solution, we impose the constraints
 \begin{equation}
\label{eq:boundednessconditions}
\lambda_{\eta}>0, \qquad{} \lambda_{\chi}>0, \qquad \lambda_{\eta\chi}v^2>\mu^2_{\chi}>0.
\end{equation}
 Under these conditions, the global minima are $\eta = \pm{}v$, $\chi=0$. To find a kink-lump solution, we need to find solutions for $\eta$ and $\chi_{1}$ to the Euler-Lagrange equations subject to the boundary conditions
 \begin{equation}
\begin{aligned}
\label{eq:singlekinklumpbc}
\eta{}(y=\pm{}\infty)&=\pm{}v, \\
\chi_{1}(y=\pm{}\infty)&=0.
\end{aligned}
\end{equation}

\begin{figure}[h]
\includegraphics[scale=1.0]{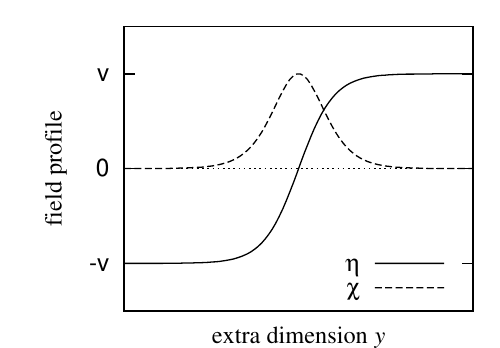}
\caption{A plot of the profiles for $\eta$ and $\chi_{1}$.}
\label{fig:singlekinklumpplot}
\end{figure}

 For a finite region of parameter space, numerical solutions with kink-like profiles exist. For the special parameter choice 
 \begin{equation}
\label{eq:singlekinklumpconditions}
2\mu^2_{\chi}(\lambda_{\eta\chi}-\lambda_{\chi})+(\lambda_{\eta}\lambda_{\chi}-\lambda^2_{\eta\chi})v^2=0,
\end{equation}
one finds the analytic solution
\begin{equation}
\begin{aligned}
\label{eq:singlekinklumpsolution}
\eta(y)&=v\tanh{(ky)}, \\
\chi_{1}(y)&=A\sech{(ky)},
\end{aligned}
\end{equation}
where $k^2=\mu^2_{\chi}$ and $A^2=\frac{\lambda_{\eta\chi}v^2-2\mu^2_{\chi}}{\lambda_{\chi}}$. A plot of this solution is shown in Fig.\ \ref{fig:singlekinklumpplot}.

 Hence, we have successfully generated a kink-lump solution in which the underlying $SU(2)$ gauge symmetry is unbroken in the bulk but spontaneously broken in the interior of the topological defect. Hence, since $SU(2)$ is confining in the bulk, the Dvali-Shifman mechanism is induced, localizing the unbroken $U(1)$ photon to the domain-wall. We assume, if non-Abelian theories are confining in the bulk (which is not generally known in dimensions higher than four, although as noted in the introduction there is some evidence that a confining phase exists in 4+1D gauge theories \cite{5dconfinement, damiengeorgephd}), that this can generalize to higher gauge groups. Indeed, the model proposed in Ref.~\cite{firstpaper} localizes the entire Standard Model gauge group by choosing $G=SU(5)$ and then choosing parameters such that the hypercharge component of $\chi$ condenses on the domain wall.  

  In single domain-wall models, we can generalize the Dvali-Shifman mechanism to the clash-of-symmetries (CoS) mechanism. Several applications of the CoS mechanism were given in Refs.~\cite{o10kinks, abeliankinkscos, clashofsymmetries, e6domainwallpaper} and for a more detailed treatment of the underlying group theory behind the CoS mechanism, see Ref.~\cite{clashofsymgrouptheorypaper}. Some other papers in which the CoS mechanism is utilized but with different motivations to ours are given in Refs.~\cite{pogosianvachaspaticos, vachaspaticos2, pogosianvachaspaticos3}. Under the CoS mechanism, only the field generating the kink, $\eta$, is retained and it assigned to the adjoint representation of the gauge group $G$ rather than being a singlet. To employ this mechanism we require a disconnected vacuum manifold and the way we achieve this is to ensure that the discrete $\mathbb{Z}_{2}$ symmetry is \emph{outside} the gauge group. Hence, the full symmetry group is $G\times{}\mathbb{Z}_{2}$. In the CoS mechanism, $\eta$ attains a vacuum expectation value towards spatial infinity on each side of the wall except this time these vacua spontaneously break $G$. In general, $\eta$ can break $G$ to two \emph{differently embedded} but isomorphic subgroups $H$ and $H'$ on each side of the wall. On such a CoS domain wall, there is a further breaking in the interior to the subgroup $H\cap{}H'$. Assuming the $H$ and $H'$ respecting bulks are confining, there should be a similar Dvali-Shifman mechanism localizing the gauge fields of $H\cap{}H'$.

  Whether the full clashing group $H\cap{}H'$ or only some of its factor groups are localized on the wall depends on how they are embedded within the subgroups of $G$ respected on each wall. Generically, the subgroups on each side of the wall, $H$ and $H'$, will be semi-simple and may be written in the form
  \begin{equation}
  \begin{aligned}
  \label{eq:semisimple}
  H &= N_{1}\times{}N_{2}\times{}N_{3}\times{}...\times{}N_{k-1}\times{}N_{k}\times{}U(1)_{Q_{1}}\times{}U(1)_{Q_{2}}\times{}U(1)_{Q_{3}}...U(1)_{Q_{l-1}}\times{}U(1)_{Q_{l}}, \\
  H' &= N'_{1}\times{}N'_{2}\times{}N'_{3}\times{}...\times{}N'_{k'-1}\times{}N'_{k'}\times{}U(1)_{Q'_{1}}\times{}U(1)_{Q'_{2}}\times{}U(1)_{Q'_{3}}...U(1)_{Q'_{l'-1}}\times{}U(1)_{Q'_{l'}},
  \end{aligned}
  \end{equation}
  where the $N_{i}$ and $N'_{i}$ denote the non-Abelian factor groups and the $Q_{i}$ and $Q'_{i}$ denote the generators of the Abelian factor groups belonging to $H$ and $H'$ respectively. 
  Since, $H$ and $H'$ are semi-simple, $H\cap{}H'$ is also semi-simple. We will denote its non-Abelian factor groups as $n_{i}$ and the generators of its Abelian factor groups as $q_{i}$ and write
  \begin{equation}
  \label{eq:intersectingsemisimple}
  H\cap{}H' = n_{1}\times{}n_{2}\times{}n_{3}\times{}...\times{}n_{r-1}\times{}n_{r}\times{}U(1)_{q_{1}}\times{}U(1)_{q_{2}}\times{}U(1)_{q_{3}}\times{}...\times{}U(1)_{q_{s-1}}\times{}U(1)_{q_{s}}.
  \end{equation}
  
   The above is the general form of the entire $H\cap{}H'$ group respected on the domain wall at the level of symmetries. In general, not all of the factor groups, both Abelian and non-Abelian, of $H\cap{}H'$ will be fully localized to the wall. For a factor group of $H\cap{}H'$ to be localized, it must be fully embedded in the non-Abelian factor groups of both $H$ and $H'$ respected in each semi-infinite region of the bulk, since for a gauge group to be localized via a Dvali-Shifman mechanism, it must lie inside a larger non-Abelian group which is confining in the bulk.
   
   In the non-Abelian case, this means that a non-Abelian factor of $H\cap{}H'$, $n_{i}$ ($1\leq{}i\leq{}r$), is localized only if it is a \emph{proper} subgroup of simple, non-Abelian factors $N_{a}$ and $N'_{b}$ of both $H$ and $H'$ respectively. In other words, we require
   \begin{equation}
   \begin{gathered}
  \label{eq:nonabelianlocclashofsym}
  n_{i}\subset{}N_{a} \; \mathrm{and} \; n_{i}\subset{}N'_{b},
  \end{gathered}
   \end{equation}
  for some $1\leq{}a\leq{}k$ and $1\leq{}b\leq{}k'$. If for any $a$, $n_{i}$ is precisely equal to $N_{a}$ but is still a proper subgroup of $N'_{b}$ for some $b$, there will be no Dvali-Shifman mechanism taking place in the $H$-respecting part of the bulk and thus the gauge bosons of $n_{i}$ will be semi-delocalized. Likewise, if $n_{i}\subset{}N_{a}$ but $n_{i}=N'_{b}$, $n_{i}$ will be semi-delocalized and its gauge bosons will be able to propagate into the $H'$-respecting bulk. If $n_{i}=N_{a}=N'_{b}$ for some $a$ and $b$, then there is no Dvali-Shifman mechanism acting on $n_{i}$ on either side of the bulk and it is thus fully delocalized: its gauge bosons are able to propagate through the whole bulk. 
  
  The Abelian case is a little more complicated but follows similar principles. All the generators $q_{i}$ from Eq.~\ref{eq:intersectingsemisimple} which are preserved on the wall at the level of symmetries must be linear combinations of generators residing in both $H$ and $H'$. Obviously, the respective $U(1)$ generators $Q_{i}$ and $Q'_{i}$ can contribute to both these linear combinations, but there are also generators that belong to the non-Abelian factor groups $N_{a}$ and $N'_{b}$ which lie \emph{outside} the resultant non-Abelian factors $n_{i}$ of the clash. For example, suppose we had for some $a$ and $b$ the factors $N_{a} = SU(4)$ and $N'_{b} = SU(3)$ and the resultant clash was a group $n_{i} = SU(2)$. Then there exists a generator $T=\rm{diag}(+1, +1, -1, -1)$ in $N_{a}$ for which the first two eigenvalues act on components transforming under $n_{i}$ and the latter two act on the two components which do not. Because this generator acts non-trivially on components not acted upon by the resultant $SU(2)$ subgroup, it is outside $n_{i}$. Similarly $N'_{b}$ will have some generator $T'=\rm{diag}(-2, +1, +1)$ in which the latter two components act on $n_{i}$ which is also outside $n_{i}$. We will label these generators $T_{i}$ and $T'_{i}$ for $H$ and $H'$ respectively. Hence, for a generator $q_{i}$ to be a preserved generator on the domain wall at the level of symmetries, it must be that
  \begin{equation}
  \begin{aligned}
  \label{eq:abeliangeneratorsymmetrycondition}
  q_{i} &= \sum^{l}_{i=1}\alpha_{i}Q_{i}+\sum^{m}_{i=1}\beta_{i}T_{i}, \\
        &= \sum^{l'}_{i=1}\alpha'_{i}Q'_{i}+\sum^{m'}_{i=1}\beta'_{i}T'_{i}, 
  \end{aligned}
  \end{equation}
  where all the $\alpha_{i}$, $\beta_{i}$, $\alpha'_{i}$ and $\beta'_{i}$ are real numbers and $m$ and $m'$ are some non-negative integers. 
  
  Equation \ref{eq:abeliangeneratorsymmetrycondition} is just the condition for the generator to be respected at the level of symmetries: the condition for the Abelian generator to be localized is more stringent. For an Abelian generator $q_{i}$ to be fully localized to the domain wall, it must be always embedded inside non-Abelian subgroups of both $H$ and $H'$ for the photon to experience the Dvali-Shifman mechanism from both sides of the bulk. This means that it cannot contain any partition proportional to one of the $Q_{i}$ or $Q'_{i}$ in either of the linear combinations describing $q_{i}$ in terms of generators from $H$ and $H'$, otherwise it will be delocalized in at least one part of the bulk. This means that the condition for full \emph{localization} of an Abelian generator $q_{i}$ to the domain wall is 
  \begin{equation}
  \label{eq:abeliangeneratorlocalizationcondition}
  q_{i} = \sum^{m}_{i=1}\beta_{i}T_{i} = \sum^{m'}_{i=1}\beta'_{i}T'_{i}, \qquad{} \alpha_{i} = \alpha'_{i'}  = 0\; \forall{} \; i,i'.\\
  \end{equation}
  If some $\alpha_{i}$ are non-zero but all the $\alpha'_{i}$ are zero, then $q_{i}$ is free to propagate and leak into the $H$-respecting side of the bulk. Likewise, if all the $\alpha_{i}$ are zero but some $\alpha'_{i}$ are non-zero, $q_{i}$ is semi-delocalized with respect to the $H'$-respecting side of the bulk. If there exist some $\alpha_{a}$ and some $\alpha'_{b}$ which are non-zero, the photon corresponding to $q_{i}$ is free to propagate in both sides of the bulk and is thus fully delocalized.

  Several attempts have been made at constructing a realistic model via the CoS mechanism \cite{e6domainwallpaper}. In this paper, the authors first mentioned an attempt to construct a model based on $SO(10)$, as noted in the paragraphs above. Notwithstanding some issues with the energetics, this model fails because the resultant photon is semi-delocalized. Here, on one side of the wall $H=SU(5)\times{}U(1)$ and on the other $H'=SU(5)'\times{}U(1)'$. Depending on the vacua at the two ends at spatial infinity, there are three possible outcomes for $H\cap{}H'$: $SU(5)\times{}U(1)$, $SU(3)\times{}SU(2)\times{}U(1)\times{}U(1)$ and $SU(4)\times{}U(1)\times{}U(1)$. Obviously, it is the second of these two outcomes which is potentially the desirable one. It turned out that in the region of parameter space that was assumed in that paper to generate analytic solutions, the third option was the most energetically favorable one, that is it minimized the domain-wall tension. However, the authors continued the analysis assuming the second outcome on the basis that there existed a different region of parameter space where the second outcome was the most energetically favorable. If we do this we immediately notice that the $SU(3)$ color and $SU(2)$ weak isospin subgroups are localized to the domain wall since these groups are contained in both $SU(5)$ of $H$ and $SU(5)'$ of $H'$. Where even the second outcome fails is in considering the localization of the hypercharge generator $Y$. Since the hypercharge generator can be embedded entirely in an $SU(5)$ subgroup, we can choose it to be embedded in either $SU(5)$ or $SU(5)'$. Without loss of generality, we will assume that $Y$ is contained in $SU(5)$ subgroup of $H$. However, since $SU(5)'$ is a differently embedded subgroup of $SO(10)$, it cannot be that the analogous generator $Y'$ is equal to $Y$. Hence, the hypercharge generator $Y$ must be a non-trivial linear combination of the $Y'$ and the generator of the $U(1)'$ subgroup of $H'$. From the analysis above, it follows that the hypercharge generator is semi-delocalized (the other $U(1)$ of $H\cap{}H'$ will also be semi-delocalized).
  
  There are several approaches that one could take to get around the problem of semi-delocalized photons in generating a theory in which the Standard Model is reproduced on the domain-wall, or as it turns out in the different Clash-of-Symmetries mechanism on a domain-wall intersection in the 6D model that we will discuss in the rest of the paper. One approach is to localize the gauge fields corresponding to a grand unification group containing the Standard Model on the domain wall instead of just the Standard Model gauge group (plus some additional $U(1)$'s perhaps). This is indeed the approach taken in Ref.~\cite{e6domainwallpaper} in which the authors utilize the gauge group $E_{6}$ instead of $SO(10)$ and break it down to $H=SO(10)\times{}U(1)$ and $H'=SO(10)'\times{}U(1)'$ on each side of the wall. One particular outcome for the clash is $H\cap{}H'=SU(5)\times{}U(1)\times{}U(1)$ for which the $SU(5)$ subgroup is always localized since it is contained in both $SO(10)$ and $SO(10)'$. Assuming there is a region of parameter space where this is the most stable configuration, to reproduce an acceptable model it is just a case of breaking the localized $SU(5)$ subgroup to the Standard Model as well as breaking the additional $U(1)$ subgroups and localizing the required matter content to the wall. 
  
  A second approach, the one we will take when we utilize the Clash-of-Symmetries mechanism for intersecting domain walls in a theory based on $SU(7)$, is to employ a gauge group which is large enough to generate and localize the $SU(3)$ color and $SU(2)$ weak isospin subgroups and at the same type generate more contributing $U(1)$ generators of the second type described in this section, those that initially belong to non-Abelian subgroups respected in the bulk. If at the very least one of the clashing subgroups contained at least two $U(1)$ generators coming from non-Abelian groups and the other at least one, then as noted above if there exists a $U(1)$ generator which is a linear combination of $U(1)$ generators derived solely from non-Abelian subgroups of both the subgroups of $G$ which clash, then this photon will be localized. This is exactly how the $SU(7)$ theory localizes a generator containing the correct hypercharge quantum numbers for the Standard Model components, along with quantum numbers of $\pm{}2$ for non-SM components (so we get the Standard Model along with some exotics with $Y=\pm{}2$). Before discussing the Clash-of-Symmetries mechanism for intersecting domain walls, we will discuss the generation of intersecting kink-lump solutions in the next section.

\section{Intersecting Kink-Lump solutions in a $\mathbb{Z}_{2}\times{}\mathbb{Z}_{2}$ scalar field theory}
\label{sec:intersectingdwsolution}

 In this section, we review the intersecting domain-wall solution of a $\mathbb{Z}_{2}\times{}\mathbb{Z}_{2}$-invariant scalar field theory proposed in Ref.~\cite{intersectingdwpaper}. This scalar field theory has four scalar fields: $\eta_{1}$ and $\eta_{2}$ form the domain-wall kinks while the fields $\chi_{1}$ and $\chi_{2}$ form lump-like profiles parallel to each domain wall. The parities assigned for these fields under the $\mathbb{Z}_{2}\times{}\mathbb{Z}_{2}$ are
 \begin{equation}
\begin{aligned}
\label{eq:z2xz2parities}
\eta_1 &\sim (-, +) \qquad{} \chi_1 \sim (-, +), \\
\eta_2 &\sim (+, -) \qquad{} \chi_2 \sim (+, -),
\end{aligned}
\end{equation}
and thus the most general scalar potential under these discrete symmetries is 
\begin{equation}
\begin{aligned}
\label{eq:scalarpotential}
V_{DW} &= \frac{1}{4}\lambda_{\eta_1}(\eta^2_1-v^2_1)^2+\frac{1}{2}\lambda_{\eta_1\chi_1}(\eta^2_1-v^2_1)\chi^2_1+\frac{1}{2}\mu^2_{\chi_1}\chi^2_1 \\
       &+\frac{1}{4}\lambda_{\chi_1}\chi^4_{1}+g_{\eta_1\chi_1}\eta^3_1\chi_1+h_{\eta_1\chi_1}\eta_1\chi^3_1 \\
       &+ \frac{1}{4}\lambda_{\eta_2}(\eta^2_2-v^2_2)^2+\frac{1}{2}\lambda_{\eta_2\chi_2}(\eta^2_2-v^2_2)\chi^2_2+\frac{1}{2}\mu^2_{\chi_2}\chi^2_2 \\
       &+\frac{1}{4}\lambda_{\chi_2}\chi^4_{2}+g_{\eta_2\chi_2}\eta^3_2\chi_2+h_{\eta_2\chi_2}\eta_2\chi^3_2 \\
       &+\frac{1}{2}\lambda_{\eta_1\eta_2}(\eta^2_1-v^2_1)(\eta^2_2-v^2_2)+\frac{1}{2}\lambda_{\eta_1\chi_2}(\eta^2_1-v^2_1)\chi^2_2 \\
       &+\frac{1}{2}\lambda_{\chi_1\eta_2}\chi^2_1(\eta^2_2-v^2_2)+\frac{1}{2}\lambda_{\chi_1\chi_2}\chi^2_1\chi^2_2  \\
       &+\frac{1}{2}\lambda_{\eta_1\eta_2\chi_2}\eta^2_1\eta_2\chi_2+\frac{1}{2}\lambda_{\chi_1\eta_2\chi_2}\chi^2_1\eta_2\chi_2 \\
       &+\frac{1}{2}\lambda_{\eta_1\chi_1\eta_2}\eta_1\chi_1\eta^2_2+\frac{1}{2}\lambda_{\eta_1\chi_1\chi_2}\eta_1\chi_1\chi^2_2 \\
       &+\lambda_{\eta_1\chi_1\eta_2\chi_2}\eta_1\chi_1\eta_2\chi_2.
\end{aligned}
\end{equation}
We wish to find a solution with two stable, perpendicular kink-lump solutions. This requires choosing a potential which is bounded from below, has four discrete and degenerate minima, and has the fields $\chi_{1}$ and $\chi_{2}$ attaining tachyonic masses in the centers of each wall generated respectively by $\eta_{1}$ and $\eta_{2}$. To ensure this we impose the parameter conditions $\lambda_{\eta_{1,2}}>0$, $\lambda_{\chi_{1,2}}>0$, $\lambda_{\eta_{1}\eta_{2}}>0$, $\lambda_{\eta_{1}\chi_{2}}>0$, $\lambda_{\chi_{1}\eta_{2}}>0$, $\lambda_{\chi_{1}\chi_{2}}>0$, $\lambda_{\eta_{1}\chi_{1}}v^{2}_{1}>\mu^2_{\chi_{1}}$ and $\lambda_{\eta_{2}\chi_{2}}v^{2}_{2}>\mu^2_{\chi_{2}}$. To set up the background intersecting domain walls and the corresponding lumps, we need to find solutions to the Euler-Lagrange equations for $\eta_{1}$, $\eta_{2}$, $\chi_{1}$ and $\chi_{2}$ subject to some boundary conditions which interpolate amongst the four degenerate vacua $\eta_{1}=\pm{}v_{1}, \eta_{2}=\pm{}v_2, \chi_{1}=\chi_{2}=0$.
 To generate an intersection wall solution, generally one can consider the boundary conditions of the fields at infinity to interpolate amongst all four vacua along the edge of some two-dimensional object of infinite size, ideally a square. Thus, if we desire perpendicular kink-lump solutions, we can impose the boundary conditions 
\begin{equation}
\begin{aligned}
\label{eq:perpendicularsolboundaryconditions}
&\eta_{1}(y=\pm{}\infty{}, z)=\pm{}v_1, \,{} \eta_{1}(y, z=\pm{}\infty{})=v_1\tanh{(ky)}, \\
&\eta_{2}(y=\pm{}\infty{}, z)=v_2\tanh{(lz)}, \,{} \eta_{2}(y, z=\pm{}\infty{})=\pm{}v_2, \\
&\chi_{1}(y=\pm{}\infty{}, z)=0, \,{} \chi_{1}(y, z=\pm{}\infty{})=A_{1}\sech{(ky)}, \\
&\chi_{2}(y=\pm{}\infty{}, z)=A_{2}\sech{(lz)}, \,{} \chi_{2}(y, z=\pm{}\infty{}) = 0.
\end{aligned}
\end{equation}
The above conditions are basically one-dimensional kink-lump solutions interpolating from one vacuum to another vacuum along all four corners of a rectangle at infinity. Upon taking the parameter choice
\begin{equation}
\begin{gathered}
\label{eq:parameterconditionsforasol}
\lambda_{\eta_1\eta_2\chi_2} = \lambda_{\chi_1\eta_2\chi_2} = \lambda_{\eta_1\chi_1\eta_2} = \lambda_{\eta_1\chi_1\chi_2} = \lambda_{\eta_1\chi_1\eta_2\chi_2} = 0, \\
g_{\eta_1\chi_1}=h_{\eta_1\chi_1}=g_{\eta_2\chi_2}=h_{\eta_2\chi_2}=0, \\
\lambda_{\eta_1\eta_2}v^2_1=\lambda_{\chi_1\eta_2}A^2_{1}, \quad{} \lambda_{\eta_1\eta_2}v^2_2=\lambda_{\eta_1\chi_2}A^2_{2}, \\
\lambda_{\eta_1\chi_2}v^2_1=\lambda_{\chi_1\chi_2}A^2_{1}, \quad{} \lambda_{\chi_1\eta_2}v^2_2=\lambda_{\chi_1\chi_2}A^2_{2}, \\
2\mu^2_{\chi_1}(\lambda_{\eta_1\chi_1}-\lambda_{\chi_1})+(\lambda_{\eta_1}\lambda_{\chi_1}-\lambda^2_{\eta_1\chi_1})v^2=0, \\
2\mu^2_{\chi_2}(\lambda_{\eta_2\chi_2}-\lambda_{\chi_2})+(\lambda_{\eta_2}\lambda_{\chi_2}-\lambda^2_{\eta_2\chi_2})v^2=0,
\end{gathered}
\end{equation}
one can show that the solution to the Euler-Lagrange equations satisfying Eq.~\ref{eq:perpendicularsolboundaryconditions} is
\begin{equation}
\begin{gathered}
\label{eq:perpendicularsolution}
\eta_{1}(y) = v_{1}\tanh{(ky)}, \quad{} \chi_{1}(y) = A_{1}\sech{(ky)}, \\
\eta_{2}(z) = v_{2}\tanh{(lz)}, \quad{} \chi_{2}(z) = A_{2}\sech{(lz)}, \\
\end{gathered}
\end{equation}
where $k^2=\mu^2_{\chi_1}$, $l^2=\mu^2_{\chi_2}$, $A^2_{1} = \frac{\lambda_{\eta_1\chi_1}v^2_1-2\mu^2_{\chi_1}}{\lambda_{\chi_1}}$, $A^2_{2} = \frac{\lambda_{\eta_2\chi_2}v^2_2-2\mu^2_{\chi_2}}{\lambda_{\chi_2}}$.

 One can also show that under the conditions of Eq.~\ref{eq:parameterconditionsforasol} that there exists a class of solutions of the form 
\begin{equation}
\begin{gathered}
\label{eq:angledsolution}
\eta_{1}(y) = v_{1}\tanh{(ky)}, \quad{} \chi_{1}(y) = A_{1}\sech{(ky)}, \\
\eta_{2}(y, z) = v_{2}\tanh{(lu(y, z))}, \quad{} \chi_{2}(y, z) = A_{2}\sech{(lu(y, z))}, \\
\end{gathered}
\end{equation}
where $u(y, z)=\cos{\theta}y+\sin{\theta}z$ and $0\leq{}\theta\leq{}\pi{}/2$. Solutions with different $\theta$ will satisfy different boundary conditions and in particular all solutions with $\theta<\pi{}/2$ will satisfy boundary conditions which are different from those in Eq.~\ref{eq:perpendicularsolboundaryconditions}. The solution with $\theta=0$ has the two walls parallel and can be thought of as a single wall between the vacua $\eta_{1}=-v_{1}, \eta_{2}=-v_{2}, \chi_{1}=\chi_{2}=0$ and $\eta_{1}=+v_{1}, \eta_{2}=+v_{2}, \chi_{1}=\chi_{2}=0$, while the solution with $\theta=\pi{}/2$ is obviously the perpendicular solution. All other solutions, with $0<\theta<\pi{}/2$, describe walls that intersect at an angle less than ninety degrees. 
 
  There exists a conserved topological charge defined by 
\begin{equation}
\label{eq:intersectingbranetopologicalcurrent}
Q^{ABC} = \int{}d^6x J^{0ABC},
\end{equation}
where the associated current $J^{ABCD}$ is defined by
\begin{equation}
\label{eq:intersectingbranetopologicalcurrent}
J^{MNOP} = \epsilon^{MNOPQR}\epsilon^{ij}\partial_{Q}\eta_{i}\partial_{R}\eta_{j},
\end{equation}
which is zero for the $\theta=0$ solution and equal to $4v_1v_2$ for $0<\theta\leq{}\pi{}/2$. Hence, the perpendicular solution in Eq.~\ref{eq:perpendicularsolution} as well as the intersecting solutions with a non-zero intersection angle less than ninety degrees cannot decay or evolve into the solution where the walls are parallel. 

 We have good reason to believe, at least in sections of parameter space which are small deviations away from the conditions contained in Eq.~\ref{eq:parameterconditionsforasol}, that the perpendicular solution also cannot evolve to a solution with $\theta$ less than ninety degrees \cite{intersectingdwpaper}. It was concluded in \cite{intersectingdwpaper} that further (numerical) analysis was required to show this explicitly. For the purposes of this paper, in which our main focus is on adding a gauge structure and reproducing the smaller subgroups and the Standard Model on the domain-wall intersection via a Clash-of-Symmetries realization of the Dvali-Shifman mechanism, we will assume that the perpendicular solution can always be chosen and is stable.

\section{The Clash-of-Symmetries Mechanism from Intersecting Kink-Lump Solutions}
\label{sec:intersectingcos}

\begin{figure}
\includegraphics[scale=0.7]{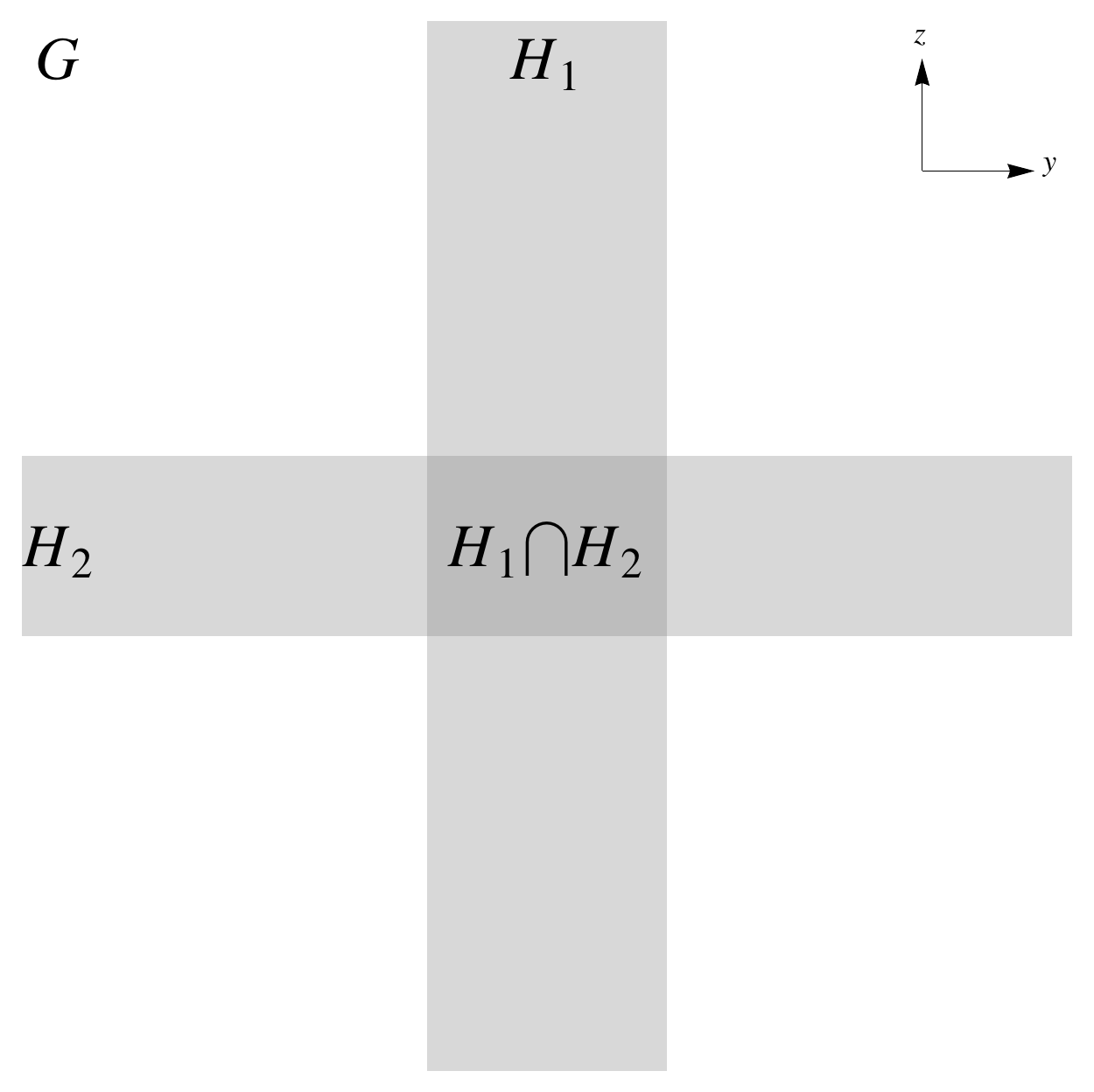}
\caption{\textrm{A picture of the intersecting Clash-of-Symmetries mechanism in the y-z plane. The gauge group $G$ is spontaneously broken to subgroups $H_{1}$ and $H_{2}$ along the walls parallel to the y and z axes respectively. Further symmetry breaking occurs in the intersection region of the walls where the total symmetry respected is $H_{1}\cap{}H_{2}$. If $H_{1}\cap{}H_{2}$ is semi-simple, then provided each factor subgroup is entirely contained in a non-Abelian subgroup or factor group of each of $H_{1}$ and $H_{2}$, it will be completely localized to the intersection. Otherwise there is at least a subgroup of $H_{1}\cap{}H_{2}$ which will be semi-delocalized along one of the domain walls.}}
\label{fig:intersectingclashofsymgraph}
\end{figure}

 We now give an outline for a new Clash-of-Symmetries mechanism applicable in the context of the intersecting domain-wall model treated in the previous section, which is the main purpose of this paper. We now add a gauge group $G$ and give the fields which form lumps, $\chi_{1}$ and $\chi_{2}$, gauge charges. When these fields condense in the interior of each of the respective domain walls $\eta_{1}$ and $\eta_{2}$, they break $G$ to subgroups $H_{1}$ and $H_{2}$ on each wall. Now consider what happens on the intersection of the domain walls. Naturally, we assume $G$ is again confining in the bulk, just as it usually is in the single-wall case. Then by the Dvali-Shifman mechanism, $H_{1}$ is localized to the domain wall described by $\eta_{1}$ and $H_{2}$ is localized to the domain wall described by $\eta_{2}$. In general $H_{1}$ and $H_{2}$ are not the same group, so in the intersection these groups will \emph{clash} and the subgroup respected on the intersection will be $H_{1}\cap{}H_{2}$, analogously to the single-wall CoS mechanism. A graph of this scenario is shown in Fig. \ref{fig:intersectingclashofsymgraph}.
 
 Unlike the single-wall CoS mechanism, $H_{1}$ and $H_{2}$ need not be differently embedded isomorphic subgroups of $G$. This is because $\chi_{1}$ and $\chi_{2}$ are independent fields and so they potentially can attain vacuum expectation values which break $G$ to two different \emph{non-isomorphic} subgroups. Furthermore, $\chi_{1}$ and $\chi_{2}$ need not be in the adjoint representation nor do they need to be in the same representation. These phenomena open up a whole new set of theoretical possibilities for the CoS mechanism. For instance, consider $G=SU(4)$. With an adjoint, we can break $G$ to $SU(3)\times{}U(1)$ or to $SU(2)\times{}SU(2)\times{}U(1)$. Unlike the single wall case where we only had one adjoint field, here we have two adjoint fields so we could break $G$ to $SU(3)\times{}U(1)$ on one wall and to $SU(2)\times{}SU(2)\times{}U(1)$ on the other, leading to possible CoS groups which are isomorphic to $SU(2)\times{}U(1)\times{}U(1)$. On the other hand we could make, say, $\chi_{2}$ transform under the fundamental representation which always breaks $G$ to $SU(3)$ and consider the possible CoS groups when $\chi_{1}$ breaks $G$ to $SU(3)\times{}U(1)$ or when it breaks $G$ to $SU(2)\times{}SU(2)\times{}U(1)$. Yet another possibility is the case where both $\eta_{1}$ and $\eta_{2}$ are fundamentals, leading to both $H_{1}$ and $H_{2}$ being isomorphic to $SU(3)$. In fact, for the case where $G=SU(7)$ it turns out that there is a phenomenologically acceptable solution which breaks to the Standard Model (plus two $U(1)$ gauge groups) which results from a clash between non-isomorphic subgroups, with $H_{1}=SU(5)\times{}SU(2)\times{}U(1)$ and $H_{2}=SU(4)\times{}SU(3)\times{}U(1)$. We will discuss all these possibilities in further detail in the sections that follow.
 
 When interactions between these fields are switched on, the configuration of vacua attained by these fields will be the one that minimizes the energy of the solution. This is not necessarily the one where both vacua are the same and aligned. To see this one needs to see the different contributions to the energy density. The contributions will be the energy densities of each 4+1-dimensional domain wall as well as a 3+1-dimensional junction tension which is associated with the interactions between the walls. Due to the additional dimensionality, the 4+1-dimensional wall tensions will be positive and infinity larger in magnitude than the junction tension. As an illustrative example to describe this set of physics, let $\chi_{1}$ and $\chi_{2}$ transform under the adjoint representation, although this also works more generally. Each kink-lump pair can by itself break $G$ to a number of subgroups depending on the VEV pattern of the respective lump fields $\chi_{1}$ and $\chi_{2}$. Since the value of these VEVs depends on the coordinates, we can write these patterns in the form $\chi_{1}(y) = A^{a}_{1}T_{a}\chi_{A_{1}}(y)$ and $\chi_{2}(z) = A^{a}_{2}T_{a}\chi_{A_{2}}(z)$, where $\chi_{A_{1}}(y)$ and $\chi_{A_{2}}(z)$ are just one-dimensional real fields corresponding to the generators encompassed by the breaking patterns $A_{1}$ and $A_{2}$ respectively. Due to the presence (in general) of $\rm{Tr}[\chi^{4}_{1,2}]$ terms, each of the different configurations with the lumps breaking $G$ to different subgroups will generate different effective quartic self-couplings for $\chi_{A_{1}}(y)$ and $\chi_{A_{2}}(z)$ and thus affects the energy densities of these kink-lump solutions. The resultant clashing groups $H_{1}$ and $H_{2}$ will be determined by which breakings minimize the 4D brane tensions. After determining the subgroups respected on each wall up to isomorphism, since single kink-lump configurations which respect isomorphic but differently embedded subgroups will not differ in energy, it will be the minimization of the 3D junction tension energy density that will determine which particular \emph{clash} gives the minimal energy configuration and thus the intersection group $H_{1}\cap{}H_{2}$. Taking a given embedding of $H_{2}$ as a reference, then the resultant intersection of $H_{2}$ with different embeddings of $H_{1}$ will not be the same in general. It turns out that the various interaction terms between the two sets of fields generating the kink-lump pairs, like $\rm{Tr}[\chi^{2}_{1}\chi^{2}_{2}]$, $\rm{Tr}[\chi_{1}\chi_{2}\chi_{1}\chi_{2}]$ and $[\rm{Tr}(\chi_{1}\chi_{2})]^{2}$, are sensitive to the exact clash and thus the surviving subgroup resulting from the clash of $H_{1}$ and $H_{2}$. Thus the final subgroup respected on the domain-wall intersection firstly depends on the subgroups respected on each wall, which are more or less determined by the coupling constants in the $\eta_{1}-\chi_{1}$ and $\eta_{2}-\chi_{2}$ sectors, and then secondly on which particular embeddings of those subgroups minimize the junction energy density which is determined from the couplings between the $\eta_{1}-\chi_{1}$ and $\eta_{2}-\chi_{2}$ sectors. 

  The localization of the subgroups of $H_{1}\cap{}H_{2}$ in the Clash-of-Symmetries mechanism in the intersecting wall scenario follows analogously to the single domain wall case discussed in Sec.~\ref{sec:dsmechandcosmechsinglewall}. As discussed above, since $G$ is non-Abelian and confining in the bulk, $H_{1}$ and $H_{2}$ are automatically localized to the respective domain walls. Again, as in the single wall scenario, $H_{1}$ and $H_{2}$ are in general semi-simple and may be written in the form described by Eq.~\ref{eq:semisimple} and their overlap $H_{1}\cap{}H_{2}$ is also described by Eq.~\ref{eq:intersectingsemisimple}. The conditions for the full localization of non-Abelian and Abelian groups to the junction are the same as those for the single-kink Clash-of-Symmetries; a non-Abelian subgroup $n$ of $H_{1}\cap{}H_{2}$ must satisfy Eq.~\ref{eq:nonabelianlocclashofsym} and an Abelian generator $q$ must satisfy Eq.\ref{eq:abeliangeneratorlocalizationcondition}. In the case that these conditions are not satisfied, the gauge bosons are semi-delocalized and there are obvious physical differences to the single-wall case; in this case semi-delocalized photons are able to propagate along one or both walls (but not into the $G$-respecting parts of the bulk) rather than being able to propagate through one half of the bulk or through the entire bulk in the single-wall case.

  For this application of the Dvali-Shifman mechanism to work, there is a certain hierarchy of scales which needs to be respected. This hierarchy is very similar to that stated for the single-wall $SU(5)$ model of Ref. \cite{firstpaper}, and it is based on similar principles. Firstly, as our theory is a 5+1D field theory, it is non-renormalizable and a UV cutoff $\Lambda_{UV}$ must be imposed. Secondly, there are the symmetry breaking scales for $H_{1}$ and $H_{2}$ on each wall which are roughly of the order of $A^{1/2}_{1}$ and $A^{1/2}_{2}$ respectively, where here $A_{1}$ and $A_{2}$ simply denote the maximum value of the lump profiles in the components of $\chi_{1}$ and $\chi_{2}$ which condense. Due to the bulk being in confinement phase, there exists the bulk confinement scale for $G$ which we call $\Lambda_{G,\rm{conf}}$. There are also the confinement scales for the non-Abelian factor groups of $H_{1}$ and $H_{2}$, which we label collectively as $\Lambda_{H_{1},\rm{conf}}$ and $\Lambda_{H_{2},\rm{conf}}$ as well as the confinement scales of the localized non-Abelian factor groups of $H_{1}\cap{}H_{2}$, which we label $\Lambda_{H_{1}\cap{}H_{2},\rm{conf}}$. Finally, there are the inverse widths of each domain wall, $k$ and $l$. The required hierarchy is 
  \begin{equation}
  \label{eq:intersectingcosenergyscalehierarchy}
  \Lambda_{UV}>A^{1/2}_{1},A^{1/2}_{2}>\Lambda_{G,\rm{conf}}>\Lambda_{H_{1},\rm{conf}}, \Lambda_{H_{2},\rm{conf}}>k, l> \Lambda_{H_{1}\cap{}H_{2},\rm{conf}}.
  \end{equation}
  Obviously, $\Lambda_{UV}$ must be the highest scale of the theory. Next, the symmetry breaking scales $A^{1/2}_{1}$ and $A^{1/2}_{2}$ must be larger than the confinement scale in the bulk $\Lambda_{G,\rm{conf}}$ so that our background solutions for $\chi_{1}$ and $\chi_{2}$ are not destroyed by the confinement dynamics of $G$. In turn, $\Lambda_{G,\rm{conf}}$ must be higher than any of the confinement scales $\Lambda_{H_{1},\rm{conf}}$ and $\Lambda_{H_{2},\rm{conf}}$ in order to localize $H_{1}$ and $H_{2}$ by the Dvali-Shifman mechanism and ensure that there is a mass gap between the masses of the glueballs of $G$ and those of the non-Abelian factor groups of $H_{1}$ and $H_{2}$. The confinement scales $\Lambda_{H_{1},\rm{conf}}$ and $\Lambda_{H_{2},\rm{conf}}$ on each wall must be larger the the inverse widths of the domain walls $k$ and $l$ for the same reasons that the bulk confinement scale must be larger than the domain wall scale in the single wall case utilizing Dvali-Shifman, as discussed in \cite{2plus1dconfinement}. Finally, $\Lambda_{H_{1}\cap{}H_{2},\rm{conf}}$ must be lower than $\Lambda_{H_{1},\rm{conf}}$ and $\Lambda_{H_{2},\rm{conf}}$ to ensure that its gauge bosons are localized by the Dvali-Shifman mechanism. In fact, $\Lambda_{H_{1}\cap{}H_{2},\rm{conf}}$ should be the lowest scale of the theory since if we reproduce the Standard Model on the domain-wall intersection we naturally expect $\Lambda_{H_{1}\cap{}H_{2},\rm{conf}}\sim{}\Lambda_{QCD}$. All the scales except $\Lambda_{H_{1}\cap{}H_{2},\rm{conf}}$ should be above the electroweak scale.

 In the next section, we will discuss applying this realization of the Clash-of-Symmetries mechanism in practice. In Section \ref{sec:su7cos}, we will discuss how to build a realistic model from an $SU(7)$ gauge group.

\section{Some Slices of Heaven From $SU(7)$: A Construction of a Realistic Model from the Clash-of-Symmetries Mechanism}
\label{sec:su7cos}

 In this section we discuss how to build a realistic model on an $SU(7)$ gauge group. Given $SU(7)$ is not a commonly used gauge group, we give a basic overview of the representation theory of $SU(7)$ in Appendix \ref{appendix:su7reptheory}. In the forthcoming analysis, we will assume that both $\chi_{1}$ and $\chi_{2}$ are charged under the adjoint representation, that is the $48$ of $SU(7)$. Firstly, we need to consider the possible breaking patterns of a single adjoint scalar field, which can be analyzed by simply looking at the Cartan subalgebra. 
 
  We can always gauge rotate the vacuum expectation value of an adjoint scalar field $\chi$ (which could be either $\chi_{1}$ and $\chi_{2}$ here) such that it is represented by a traceless diagonal matrix, which in the case of $SU(7)$ may be written
\begin{equation}
\label{eq:diagonaladjointvev}
\chi = \rm{diag}(a_{1}, a_{2}, a_{3}, a_{4}, a_{5}, a_{6}, a_{7}),
\end{equation}
where the $a_{i}$ are numbers parametrizing the Cartan subalgebra and satisfy the traceless condition $\sum_{i=1}^{7}a_{i}=0$. From considering various values of the six independent $a_{i}$ it is possible to generate all the possible symmetry breaking patterns for a single adjoint. The most stable configuration will depend on the potential for $\chi$ in the theory. In Ref.~\cite{rueggscalarpot}, Ruegg showed that the quartic Higgs potential resulting for the $a_{i}$ after substitution for $\chi$ only has extrema (and thus minima) if at most two of the $a_{i}$ are different. Hence the possible resulting subgroups after breaking with the $48$ of $SU(7)$ are $SU(6)\times{}U(1)$, for which six of the $a_{i}$ are equal and the other differs, $SU(5)\times{}SU(2)\times{}U(1)$ when five $a_{i}$ are equal and the remaining two $a_{i}$ are equal to a different value, and $SU(4)\times{}SU(3)\times{}U(1)$ which results when one eigenvalue of $\chi$ has a multiplicity of four and the other three.

 In the context of our model with intersecting kink-lump solutions, this means that each of $\chi_{1}$ and $\chi_{2}$ break $SU(7)$ to one of these three subgroups. As a result, the possible clashes are between two different embeddings of one of the three subgroups $SU(6)\times{}U(1)$, $SU(5)\times{}SU(2)\times{}U(1)$ or $SU(4)\times{}SU(3)\times{}U(1)$ or between particular embeddings of two different choices of these groups. Most of the possibilities are physically uninteresting; a full description of all the possibilities is given in Appendix \ref{appendix:allpossiblecosgroups}. 
 
  The most physically interesting possibility with $\chi_{1}$ and $\chi_{2}$ in the adjoint representation is a clash between a particular embedding of $H_{1} = SU(5)\times{}SU(2)\times{}U(1)$ and $H_{2} = SU(4)\times{}SU(3)\times{}U(1)$. It turns out that a possible subgroup resulting from the clash contains a Standard Model gauge group, including the Abelian group generated by hypercharge, which is fully \emph{localized} to the domain-wall intersection, along with some semi-delocalized $U(1)$ gauge groups that we must break by adding additional Higgs fields in the appropriate representations. Given that the top $5\times{}5$ block of the localized Abelian generator is just the usual $SU(5)$ hypercharge generator, the Standard Model fermions can be embedded in $SU(7)$ multiplets in the most obvious way: in a combination of the anti-fundamental $\overline{7}$ representation and the anti-symmetric $21$ representation along with a couple of additional fermions in the $\overline{7}$ to ensure that the effective 3+1D field theory is anomaly-free. The main difficulty with this arrangement is ensuring that the kink-lump solution breaking $SU(7)$ to $H_{1} = SU(5)\times{}SU(2)\times{}U(1)$ is the most energetically favorable one in the $\eta_{1}$-$\chi_{1}$ sector. This cannot be generated in the parameter region with analytic solutions with a quartic potential and it seems necessary to utilize a sextic potential. 
  
  Another particular choice that we mention that works in an unusual way is that between two different embeddings of $SU(4)\times{}SU(3)\times{}U(1)$. Having looked at the possibility mentioned in the previous paragraph, it might seem perfectly reasonable to consider two different embeddings of $SU(4)\times{}SU(3)\times{}U(1)$ and particularly so since it avoids some of the problems of the previous solution in ensuring that it is energetically favorable. This choice indeed can localize a SM-like gauge group but with a localized $U(1)$ subgroup whose generator has the \emph{wrong} relative sign between the charges of the right-handed down quark and the lepton doublet! In spite of this, the Standard Model fermions can be successfully embedded in to representations of $SU(7)$, albeit in a rather unusual way: they are embedded in the combination of a $7$, a $\overline{21}$ and a $35$ rather than the more obvious combination of a $\overline{7}$ and a $21$. This means that this solution yields a Standard Model with more exotics. 

  The third possibility we mention is one between two different embeddings of $SU(6)\times{}U(1)$. Like the case with two embeddings of $SU(4)\times{}SU(3)\times{}U(1)$, one can easily choose energetically favored solutions for the two different walls. In the case of differently embedded $SU(6)\times{}U(1)$ subgroups, there will be a localized $SU(5)$ gauge group on the intersection along with two semi-delocalized $U(1)$ gauge groups. Hence, this example provides a six-dimensional realization of the approach taken in the single wall case in \cite{e6domainwallpaper} to localizing the photon along with the non-Abelian gauge bosons of the Standard Model, namely that of localizing a grand unified gauge group to the intersection containing our 3+1D universe. It then follows that one just needs to break the semi-delocalized Abelian groups and then break the $SU(5)$ group to the SM in the usual way. 
  
  The last possibility we illustrate is a case where neither $\chi_{1}$ or $\chi_{2}$ are adjoint scalars but transform instead under the totally antisymmetric $21$ and $35$ representations respectively. The $21$ can break $SU(7)$ to $H_{1} = SU(5)\times{}SU(2)$ and the $35$ can induce a breaking to $H_{2} = SU(4)\times{}SU(3)$. A particular clash between these two groups leads directly to the localization of the same Standard Model gauge group as that generated in the first example given with adjoint scalars. There are two main advantages with this situation over the one with two adjoint scalars in generating the same Standard Model gauge group. Obviously, the first is that we have a localized Standard Model without the need to break any additional semi-delocalized Abelian groups. The second is that the unlike the case with an adjoint scalar, for a particular parameter choice the arrangement on the first wall where the $21$ induces the breaking to the $SU(5)\times{}SU(2)$ subgroup can be guaranteed to be the most stable one with a quartic potential. The breaking to $SU(4)\times{}SU(3)$ on the second wall with the $35$ can also be guaranteed to be the most stable arrangement with a quartic potential.

  We discuss these four possibilities in the following four subsections.

\subsection{A fully localized Standard Model with $H_{1} = SU(5)\times{}SU(2)\times{}U(1)$ and $H_{2} = SU(4)\times{}SU(3)\times{}U(1)$ on a Domain-Wall Intersection}
\label{subsec:su5su2u1versussu4su3u1yieldingsm}

 Here we will describe firstly the group theoretic background behind the solution with $H_{1} = SU(5)\times{}SU(2)\times{}U(1)$ and $H_{2} = SU(4)\times{}SU(3)\times{}U(1)$ which localizes the Standard Model along with some $Y=\pm{}2$ exotics. Later we will discuss the energetics and parameter choices needed to ensure that such a solution is the most stable one. 
 
  Let's list all the possible subgroups resulting from a clash between an $SU(5)\times{}SU(2)\times{}U(1)$ subgroup and an $SU(4)\times{}SU(3)\times{}U(1)$ subgroup of $SU(7)$, at the level of symmetries. There are three possibilities: $H_{1}\cap{}H_{2} = SU(4)\times{}SU(2)\times{}U(1)\times{}U(1)$, $H_{1}\cap{}H_{2} = SU(3)\times{}SU(2)\times{}SU(2)\times{}U(1)\times{}U(1)$ and $H_{1}\cap{}H_{2} = SU(3)\times{}SU(2)\times{}U(1)\times{}U(1)\times{}U(1)$. The first two are physically uninteresting since, in both these cases, one of the non-Abelian subgroups is semi-delocalized due to being respected along one wall (the $SU(4)$ subgroup in the first case, the $SU(3)$ subgroup in the second). It is the last case which is interesting since here the whole Standard Model gauge group is localized. Along with the Standard Model come two $U(1)$ subgroups which are semi-delocalized and thus must be broken at a sufficiently high energy scale to avoid a leakage of energy into the bulk in the low energy field theory.
  
  As an example which yields this desired situation, consider the case where the component of $\chi_{1}$ which condenses is proportional to the matrix
  \begin{equation}
  \label{eq:firstvevsmclash}
  Q_{1} = \begin{pmatrix} 2 & 0 & 0 & 0 & 0 & 0 & 0 \\
                          0 & 2 & 0 & 0 & 0 & 0 & 0 \\
                          0 & 0 & 2 & 0 & 0 & 0 & 0 \\
                          0 & 0 & 0 & 2 & 0 & 0 & 0 \\
                          0 & 0 & 0 & 0 & 2 & 0 & 0 \\
                          0 & 0 & 0 & 0 & 0 & -5 & 0 \\
                          0 & 0 & 0 & 0 & 0 & 0 & -5 \end{pmatrix},
  \end{equation}
and the component of $\chi_{2}$ which condenses is proportional to 
\begin{equation}
  \label{eq:secondvevsmclash}
 Q'_{1} = \begin{pmatrix} 3 & 0 & 0 & 0 & 0 & 0 & 0 \\
                          0 & 3 & 0 & 0 & 0 & 0 & 0 \\
                          0 & 0 & 3 & 0 & 0 & 0 & 0 \\
                          0 & 0 & 0 & -4 & 0 & 0 & 0 \\
                          0 & 0 & 0 & 0 & -4 & 0 & 0 \\
                          0 & 0 & 0 & 0 & 0 & -4 & 0 \\
                          0 & 0 & 0 & 0 & 0 & 0 & 3 \end{pmatrix}.
\end{equation}

 The former clearly induces the breaking $SU(7)\rightarrow{}SU(5)\times{}SU(2)\times{}U(1)$ and the latter induces the breaking $SU(7)\rightarrow{}SU(4)\times{}SU(3)\times{}U(1)$. Inspecting these two matrices, one notices that there is an $SU(3)$ subgroup which preserves the top-left $3\times{}3$ blocks of these two matrices. This $SU(3)$ subgroup is the one common to the $SU(5)$ and $SU(4)$ subgroups induced by the respective vacua. Similarly, an $SU(2)$ subgroup represented by generators with non-trivial components in the $2\times{}2$ block on the intersection of the fourth and fifth rows and fourth and fifth columns preserves the fourth and fifth elements along the diagonal along both matrices, which is common to the $SU(5)$ and $SU(3)$ subgroups. Looking at the lower-right $2\times{}2$ block, one sees that the $SU(2)$ subgroup induced by the condensation of $\chi_{1}$ does not survive and is thus broken since this same group does not preserve the corresponding elements of the diagonal in the VEV pattern of $\chi_{2}$, represented by $Q'_{1}$. Hence, the non-Abelian sector surviving the clash is $SU(3)\times{}SU(2)$, which is precisely that required for a localized SM. Since both these non-Abelian subgroups are entirely contained in larger non-Abelian subgroups respected along each wall ($SU(5)$ and $SU(4)$ in the case of $SU(3)$ color, and $SU(5)$ and $SU(3)$ in the case of $SU(2)$ weak isospin), they are fully localized as required to the domain-wall intersection. 
 
 Next, we need to determine the remaining $U(1)$ subgroups respected on the wall at the level of symmetries, and then determine if any of them are localized. As is well known, any spontaneous breaking by an adjoint scalar field always preserves a $U(1)$ subgroup and the generator of this $U(1)$ subgroup is precisely equal to the generator which condenses. Hence, $Q_{1}$ and $Q'_{1}$ are the generators of these associated $U(1)$ subgroups in the case of the walls generated by $\eta_{1}$ and $\eta_{2}$ respectively. We now look at any potential leftover generators inside the non-Abelian groups respected on each wall but which are outside the smaller non-Abelian subgroups respected on the intersection (ie. $U(1)$ generators of the $T$, $T'$ type discussed previously). For $H_{1}$, one sees that the usual $SU(5)$ hypercharge generator is one of the leftover generators, 
\begin{equation}
\label{eq:su5clashingu1generator}
 T_{1} = \begin{pmatrix} +\frac{2}{3} & 0 & 0 & 0 & 0 & 0 & 0 \\
                     0 & +\frac{2}{3} & 0 & 0 & 0 & 0 & 0 \\
                     0 & 0 & +\frac{2}{3} & 0 & 0 & 0 & 0 \\
                     0 & 0 & 0 & -1 & 0 & 0 & 0 \\
                     0 & 0 & 0 & 0 & -1 & 0 & 0 \\
                     0 & 0 & 0 & 0 & 0 & 0 & 0 \\
                     0 & 0 & 0 & 0 & 0 & 0 & 0 \end{pmatrix},
\end{equation}
which lies inside the $SU(5)$ subgroup respected on the first wall but is outside both its $SU(3)$ and $SU(2)$ subgroups that survive the clash. Similarly, 
\begin{equation}
\label{eq:su2clashingu1generator}
 T_{2} = \begin{pmatrix} 0 & 0 & 0 & 0 & 0 & 0 & 0 \\
                  0 & 0 & 0 & 0 & 0 & 0 & 0 \\
                  0 & 0 & 0 & 0 & 0 & 0 & 0 \\
                  0 & 0 & 0 & 0 & 0 & 0 & 0 \\
                  0 & 0 & 0 & 0 & 0 & 0 & 0 \\
                  0 & 0 & 0 & 0 & 0 & 1 & 0 \\
                  0 & 0 & 0 & 0 & 0 & 0 & -1 \end{pmatrix},
\end{equation}
is inside the $SU(2)$ subgroup respected on the first wall and could potentially contribute to a surviving $U(1)$.
For $H_{2}$, the respective generators inside $SU(4)$ and $SU(3)$ but outside the preserved non-Abelian groups are respectively
\begin{equation}
\label{eq:su4clashingu1generator}
 T'_{1} = \begin{pmatrix} +\frac{2}{3} & 0 & 0 & 0 & 0 & 0 & 0 \\
                  0 & +\frac{2}{3} & 0 & 0 & 0 & 0 & 0 \\
                  0 & 0 & +\frac{2}{3} & 0 & 0 & 0 & 0 \\
                  0 & 0 & 0 & 0 & 0 & 0 & 0 \\
                  0 & 0 & 0 & 0 & 0 & 0 & 0 \\
                  0 & 0 & 0 & 0 & 0 & 0 & 0 \\
                  0 & 0 & 0 & 0 & 0 & 0 & -2 \end{pmatrix},
\end{equation}
and
\begin{equation}
\label{eq:su3clashingu1generator}
 T'_{2} = \begin{pmatrix} 0 & 0 & 0 & 0 & 0 & 0 & 0 \\
                  0 & 0 & 0 & 0 & 0 & 0 & 0 \\
                  0 & 0 & 0 & 0 & 0 & 0 & 0 \\
                  0 & 0 & 0 & -1 & 0 & 0 & 0 \\
                  0 & 0 & 0 & 0 & -1 & 0 & 0 \\
                  0 & 0 & 0 & 0 & 0 & 2 & 0 \\
                  0 & 0 & 0 & 0 & 0 & 0 & 0 \end{pmatrix}.
 \end{equation}

 We have listed all the possible contributing generators above. For a $U(1)$ subgroup to be respected on the wall at the level of symmetries, as discussed previously it must be a linear combination of generators satisfying Eq.~\ref{eq:abeliangeneratorsymmetrycondition}. By inspection, one can easily see from the above generators that there exists a generator $Y$ which can be written solely in terms of the $T_{i}$ and $T'_{i}$ generators: 
\begin{equation}
\label{eq:localizedhyperchargegenerator}
Y = -T_{1}-2T_{2} = -T'_{1}-T'_{2} = \begin{pmatrix} -\frac{2}{3} & 0 & 0 & 0 & 0 & 0 & 0 \\
                                 0 & -\frac{2}{3} & 0 & 0 & 0 & 0 & 0 \\
                                 0 & 0 & -\frac{2}{3} & 0 & 0 & 0 & 0 \\
                                 0 & 0 & 0 & +1 & 0 & 0 & 0 \\
                                 0 & 0 & 0 & 0 & +1 & 0 & 0 \\
                                 0 & 0 & 0 & 0 & 0 & -2 & 0 \\
                                 0 & 0 & 0 & 0 & 0 & 0 & +2 \end{pmatrix}.
\end{equation}
Thus, $Y$ satisfies Eq.~\ref{eq:abeliangeneratorlocalizationcondition} and is thus localized to the domain-wall intersection. Furthermore, the upper left $5\times{}5$ block of $Y$ is precisely the usual hypercharge generator so it has the desired properties of a localized Abelian generator on the components which transform under the $SU(3)$ color and $SU(2)$ isospin subgroups. Hence, this configuration successfully localizes the Standard Model gauge group. 

 Along with the localized Standard Model, we also get a couple of semi-delocalized $U(1)$ gauge groups. The generators of these Abelian groups may be taken to be
 \begin{equation}
\label{eq:firstsemidelocU1}
A = 4Q_{1}+7T_{1}-6T_{2} = 2Q'_{1}+10T'_{1}+9T'_{2} = \begin{pmatrix} \frac{38}{3} & 0 & 0 & 0 & 0 & 0 & 0 \\
                                                                       0 & \frac{38}{3} & 0 & 0 & 0 & 0 & 0 \\
                                                                       0 & 0 & \frac{38}{3} & 0 & 0 & 0 & 0 \\
                                                                       0 & 0 & 0 & +1 & 0 & 0 & 0 \\
                                                                       0 & 0 & 0 & 0 & +1 & 0 & 0 \\
                                                                       0 & 0 & 0 & 0 & 0 & -26 & 0 \\
                                                                       0 & 0 & 0 & 0 & 0 & 0 & -14 \end{pmatrix},
 \end{equation}
 and 
 \begin{equation}
\label{eq:firstsemidelocU1}
B = -3Q_{1}+12T_{1}+12T_{2} = \frac{3}{2}Q'_{1}-\frac{3}{8}T'_{1}-15'T_{2} = \begin{pmatrix}  2 & 0 & 0 & 0 & 0 & 0 & 0 \\
                                                                       0 & 2 & 0 & 0 & 0 & 0 & 0 \\
                                                                       0 & 0 & 2 & 0 & 0 & 0 & 0 \\
                                                                       0 & 0 & 0 & -18 & 0 & 0 & 0 \\
                                                                       0 & 0 & 0 & 0 & -18 & 0 & 0 \\
                                                                       0 & 0 & 0 & 0 & 0 & 27 & 0 \\
                                                                       0 & 0 & 0 & 0 & 0 & 0 & 3 \end{pmatrix}.
 \end{equation}
Evidently, both $A$ and $B$ satisfy Eq.~\ref{eq:abeliangeneratorsymmetrycondition} but not Eq.~\ref{eq:abeliangeneratorlocalizationcondition}, as one expects for semi-delocalized generators. The resultant photons are able to propagate along both walls and thus these Abelian groups must be broken on the domain-wall intersection as the existence of massless 5D states coupling to the Standard Model fermions would obviously be disastrous.  

Since the lower right $2\times{}2$ block is proportional to twice the third Pauli matrix, once we include the fermionic particle content and Higgs fields required for electroweak symmetry breaking, we expect exotic scalars and fermions. If we embed the right-handed down quark and the lepton doublet in a $\overline{7}$ for instance, there will be exotics with hypercharge $Y=\pm{}2$. Thus, to construct realistic models, we need to ensure that the masses of the localized modes for these exotics are sufficiently more massive than those corresponding to the SM particle content. The exact breakdowns of the $\overline{7}$, $21$ and $35$ representations in terms of the full $SU(3)_{c}\times{}SU(2)_{I}\times{}U(1)_{Y}\times{}U(1)_{A}\times{}U(1)_{B}$ subgroup preserved at the level of symmetries on the domain-wall intersection are
\begin{equation}
\label{eq:antifundamentalYSM}
\overline{7} = (\overline{3}, 1, +\frac{2}{3}, -\frac{38}{3}, -2)+(1, 2, -1, -1, +18)+(1, 1, +2, +26, -27)+(1, 1, -2, +14, -3),
\end{equation}
\begin{equation}
\begin{aligned}
\label{eq:antisymmetricYSM}
21 &= (\overline{3}, 1, -\frac{4}{3}, +\frac{76}{3}, +4)+(3, 2, +\frac{1}{3}, +\frac{41}{3}, -16)+(1, 1, +2, +2, -36)+(3, 1, -\frac{8}{3}, -\frac{40}{3}, +29) \\
   &+(3, 1, +\frac{4}{3}, -\frac{4}{3}, +5)+(1, 2, -1, -25, +9)+(1, 2, +3, -13, -15)+(1, 1, 0, -40, +30),
\end{aligned}
\end{equation}
and
\begin{equation}
\begin{aligned}
\label{eq:antisymmetricYSM}
35 &=(3, 1, +\frac{4}{3}, +\frac{44}{3}, -34)+(\overline{3}, 2, -\frac{1}{3}, +\frac{79}{3}, -14)+(1, 1, -2, +38, +6)+(\overline{3}, 1, -\frac{10}{3}, -\frac{2}{3}, +31) \\
   &+(3, 2, -\frac{5}{3}, -\frac{37}{3}, +11)+(1, 1, 0, -24, -9)+(\overline{3}, 1, +\frac{2}{3}, +\frac{34}{3}, +7)+(3, 2, +\frac{7}{3}, -\frac{1}{3}, -13) \\
   &+(1, 1, +4, -12, -33)+(3, 1, -\frac{2}{3}, -\frac{82}{3}, +32)+(1, 2, +1, -39, +12). 
\end{aligned}
\end{equation}
Thus we can easily see that we can embed the Standard Model fermions in the most obvious way with the charge conjugate of the right-chiral down quark $(d_{R})^{c}$ and the lepton doublet $L$ embedded in the $\overline{7}$, and the charge conjugates of the right-chiral up quark $(u_{R})^{c}$ and of the right-chiral electron $(e_{R})^{c}$ along with the quark doublet $Q$ embedded in the $21$. There is also a component which is a singlet under the SM, the $(1, 1, 0, -40, +30)$ component, inside the $21$ which could be potentially used as a right-chiral neutrino or its charge conjugate. 

 One thing that is not completely clear is what is the \emph{minimal} content necessary for anomaly cancellation. Fermion localization in the model described in Sec.~\ref{sec:intersectingdwsolution} was treated in Ref.~\cite{intersectingdwpaper}, and it was shown that a single chiral zero mode was reproduced on the intersection when a 5+1D \emph{Dirac} fermion was coupled to the background scalar fields through scalar and pseudoscalar Yukawa couplings. The fact that we use full eight-component Dirac spinors to embed 3+1D chiral zero modes is important since this means that the underlying 5+1D theory is vector-like and is thus free from both 5+1D gravitational and gauge anomalies. However, the effective 3+1D theory reproduced on the intersection is in general chiral since each 5+1D Dirac fermion produces a single chiral zero mode. Hence, one may plausibly reproduce an \emph{anomalous} 3+1D theory from an \emph{anomaly-free} 5+1D theory, as would be the case if we chose the only fermionic content to be a single 5+1D Dirac fermion in the $\overline{7}$ representation and another in the $21$ representation to embed each generation of the SM fermions. In an $SU(7)$ theory in 3+1D with chiral fermions, $\overline{7}+21$ is anomalous and the minimal anomaly-free combination is in fact a left-chiral fermion in the $21$ representation along with \emph{three} transforming as a $\overline{7}$. This phenomenon of an anomalous lower dimensional theory reproduced from an anomaly-free one in higher dimensions has been noted previously \cite{daemishaposhnikov, anomaliesed} and in some cases the anomalies of the lower dimensional theory have been shown to be canceled by effects coming from the bulk \cite{daemishaposhnikov}. It is not clear to us if this is the case in our model and that bulk effects will protect our 3+1D theory from anomalies if we simply choose a single $\overline{7}$ Dirac fermion and a $21$ Dirac fermion in 5+1D for each SM generation. Nevertheless, we can always make the safe choice and include the full $\overline{7}+\overline{7}+\overline{7}+21$ combination for our initial 5+1D Dirac fermion content. Alternatively, there is the next-to-minimal choice $\overline{7}+21+\overline{35}$.

 With regards to the Higgs sector, we not only need a Higgs field in which to embed the electroweak Higgs doublet but we also need to include the requisite Higgs fields to break the semi-delocalized Abelian groups generated by $A$ and $B$. Both the Abelian groups $U(1)_{A}$ and $U(1)_{B}$ must be broken without breaking $U(1)_{Y}$ so the required Higgs fields must contain components which transform as singlets under the Standard Model but are charged under the semi-delocalized Abelian groups. The obvious candidates are the $21$ and the $35$ since the $21$ contains a component transforming as $(1, 1, 0, -40, +30)$ and the $35$ contains a component transforming as $(1, 1, 0, -24, -9)$. Furthermore, it is obvious that these two different components will completely break $U(1)_{A}\times{}U(1)_{B}$ since each component will preserve different linear combinations of $A$ and $B$ after attaining a VEV. Hence the $21+35$ combination will do the job. For embedding the electroweak Higgs, one might first consider the anti-fundamental $\overline{7}$. While a scalar transforming as a $\overline{7}$ can form a gauge invariant Yukawa coupling with fermion in the $\overline{7}$ and another in the $21$, it cannot form a Yukawa coupling with the $21$ fermion and its charge conjugate, which we need to get a mass matrix for the up-like quarks. Neither can a scalar in the $21$ representation, even though it contains a component transforming as an electroweak doublet. On the other hand, the $\overline{35}$ representation can both form a Yukawa coupling between a $\overline{7}$ and a $21$ as well as a gauge invariant Yukawa coupling between the $21$ and its charge conjugate. Further, the $\overline{35}$ contains a component transforming as an electroweak doublet, the $(1, 2, -1, +39, -12)$ component, and thus it is necessary to embed the electroweak Higgs in this representation. Although a phenomenological analysis of the fermion and scalar modes is beyond the scope of this paper, it would be interesting to see if we can embed the electroweak Higgs field along with the SM singlet required to break one of the semi-delocalized Abelian groups in the same $\overline{35}$ scalar and choose parameters such that these components attain tachyonic masses on the intersection while all other components attain positive definite squared masses.

 To ensure that we get this configuration, we need to ensure that it is the most energetically favorable and stable one. The most general $\mathbb{Z}_2\times{}\mathbb{Z}_2$-invariant quartic potential for $\eta_{1}$, $\chi_{1}$, $\eta_{2}$, and $\chi_{2}$ with $\chi_{1}$ and $\chi_{2}$ as adjoint scalar fields is
 \begin{equation}
 \label{eq:adjointadjointscalarpot}
 V = V_{\eta_{1}\chi_{1}}+V_{\eta_{2}\chi_{2}}+V_{\eta_{1}\chi_{1}\eta_{2}\chi_{2}},
 \end{equation}
where $V_{\eta_{i}\chi_{i}}$ for $i=1,2$ are the self interaction potentials 
\begin{equation}
\begin{aligned}
\label{eq:dvalishifmanmodeletaichii}
V_{\eta_{i}\chi_{i}} &= \frac{1}{4}\lambda_{\eta_{i}}(\eta_{i}^2-v_{i}^2)^2+\lambda_{\eta_{i}\chi_{i}}(\eta_{i}^2-v_{i}^2)\rm{Tr}[\chi_{i}^2]+\mu^2_{\chi_{i}}\rm{Tr}[\chi_{i}^2]+h_{\eta_{i}\chi_{i}}\eta_{i}\rm{Tr}[\chi^{3}_{i}] \\
              &+\lambda^{1}_{\chi_{i}}\rm{Tr}[\chi_{i}^2]^2+\lambda^{2}_{\chi_{i}}\rm{Tr}[\chi_{i}^4],
\end{aligned}
\end{equation}
for the $\eta_{1}$-$\chi_{1}$ and $\eta_{2}$-$\chi_{2}$ sectors respectively and $V_{\eta_{1}\chi_{1}\eta_{2}\chi_{2}}$ is the interaction potential between these two sectors, which may be written
\begin{equation}
\begin{aligned}
\label{eq:eta1chi1eta2chi2intpotadjointcase}
V_{\eta_{1}\chi_{1}\eta_{2}\chi_{2}} &= \frac{1}{2}\lambda_{\eta_{1}\eta_{2}}(\eta_{1}^2-v_{1}^2)(\eta_{2}^2-v_{2}^2)+\lambda_{\eta_{1}\chi_{2}}(\eta_{1}^2-v_{1}^2)\rm{Tr}[\chi_{2}^2]+\lambda_{\chi_{1}\eta_{2}}(\eta_{2}^2-v_{2}^2)\rm{Tr}[\chi_{1}^2] \\
 &+2\lambda^{1}_{\chi_{1}\chi_{2}}\rm{Tr}[\chi_{1}^2]\rm{Tr}[\chi_{2}^2]+2\lambda^{2}_{\chi_{1}\chi_{2}}\big[\rm{Tr}[\chi_{1}\chi_{2}]\big]^2+2\lambda^{3}_{\chi_{1}\chi_{2}}\rm{Tr}[\chi_{1}^2\chi_{2}^2]+2\lambda^{4}_{\chi_{1}\chi_{2}}\rm{Tr}[\chi_{1}\chi_{2}\chi_{1}\chi_{2}] \\
 &+\lambda_{\eta_{1}\chi_{1}\chi_{2}}\eta_{1}\rm{Tr}[\chi_{1}\chi^{2}_{2}]+\lambda_{\chi_{1}\eta_{2}\chi_{2}}\eta_{2}\rm{Tr}[\chi^{2}_{1}\chi_{2}]+\lambda_{\eta_{1}\chi_{1}\eta_{2}\chi_{2}}\eta_{1}\eta_{2}\rm{Tr}[\chi_{1}\chi_{2}].
\end{aligned}
\end{equation}
 
 Firstly, we need to ensure that the configurations on the boundary leading to the desired subgroups being respected on each wall are the most stable. This involves analyzing the respective one-dimensional kink-lump solutions which we utilize as the boundary conditions generated by the self-interaction potentials $V_{\eta_{i}\chi_{i}}$ given in Eq.~\ref{eq:dvalishifmanmodeletaichii}. At the boundaries, we obviously set $\eta_{i}(y_{i}\rightarrow{}\pm{}\infty) = \pm{}v_{i}$ (here, $y_{1} = y$, $y_{2} = z$), and here the corresponding $\chi_{i}$ must be zero since here it experiences a potential bounded from below with a positive definite squared mass. At some point, on the wall where $\eta_{i}$ traverses from one vacuum to the other, $\eta_{i}=0$ and here the squared mass of $\chi_{i}$ becomes tachyonic and is thus expected to condense. In this region, $\chi_{i}$ experiences a standard quartic symmetry-breaking potential for an adjoint scalar. In generating analytic solutions, we normally set the coupling constant for the $\eta_{i}\rm{Tr}[\chi^{3}_{i}]$ term to zero (and other terms involving odd powers of $\eta_{i}$ and $\chi_{i}$ in the full potential for similar reasons). This means that the resultant effective quartic potential experienced by $\chi_{i}$ in the region where it is tachyonic has a $\mathbb{Z}_{2}$ symmetry, with its breaking patterns determined by Li \cite{lingfongli74}. Since all generators are normalized to $1/2$, the $[\rm{Tr}(\chi^{2}_{i})]^2$ always yields a quartic self-interaction term which has the same strength no matter which breaking pattern is chosen. On the other hand the value of $\rm{Tr}[\chi^{4}_{i}]$ differs depending on the VEV pattern chosen. Hence, the real components of $\chi_{i}$ corresponding to different symmetry breaking patterns experience different effective quartic self-couplings, which will be linear combinations of $\lambda^{1}_{\chi_{i}}$ and $\lambda^{2}_{\chi_{i}}$. If we write the effective $\lambda_{\chi_{i}eff}$ coupling constants for these different components with the normalization given in Eq.~\ref{eq:scalarpotential} in terms of $\lambda^{1}_{\chi_{i}}$ and $\lambda^{2}_{\chi_{i}}$, then for an $SU(6)\times{}U(1)$ breaking pattern the effective coupling is $\lambda^{1}_{\chi_{i}}+31\lambda^{2}_{\chi_{i}}/42$, for an $SU(5)\times{}SU(2)\times{}U(1)$ breaking pattern it is $\lambda^{1}_{\chi_{i}}+19\lambda^{2}_{\chi_{i}}/70$ and for $SU(4)\times{}SU(3)\times{}U(1)$ it is $\lambda^{1}_{\chi_{i}}+13\lambda^{2}_{\chi_{i}}/84$. Since the energy of the effective potential for $\chi_{i}$ at the respective vacuum is $-\mu^{4}_{\chi_{i}}/4\lambda_{\chi_{i}eff}$, the configuration with the lowest effective quartic coupling will have the lowest energy and consequently the most stable vacuum. Thus, of the three breaking patterns, for $\lambda^{2}_{\chi_{i}}>0$ the $SU(4)\times{}SU(3)\times{}U(1)$ is the most stable and for $\lambda^{2}_{\chi_{i}}<0$ the $SU(6)\times{}U(1)$ vacuum is the most stable (provided $\lambda^{1}_{\chi_{i}}+\lambda^{2}_{\chi_{i}}>0$ to ensure the potential is bound from below); these results agree with Ref.~\cite{lingfongli74}. 
 
 The energy of the effective potential for an $SU(5)\times{}SU(2)\times{}U(1)$ symmetry breaking pattern thus always lies in between that for the $SU(6)\times{}U(1)$ and $SU(4)\times{}SU(3)\times{}U(1)$ symmetry breaking patterns in the case that the effective potential for $\chi_{i}$ has the $\mathbb{Z}_{2}$ symmetry (not the ones initially imposed). This means that for a quartic potential with the $\eta_{1}\chi^{3}_{1}$ term set to zero, that the configuration where the component of $\chi_{1}$ which is proportional to $Q_{1}$ condenses to form the lump is never the most stable one. We can ensure that the $SU(4)\times{}SU(3)\times{}U(1)$ breaking pattern is the most stable one in the $\eta_{2}-\chi_{2}$ sector, but we need some way to ensure that $SU(5)\times{}SU(2)\times{}U(1)$ breaking kink-lump configuration is the most stable one for the $\eta_{1}-\chi_{1}$ sector if we are to generate the desired outcome with a localized Standard Model outlined in this section. 
 
  There are several ways around the problem in the previously mentioned paragraph. One might first think that one of these would be to switch on the $\eta_{1}\rm{Tr}[\chi^{3}_{1}]$ term. However, there will still be a point where $\eta_{1}=0$ and thus around this point one of the other breaking patterns would still be expected to be more stable. Furthermore, this term will affect the localization of the lump and once again the effective coupling for this interaction is largest in magnitude for the $SU(6)\times{}U(1)$ breaking pattern followed by that for $SU(5)\times{}SU(2)\times{}U(1)$ followed again by $SU(4)\times{}SU(3)\times{}U(1)$. By examining this term, we have noticed that it generally lowers the energy density of the solutions and the more the lump is delocalized from the center of the kink, the more the energy density lowers. Thus, one initially thinks that there may be a way to make the $SU(5)\times{}SU(2)\times{}U(1)$ preserving configuration the lowest in energy. The magnitude of its effective coupling constant for this term is greater than the $SU(4)\times{}SU(3)\times{}U(1)$ one, so if we choose $\lambda^{2}_{\chi_{1}}>0$ initially and then slowly increase $h_{\eta_{1}\chi_{1}}$ from zero, one may expect the energy density of the $SU(5)\times{}SU(2)\times{}U(1)$ to become lower. Unfortunately, in the exploration of the parameter space that we have done, it seems that the energy density of the $SU(6)\times{}U(1)$ decreases too rapidly for there to be some point at which $SU(5)\times{}SU(2)\times{}U(1)$ becomes the most stable one. Thus, the $\eta_{1}\rm{Tr}[\chi^{3}_{1}]$ term seems unlikely to solve this problem.
  
   In terms of the cubic invariant, what one would really like is just a bare cubic term of the form $d_{\chi_{1}}\rm{Tr}[\chi^{3}_{1}]$. Let us first mention that in Ref.~\cite{rueggscalarpot}, Ruegg also showed that when $\lambda^{2}_{\chi_{i}}>0$, as the ratio between $d_{\chi_{1}}$ and $\lambda^{2}_{\chi_{i}}$ increases from zero to infinity the most stable breaking pattern cascades from $SU(N-n)\times{}SU(n)\times{}U(1)$, where $n=\lfloor{}N/2\rfloor{}$, to $SU(N-n+1)\times{}SU(n-1)\times{}U(1)$, then to $SU(N-n+2)\times{}SU(n-2)\times{}U(1)$ and so on up to $SU(N-1)\times{}U(1)$. Hence, in the case of $SU(7)$, there would exist a parameter region where the configuration breaking to $SU(5)\times{}SU(2)\times{}U(1)$ would become the most stable one if we had a bare cubic term for $\chi_{1}$. The main difficulty would then be ensuring that this cubic term would be allowed, as it is not under the current symmetries and parities imposed in our theory. One could imagine changing the parity of $\chi_{1}$ to $(+, +)$ under the $\mathbb{Z}_2\times{}\mathbb{Z}_2$ symmetry or perhaps utilizing a different discrete symmetry with which to form a domain wall between discrete vacua so that such a cubic term is allowed. Provided $\chi_{1}$ could then be coupled to scalars and fermions in an acceptable way, this would be an ideal approach.

  Another obvious solution is to go to a sextic potential. Resorting to a sextic potential in our extra-dimensional theory is not a problem since any interacting field theory in a spacetime with dimension more than four is non-renormalizable anyway. One of the problems we had was ensuring that there were enough different invariant operators, and hence parameters, for $\chi_{1}$ to permit greater freedom in symmetry breaking outcomes. For the sake of simplicity and as an example, make the quartic self-couplings for $\chi_{1}$ and any sextic term involving $\eta_{1}$ zero, with just the sextic self-couplings for $\chi_{1}$ non-zero. In this case, the effective potential for $\chi_{1}$ where $\eta_{1}=0$ is just a tachyonic mass term with a positive definite sextic term. Just as before with the quartic case, the symmetry breaking pattern will be determined by the effective sextic coupling and the configuration with the lowest effective sextic coupling will be the most stable. Unlike the quartic case, there are more invariants to play with since we can have $\rm{Tr}[\chi^{6}_{1}]$, $(\rm{Tr}[\chi^2_{1}])^3$, $\rm{Tr}[\chi^2_{1}]\rm{Tr}[\chi^4_{1}]$ and $(\rm{Tr}[\chi^3_{1}])^2$. With this number of invariants, one can easily manipulate the parameters such that the $SU(5)\times{}SU(2)\times{}U(1)$ respecting configuration has the lowest effective sextic coupling and thus yields the kink-lump solution where the corresponding component for $Q_{1}$ condenses is the most stable. A potential difficulty with this approach is that the theorem proven by Ruegg \cite{rueggscalarpot} where any extrema and thus minima of the potential for an adjoint scalar exist only if at most two of the eigenvalues of the VEV of the adjoint scalar differ may not apply here since we are dealing with a sextic potential and the aforementioned theorem was only proven for a quartic potential. Thus, with a sextic there may be configurations where the VEV pattern has more than two distinct eigenvalues and one would need to check through these to ensure that the desired configuration is the most stable one. 
  
   Once one has ensured that one wall generating $SU(5)\times{}SU(2)\times{}U(1)$ is stable and has chosen parameters such that the other wall breaks $SU(7)$ to $SU(4)\times{}SU(3)\times{}U(1)$, we need to determine the possible symmetries and localized groups on the intersection under the clash-of-symmetries mechanism. As we stated in the previous section, the most stable clash-of-symmetries arrangement will be the one that minimizes the 3+1D junction energy density. Just as there existed effective quartic self-couplings for the components of $\chi_{1}$ and $\chi_{2}$ chosen to condense after computing the traces of the powers of the respective generators involved, so there will exist other effective coupling constants describing interactions between these different components. In fact, each different configuration will lead to a different effective scalar potential of the form given in Eq.~\ref{eq:scalarpotential}. For the analytic solution given in Eq.~\ref{eq:perpendicularsolution} yielded by the parameter conditions in Eq.~\ref{eq:parameterconditionsforasol}, only the terms in $V_{\eta_{1}\chi_{1}\eta_{2}\chi_{2}}$ contribute to the junction energy density. For parameters not satisfying Eq.~\ref{eq:parameterconditionsforasol}, the self-interaction potentials $V_{\eta_{i}\chi_{i}}$ will in general make a small contribution. Fortunately, there is a way to extract the energy density by defining the fields $\overline{\eta_{1}}$, $\overline{\chi_{1}}$, $\overline{\eta_{2}}$ and $\overline{\chi_{2}}$ as differences between the real two-dimensional interacting kink-lump solutions and the one-dimensional kink-lump solutions which are used as the boundary conditions. In other words, these fields are defined as
   \begin{equation}
   \begin{gathered}
   \label{eq:perturbationsaboutbcs}
   \overline{\eta_{1}}(y, z) = \eta_{1}(y, z)-\eta^{1d}_{1}(y) = \eta_{1}(y, z)-v_{1}\tanh{(ky)}, \\
   \overline{\chi_{1}}(y, z) = \chi_{1}(y, z)-\chi^{1d}_{1}(y) = \chi_{1}(y, z)-A_{1}\sech{(ky)}, \\
   \overline{\eta_{2}}(y, z) = \eta_{2}(y, z)-\eta^{1d}_{2}(z) = \eta_{2}(y, z)-v_{2}\tanh{(lz)}, \\
   \overline{\chi_{2}}(y, z) = \chi_{2}(y, z)-\chi^{1d}_{2}(z) = \chi_{2}(y, z)-A_{2}\sech{(lz)}.
   \end{gathered}
   \end{equation}
  Given the boundary conditions in Eq.~\ref{eq:perpendicularsolboundaryconditions} for the full fields $\eta_{1}$, $\eta_{2}$, $\chi_{1}$ and $\chi_{2}$, one can show that  
  $\overline{\eta_{1}}$, $\overline{\chi_{1}}$, $\overline{\eta_{2}}$ and $\overline{\chi_{2}}$ all vanish along the entire two-dimensional boundary at infinity. Since for a sensible solution the deviations from the 1D solutions should be largest on the intersection with the solutions for $\eta_{1}$, $\eta_{2}$, $\chi_{1}$ and $\chi_{2}$ asymptoting to the 1D solutions out at infinity, it should also be the case that $\overline{\eta_{1}}$, $\overline{\chi_{1}}$, $\overline{\eta_{2}}$ and $\overline{\chi_{2}}$ should decay to zero faster than $1/y$ and $1/z$ in both directions towards infinity. Given this, since $\eta_{1}$, $\eta_{2}$, $\chi_{1}$ and $\chi_{2}$ are all bounded functions, when we expand the potential $V$ in terms of $\overline{\eta_{1}}$, $\overline{\chi_{1}}$, $\overline{\eta_{2}}$ and $\overline{\chi_{2}}$ and $\eta^{1d}_{1}$, $\chi^{1d}_{1}$, $\eta^{1d}_{2}$ and $\chi^{1d}_{2}$, any term proportional to any power of $\overline{\eta_{1}}$, $\overline{\chi_{1}}$, $\overline{\eta_{2}}$ or $\overline{\chi_{2}}$ should be integrable over the $y-z$ plane and should thus give a finite contribution to the junction energy density. 
  
   If we make choices consistent with those of Eq.~\ref{eq:parameterconditionsforasol} and set $\lambda_{\eta_{1}\chi_{1}\chi_{2}}=\lambda_{\chi_{1}\eta_{2}\chi_{2}}=\lambda_{\eta_{1}\chi_{1}\eta_{2}\chi_{2}}=0$, then the most important terms in $V_{\eta_{1}\chi_{1}\eta_{2}\chi_{2}}$ which decide which clash-of-symmetries solutions are most energetically favorable are the quartic couplings between $\chi_{1}$ and $\chi_{2}$ which are $\rm{Tr}[\chi^2_{1}]\rm{Tr}[\chi^2_{2}]$, $\rm{Tr}[\chi_{1}\chi_{2}]^2$, $\rm{Tr}[\chi^2_{1}\chi^2_{2}]$ and $\rm{Tr}[\chi_{1}\chi_{2}\chi_{1}\chi_{2}]$. For a given solution, after we take the relevant traces of these operators, we obtain an effective quartic coupling between the components of $\chi_{1}$ and $\chi_{2}$. After integrating this effective term over the $y-z$ plane, we should obtain its contribution to the junction energy density. Since this effective term is proportional to the squares of the condensing components of $\chi_{1}$ and $\chi_{2}$, if the effective coupling constant for a particular solution is positive, the contribution to the junction energy density will be positive. Furthermore, if we compare it with the contributions coming from the perturbations to the fields as a result of turning on interactions, the former will be proportional to $A^{2}_{1}A^{2}_{2}\sech^{2}(ky)\sech^{2}(lz)$ but the latter will be proportional to say (at first order) $v^3_{1}\tanh{(ky)}\sech^{2}{(ky)}\overline{\eta_{1}}(y,z)$. The $v_{i}$ and $A_{i}$ ($i=1,2$) should be roughly the same order and they will be associated with a high energy scale (typically $\Lambda_{GUT}$) and given we expect the perturbations $\overline{\eta_{1}}$, $\overline{\chi_{1}}$, $\overline{\eta_{2}}$ and $\overline{\chi_{2}}$ to be small, the contribution coming from the background dependent terms arising from the quartic couplings of $V_{\eta_{1}\chi_{1}\eta_{2}\chi_{2}}$ are naturally expected to be one power of this energy scale larger and will dominate the overall contribution to the junction energy density. It then follows that the clash-of-symmetries solution with the lowest effective coupling between the components of $\chi_{1}$ and $\chi_{2}$ which condense will minimize the energy density and thus be the most stable intersecting kink-lump solution. 
   
    We now have to determine the effective quartic couplings between $\chi_{1}$ and $\chi_{2}$ for each of the different clash-of-symmetries solutions. If $\chi_{1a}$ and $\chi_{2b}$ are the components which condense, we define the effective quartic coupling between them to have the same normalization as the $\chi^2_{1}\chi^{2}_{2}$ term in the original intersecting kink-lump model given in Eq.~\ref{eq:scalarpotential}. That is, after computing the relevant traces of the generators in which $\chi_{1}$ and $\chi_{2}$ condense, the effective coupling $\lambda^{eff}_{\chi_{1}\chi_{2}}$ is defined such that the quartic term appears in the effective potential as $\frac{1}{2}\lambda^{eff}_{\chi_{1}\chi_{2}}\chi^{2}_{1a}\chi^{2}_{2b}$. 
    
    There are three possible clash-of-symmetries solutions coming from $H_{1} = SU(5)\times{}SU(2)\times{}U(1)$ and $H_{2} = SU(4)\times{}SU(3)\times{}U(1)$. The other patterns along with the one we have discussed in this section can be found in the appendix. We will label these resultant CoS groups $X_{1} = SU(4)\times{}SU(2)\times{}U(1)\times{}U(1)$, $X_{2} = SU(3)_{c}\times{}SU(2)_{I}\times{}U(1)_{Y}\times{}U(1)\times{}U(1)$ and $X_{3} = SU(3)\times{}SU(2)\times{}SU(2)\times{}U(1)\times{}U(1)$. Obviously, the solution with $X_{2}$ is the one we have discussed and the one we desire to be the most stable. It turns out that the effective $\chi_{1}-\chi_{2}$ couplings for the three breaking patterns are 
    \begin{equation}
    \begin{aligned}
    \label{eq:effectivechi1chi2coupling}
    \lambda^{X_{1}}_{\chi_{1}\chi_{2}} &= \lambda^{1}_{\chi_{1}\chi_{2}}+\frac{1}{15}\lambda^{2}_{\chi_{1}\chi_{2}}+\frac{6}{35}(\lambda^{3}_{\chi_{1}\chi_{2}}+\lambda^{4}_{\chi_{1}\chi_{2}}), \\
    \lambda^{X_{2}}_{\chi_{1}\chi_{2}} &= \lambda^{1}_{\chi_{1}\chi_{2}}+\frac{1}{120}\lambda^{2}_{\chi_{1}\chi_{2}}+\frac{41}{280}(\lambda^{3}_{\chi_{1}\chi_{2}}+\lambda^{4}_{\chi_{1}\chi_{2}}), \\
   \lambda^{X_{3}}_{\chi_{1}\chi_{2}} &= \lambda^{1}_{\chi_{1}\chi_{2}}+\frac{3}{10}\lambda^{2}_{\chi_{1}\chi_{2}}+\frac{407}{5880}(\lambda^{3}_{\chi_{1}\chi_{2}}+\lambda^{4}_{\chi_{1}\chi_{2}}).
    \end{aligned}
    \end{equation}
  From this it follows that the solution generating the Standard Model that we have discussed above has the lowest $\lambda^{eff}_{\chi_{1}\chi_{2}}$ and is thus the most stable CoS solution if the parameter conditions $\lambda^{2}_{\chi_{1}\chi_{2}}>0$ and $-\frac{7}{3}\lambda^{2}_{\chi_{1}\chi_{2}}<\lambda^{3}_{\chi_{1}\chi_{2}}+\lambda^{4}_{\chi_{1}\chi_{2}}<\frac{1715}{454}\lambda^{2}_{\chi_{1}\chi_{2}}$ are imposed. We also impose $\lambda^{1}_{\chi_{1}\chi_{2}}+\lambda^{2}_{\chi_{1}\chi_{2}}+\lambda^{3}_{\chi_{1}\chi_{2}}+\lambda^{4}_{\chi_{1}\chi_{2}}>0$ to ensure that the potential is bounded from below. 
  
  After doing the above analysis, one notices that there is actually another solution to the problem of making the kink-lump generating the $SU(5)\times{}SU(2)\times{}U(1)$ subgroup stable, although it involves a fine-tuning that is not ideal. If we fine-tune the self-coupling $\lambda^{2}_{\chi_{1}}$ to zero, then all three solutions generating the respective subgroups $SU(6)\times{}U(1)$, $SU(5)\times{}SU(2)\times{}U(1)$ and $SU(4)\times{}SU(3)\times{}U(1)$ become degenerate. The other reason this is problematic is that it introduces an accidental $O(48)$ symmetry amongst the components of $\chi_{1}$ in the potential $V_{\eta_{1}\chi_{1}}$ and thus we would naturally expect these solutions to fluctuate. However, the interactions in $V_{\eta_{1}\chi_{1}\eta_{2}\chi_{2}}$ do not respect this $O(48)$ symmetry, breaking it explicitly back to $SU(7)$. The resultant possible solutions then are not only the three with $H_{1} = SU(5)\times{}SU(2)\times{}U(1)$ and $H_{2} = SU(4)\times{}SU(3)\times{}U(1)$, but also those where $H_{1} = SU(6)\times{}U(1)$ and $H_{2} = SU(4)\times{}SU(3)\times{}U(1)$ as well as those coming from $H_{1} = SU(4)\times{}SU(3)\times{}U(1)$ and $H_{2} = SU(4)'\times{}SU(3)'\times{}U(1)'$ (which includes the alternate SM we discuss in the next subsection). In other words, making the fine-tuning $\lambda^{2}_{\chi_{1}}=0$, our desired solution simply has more competitors. Amazingly, when one computes all the effective $\chi_{1}-\chi_{2}$ couplings of the additional CoS solutions, it is still possible to make the solution with the SM discussed in this subsection the most stable one. This is largely due to the very small coefficient in front of the $\lambda^{2}_{\chi_{1}\chi_{2}}$ coupling constant. One finds that the solution discussed in this subsection is still the most stable in this scenario if we tighten the parameter conditions to $\lambda^{2}_{\chi_{1}\chi_{2}}>0$ and $-\frac{7}{3}\lambda^{2}_{\chi_{1}\chi_{2}}<\lambda^{3}_{\chi_{1}\chi_{2}}+\lambda^{4}_{\chi_{1}\chi_{2}}<\frac{98}{383}\lambda^{2}_{\chi_{1}\chi_{2}}$.

  \subsection{A Rather Non-Standard Standard Model from $H_{1} = SU(4)\times{}SU(3)\times{}U(1)$ and $H_{2} = SU(4)'\times{}SU(3)'\times{}U(1)'$}
 \label{subsec:su4su3vsu4su3smmodel}
 
  In the last subsection, we described a scenario which produced a Standard Model-like gauge group with the correct hypercharge quantum numbers for the known SM field content along with some $Y=\pm{}2$ exotics from a clash between $SU(5)\times{}SU(2)\times{}U(1)$ and $SU(4)\times{}SU(3)\times{}U(1)$. As noted above, there are some problems in ensuring that the arrangement where we have an $SU(5)\times{}SU(2)\times{}U(1)$ subgroup as one of the clashing groups is the most stable one for one kink-lump pair. One naturally might then be motivated to consider obtaining a Standard Model-like gauge group from a clash between two differently embedded copies of $SU(4)\times{}SU(3)\times{}U(1)$. Firstly, this has the advantage that we can ensure the most stable arrangement for each kink-lump pair from a one-dimensional point of view is the one generating a $SU(4)\times{}SU(3)\times{}U(1)$ subgroup, since to do this we simply choose $\lambda^{2}_{\chi_{1}}>0$ and $\lambda^{2}_{\chi_{2}}>0$ in each sector. Furthermore, it is obvious that we can obtain the non-Abelian part of the Standard Model gauge group since if we call the second group $H_{2} = SU(4)'\times{}SU(3)'\times{}U(1)'$, we can easily choose the embeddings such that $SU(4)\cap{}SU(4)'\supset{}SU(3)_{c}$ and $SU(3)\cap{}SU(3)'\supset{}SU(2)_{I}$. One also suspects that we can get a localized $U(1)$ in this case since like the case in the previous section, there will be four leftover diagonal generators from all four non-Abelian groups involved in the clash. Indeed, it turns out that this is the case. In this case, we obtain a rather different localized hypercharge generator, one that makes it seem like a successful embedding of the Standard Model fermion content is not possible
  
   To realize the above described situation, we make $\chi_{1}$ condense in a component proportional to the Abelian generator 
 \begin{equation}
  \label{eq:su4su3u1vsu4su3u1firstvevsmclash}
  Q_{1} = \begin{pmatrix} 3 & 0 & 0 & 0 & 0 & 0 & 0 \\
                          0 & 3 & 0 & 0 & 0 & 0 & 0 \\
                          0 & 0 & 3 & 0 & 0 & 0 & 0 \\
                          0 & 0 & 0 & -4 & 0 & 0 & 0 \\
                          0 & 0 & 0 & 0 & -4 & 0 & 0 \\
                          0 & 0 & 0 & 0 & 0 & -4 & 0 \\
                          0 & 0 & 0 & 0 & 0 & 0 & 3 \end{pmatrix},
  \end{equation}
and let the component of $\chi_{2}$ which condenses be proportional to 
 \begin{equation}
  \label{eq:su4su3u1vsu4su3u1secondvevsmclash}
  Q'_{1} = \begin{pmatrix} 3 & 0 & 0 & 0 & 0 & 0 & 0 \\
                          0 & 3 & 0 & 0 & 0 & 0 & 0 \\
                          0 & 0 & 3 & 0 & 0 & 0 & 0 \\
                          0 & 0 & 0 & -4 & 0 & 0 & 0 \\
                          0 & 0 & 0 & 0 & -4 & 0 & 0 \\
                          0 & 0 & 0 & 0 & 0 & 3 & 0 \\
                          0 & 0 & 0 & 0 & 0 & 0 & -4 \end{pmatrix}.
  \end{equation}
  From this we easily see that the groups preserved by the clash are, as noted in the first paragraph of this section, $SU(3)_{c}\subset{}SU(4)\cap{}SU(4)'$ and $SU(2)_{I}\subset{}SU(3)\cap{}SU(3)'$. The leftover generators from $SU(4)$, $SU(3)$ from $H_{1}$ and $SU(4)'$, $SU(3)'$ from $H_{2}$ are respectively
 \begin{equation}
 \label{eq:su4oneclashingu1generator}
 T_{1} = \begin{pmatrix} +\frac{2}{3} & 0 & 0 & 0 & 0 & 0 & 0 \\
                     0 & +\frac{2}{3} & 0 & 0 & 0 & 0 & 0 \\
                     0 & 0 & +\frac{2}{3} & 0 & 0 & 0 & 0 \\
                     0 & 0 & 0 & 0 & 0 & 0 & 0 \\
                     0 & 0 & 0 & 0 & 0 & 0 & 0 \\
                     0 & 0 & 0 & 0 & 0 & 0 & 0 \\
                     0 & 0 & 0 & 0 & 0 & 0 & -2 \end{pmatrix},
 \end{equation}
 \begin{equation}
 \label{eq:su3oneclashingu1generator}
 T_{2} = \begin{pmatrix} 0 & 0 & 0 & 0 & 0 & 0 & 0 \\
                         0 & 0 & 0 & 0 & 0 & 0 & 0 \\
                         0 & 0 & 0 & 0 & 0 & 0 & 0 \\
                         0 & 0 & 0 & +1 & 0 & 0 & 0 \\
                         0 & 0 & 0 & 0 & +1 & 0 & 0 \\
                         0 & 0 & 0 & 0 & 0 & -2 & 0 \\
                         0 & 0 & 0 & 0 & 0 & 0 & 0 \end{pmatrix},
 \end{equation}
 \begin{equation}
 \label{eq:su4twoclashingu1generator}
 T'_{1} = \begin{pmatrix} +\frac{2}{3} & 0 & 0 & 0 & 0 & 0 & 0 \\
                     0 & +\frac{2}{3} & 0 & 0 & 0 & 0 & 0 \\
                     0 & 0 & +\frac{2}{3} & 0 & 0 & 0 & 0 \\
                     0 & 0 & 0 & 0 & 0 & 0 & 0 \\
                     0 & 0 & 0 & 0 & 0 & 0 & 0 \\
                     0 & 0 & 0 & 0 & 0 & -2 & 0 \\
                     0 & 0 & 0 & 0 & 0 & 0 & 0 \end{pmatrix},
 \end{equation}
 and
 \begin{equation}
 \label{eq:su3twoclashingu1generator}
 T'_{2} = \begin{pmatrix} 0 & 0 & 0 & 0 & 0 & 0 & 0 \\
                         0 & 0 & 0 & 0 & 0 & 0 & 0 \\
                         0 & 0 & 0 & 0 & 0 & 0 & 0 \\
                         0 & 0 & 0 & +1 & 0 & 0 & 0 \\
                         0 & 0 & 0 & 0 & +1 & 0 & 0 \\
                         0 & 0 & 0 & 0 & 0 & 0 & 0 \\
                         0 & 0 & 0 & 0 & 0 & 0 & -2 \end{pmatrix}.
 \end{equation}
 Again there is an Abelian generator surviving the clash which is solely a linear combination of the above four generators and thus satisfies the localization condition described in Eq.~\ref{eq:abeliangeneratorlocalizationcondition}, namely
 \begin{equation}
\label{eq:localizedhyperchargegeneratorsu4su3u1vsu4su3u1}
Y' = -T_{1}-T_{2} = -T'_{1}-T'_{2} = \begin{pmatrix} -\frac{2}{3} & 0 & 0 & 0 & 0 & 0 & 0 \\
                                 0 & -\frac{2}{3} & 0 & 0 & 0 & 0 & 0 \\
                                 0 & 0 & -\frac{2}{3} & 0 & 0 & 0 & 0 \\
                                 0 & 0 & 0 & -1 & 0 & 0 & 0 \\
                                 0 & 0 & 0 & 0 & -1 & 0 & 0 \\
                                 0 & 0 & 0 & 0 & 0 & +2 & 0 \\
                                 0 & 0 & 0 & 0 & 0 & 0 & +2 \end{pmatrix}.
\end{equation}

 Again, we also get a couple of semi-delocalized $U(1)$ gauge groups. In this case, the semi-delocalized generators $A$ and $B$ may be taken to be
 \begin{equation}
\label{eq:firstsemidelocU1Y'SM}
A = 4Q_{1}+T_{1}-T_{2} = 2Q'_{1}+10T'_{1}-9T'_{2} =   \begin{pmatrix} \frac{38}{3} & 0 & 0 & 0 & 0 & 0 & 0 \\
                                                                       0 & \frac{38}{3} & 0 & 0 & 0 & 0 & 0 \\
                                                                       0 & 0 & \frac{38}{3} & 0 & 0 & 0 & 0 \\
                                                                       0 & 0 & 0 & -17 & 0 & 0 & 0 \\
                                                                       0 & 0 & 0 & 0 & -17 & 0 & 0 \\
                                                                       0 & 0 & 0 & 0 & 0 & -14 & 0 \\
                                                                       0 & 0 & 0 & 0 & 0 & 0 & 10 \end{pmatrix},
 \end{equation}
 and 
 \begin{equation}
\label{eq:firstsemidelocU1Y'SM}
B = Q_{1}-2T_{1}+2T_{2} = -\frac{1}{2}Q'_{1}+\frac{29}{4}T'_{1}-6T'_{2} = \begin{pmatrix}  \frac{5}{3} & 0 & 0 & 0 & 0 & 0 & 0 \\
                                                                              0 & \frac{5}{3} & 0 & 0 & 0 & 0 & 0 \\
                                                                              0 & 0 & \frac{5}{3} & 0 & 0 & 0 & 0 \\
                                                                              0 & 0 & 0 & -2 & 0 & 0 & 0 \\
                                                                              0 & 0 & 0 & 0 & -2 & 0 & 0 \\
                                                                              0 & 0 & 0 & 0 & 0 & -8 & 0 \\
                                                                              0 & 0 & 0 & 0 & 0 & 0 & 7 \end{pmatrix}.
 \end{equation}

 Thus, we have a localized hypercharge generator with a relative sign between the charge for the lepton doublet and the charge for the conjugate of the right-chiral down quark which is opposite that of the usual $SU(5)$ hypercharge generator. It seems that it would be extremely difficult to pick representations containing the SM field content in a simple way, since the charges for the components in the antisymmetric $21$ representation would also be affected, which is problematic since the $21$ is the natural candidate for embedding the right-chiral up quark, right-chiral electron and the quark doublet. For instance, instead of having a hypercharge $Y=+1/3$, the component inside the $21$ that transforms as $(3, 2)$ under $SU(3)_{c}\times{}SU(2)_{I}$ now has $Y=-5/3$. This rules out using the minimal anomaly-free fermion combination of $\overline{7}+\overline{7}+\overline{7}+21$ to embed each generation of the Standard Model fermions. However, it in fact turns out that the SM fermion content can be embedded in the next-to-minimal anomaly-free fermion combination of $7+\overline{21}+35$. Under $SU(3)_{c}\times{}SU(2)_{I}\times{}U(1)_{Y'}\times{}U(1)_{A}\times{}U(1)_{B}$, the $SU(7)$ representations break down as
 \begin{equation}
 \label{eq:7Y'SMcontent}
 7 = (3, 1, -\frac{2}{3}, +\frac{38}{3}, +\frac{5}{3})+(1, 2, -1, -17, -2)+(1, 1, +2, -14, -8)+(1, 1, +2, +10, +7),
 \end{equation}
 \begin{equation}
 \begin{aligned}
 \label{eq:21Y'SMcontent}
 \overline{21} &= (3, 1, +\frac{4}{3}, -\frac{76}{3}, -\frac{10}{3})+(\overline{3}, 2, +\frac{5}{3}, +\frac{13}{3}, +\frac{1}{3})+(1, 1, +2, +34, +4)+(\overline{3}, 1, -\frac{4}{3}, +\frac{4}{3}, +\frac{19}{3}) \\
               &+(\overline{3}, 1, -\frac{4}{3}, -\frac{68}{3}, -\frac{26}{3})+(1, 2, -1, +31, +10)+(1, 2, -1, +7, -5)+(1, 1, -4, +4, +1),
 \end{aligned}
 \end{equation}
 \begin{equation}
 \begin{aligned}
 \label{eq:35Y'SMcontent}
 35 &= (3, 1, -\frac{8}{3}, -\frac{64}{3}, -\frac{7}{3})+(\overline{3}, 2, -\frac{7}{3}, +\frac{25}{3}, +\frac{4}{3})+(1, 1, -2, +38, +5)+(\overline{3}, 1, +\frac{2}{3}, +\frac{34}{3}, -\frac{14}{3}) \\
    &+(\overline{3}, 1, +\frac{2}{3}, +\frac{106}{3}, +\frac{31}{3})+(3, 2, +\frac{1}{3}, -\frac{55}{3}, -\frac{25}{3})+(3, 2, +\frac{1}{3}, +\frac{17}{3}, +\frac{20}{3})+(3, 1, +\frac{10}{3}, +\frac{26}{3}, +\frac{2}{3}) \\
    &+(1, 2, +3, -21, -3)+(1, 1, 0, -48, -12)+(1, 1, 0, -24, +3).
 \end{aligned}
 \end{equation}
 Hence, if we choose the couplings to the background scalar fields such that each of the fermion fields charged under these representations has a localized left-chiral zero mode, both the lepton doublet $L$ and the charge conjugate of the right-chiral electron $(e_{R})^{c}$ can be embedded in either the $7$ or the $\overline{21}$, the charge conjugate of the right-chiral up quark $(u_{R})^{c}$ can be embedded in the $\overline{21}$ and the quark doublet $Q$ can be embedded in the $35$. In choosing the representations in this way, the charge conjugate of the right-chiral down quark, $(d_{R})^{c}$, must be embedded in the $35$. We can even fit in the charge conjugate of the right-chiral neutrino as the $35$ contains two singlet representations. In fact, we can fit in two generations of quarks and 3 generations of charged leptons along with two right-chiral neutrinos. 
 
 The electroweak Higgs could fit into either a $7$ or a $21$. However, given both $Q$ and $(d_{R})^{c}$ are embedded in a $35$, to form a down-quark mass matrix we need an invariant between a Higgs field and the Dirac bilinear formed from a fermion field in the $35$ representation and its charge conjugate. The only choice that can do the job is a $7$ since the tensor product $35\times{}35$ contains a $\overline{7}$ but not a $\overline{21}$. Since the tensor products $7\times{}7\times{}\overline{21}$ and $\overline{7}\times{}\overline{21}\times{}35$ contain singlets, we can form mass matrices for the charged leptons and the up-type quarks with the electroweak Higgs in a $7$. With regards to breaking the semi-delocalized photons, we can utilize the $(1, 1, 0, -48, -12)$ and $(1, 1, 0, -24, +3)$ of the $35$. It would be interesting to see whether we could use both these components from the one $35$ and choose parameters such that both these components attain tachyonic masses. Otherwise, we can use two $35$'s. From there, like with the previous realization of the SM, the main task is to ensure that the profiles for the scalars and fermions are split appropriately so that the exotic states, other extra states and the semi-delocalized photons become sufficiently massive. Like before, we also need to make sure that there are no unwanted breakings coming from additional localized Higgs components.

  Lastly, we need to check that we can make the aforementioned CoS solution the most stable one. As in the previous section, the relevant operators are $\rm{Tr}[\chi^2_{1}]\rm{Tr}[\chi^2_{2}]$, $\rm{Tr}[\chi_{1}\chi_{2}]^2$, $\rm{Tr}[\chi^2_{1}\chi^2_{2}]$ and $\rm{Tr}[\chi_{1}\chi_{2}\chi_{1}\chi_{2}]$ and we need to take the relevant traces to compute $\lambda_{\chi_{1}\chi_{2}eff}$ for each different solution. There are three other clash-of-symmetries breaking patterns, the VEV patterns for which are listed in Appendix \ref{appendix:allpossiblecosgroups}, along with the one we have described. These other solutions break $SU(7)$ down to $W_{1} = SU(4)\times{}SU(3)\times{}U(1)$, $W_{2} = SU(2)\times{}SU(2)\times{}SU(2)\times{}U(1)\times{}U(1)\times{}U(1)$ and $W_{3} = SU(3)\times{}SU(3)\times{}U(1)\times{}U(1)$. After taking the relevant traces, it turns out that the effective coupling constants are in this case
\begin{equation}  
\begin{aligned}
\label{eq:lambdaeta1eta2effsu4su3vsu4su3prime}
\lambda^{SM\times{}U(1)^2}_{\chi_{1}\chi_{2}eff} &= \lambda^{1}_{\chi_{1}\chi_{2}}+\frac{25}{576}\lambda^{2}_{\chi_{1}\chi_{2}}+\frac{149}{1008}(\lambda^{3}_{\chi_{1}\chi_{2}}+\lambda^{4}_{\chi_{1}\chi_{2}}), \\
\lambda^{W_{1}}_{\chi_{1}\chi_{2}eff} &= \lambda^{1}_{\chi_{1}\chi_{2}}+\lambda^{2}_{\chi_{1}\chi_{2}}+\frac{13}{84}(\lambda^{3}_{\chi_{1}\chi_{2}}+\lambda^{4}_{\chi_{1}\chi_{2}}), \\
\lambda^{W_{2}}_{\chi_{1}\chi_{2}eff} &= \lambda^{1}_{\chi_{1}\chi_{2}}+\frac{1}{36}\lambda^{2}_{\chi_{1}\chi_{2}}+\frac{71}{1008}(\lambda^{3}_{\chi_{1}\chi_{2}}+\lambda^{4}_{\chi_{1}\chi_{2}}), \\
\lambda^{W_{3}}_{\chi_{1}\chi_{2}eff} &= \lambda^{1}_{\chi_{1}\chi_{2}}+\frac{9}{16}\lambda^{2}_{\chi_{1}\chi_{2}}+\frac{15}{112}(\lambda^{3}_{\chi_{1}\chi_{2}}+\lambda^{4}_{\chi_{1}\chi_{2}}).
\end{aligned}
\end{equation}

Again we can easily choose parameters such that $\lambda^{SM\times{}U(1)^2}_{\chi_{1}\chi_{2}eff}$ is the smallest of the effective couplings, rendering the arrangement we have described above the most stable. In fact, one can show that $\lambda^{SM\times{}U(1)^2}_{\chi_{1}\chi_{2}eff}$ is smaller than all of $\lambda^{W_{1}}_{\chi_{1}\chi_{2}eff}$, $\lambda^{W_{2}}_{\chi_{1}\chi_{2}eff}$ and $\lambda^{W_{3}}_{\chi_{1}\chi_{2}eff}$ if we choose parameters such that $\lambda^{2}_{\chi_{1}\chi_{2}}>0$ and $-551\lambda^{2}_{\chi_{1}\chi_{2}}/4<\lambda^{3}_{\chi_{1}\chi_{2}}+\lambda^{4}_{\chi_{1}\chi_{2}}<-21\lambda^{2}_{\chi_{1}\chi_{2}}/104$.

 \subsection{The GUT Approach: A Localized $SU(5)$ theory from $H_{1} = SU(6)\times{}U(1)$ and $H_{2} = SU(6)\times{}U(1)$}
 \label{subsec:su6vsu6su5model}
 
  We can also take the approach of Ref.~\cite{e6domainwallpaper} and localize a grand unification group. If we choose our clashing subgroups to be differently embedded copies of $SU(6)\times{}U(1)$, then it is clear that we can obtain a localized $SU(5)$ subgroup. Again, from what we know from Ref.~\cite{lingfongli74}, if we choose $\lambda^{2}_{\chi_{1}}<0$ and $\lambda^{2}_{\chi_{2}}<0$ then an $SU(6)\times{}U(1)$ breaking pattern will be the most stable 1D kink-lump configuration for each sector, provided we also choose parameters such that $\lambda^{1}_{\chi_{1}}+\lambda^{2}_{\chi_{1}}>0$ and $\lambda^{1}_{\chi_{2}}+\lambda^{2}_{\chi_{2}}>0$ still hold so that it is absolutely guaranteed that the potentials are bounded from below. This means that the only thing we really need to check is that the arrangement where the clash yields a localized $SU(5)$ subgroup is the most stable arrangement, which in this case just means that it is more stable than the only other arrangement where $H_{1}=H_{2}$ to give a semi-delocalized $SU(6)\times{}U(1)$. 
  
   The VEV pattern we desire is one in which $\chi_{1}$ condenses in the component corresponding to the matrix
   \begin{equation}
  \label{eq:su6u1vsu6u1firstvevsmclash}
  Q_{1} = \begin{pmatrix} 1 & 0 & 0 & 0 & 0 & 0 & 0 \\
                          0 & 1 & 0 & 0 & 0 & 0 & 0 \\
                          0 & 0 & 1 & 0 & 0 & 0 & 0 \\
                          0 & 0 & 0 & 1 & 0 & 0 & 0 \\
                          0 & 0 & 0 & 0 & 1 & 0 & 0 \\
                          0 & 0 & 0 & 0 & 0 & 1 & 0 \\
                          0 & 0 & 0 & 0 & 0 & 0 & -6 \end{pmatrix},
  \end{equation}
 and $\chi_{2}$ condenses in the component corresponding to 
 \begin{equation}
  \label{eq:su6u1vsu6u1secondvevsmclash}
  Q'_{1} = \begin{pmatrix} 1 & 0 & 0 & 0 & 0 & 0 & 0 \\
                          0 & 1 & 0 & 0 & 0 & 0 & 0 \\
                          0 & 0 & 1 & 0 & 0 & 0 & 0 \\
                          0 & 0 & 0 & 1 & 0 & 0 & 0 \\
                          0 & 0 & 0 & 0 & 1 & 0 & 0 \\
                          0 & 0 & 0 & 0 & 0 & -6 & 0 \\
                          0 & 0 & 0 & 0 & 0 & 0 & 1 \end{pmatrix}.
  \end{equation}
 Clearly, $SU(6)\cap{}SU(6)'=SU(5)$. The leftover generators coming from inside the $SU(6)$ and $SU(6)'$ generators are 
 \begin{equation}
  \label{eq:su6u1vsu6u1firstleftover}
  T_{1} = \begin{pmatrix} 1 & 0 & 0 & 0 & 0 & 0 & 0 \\
                          0 & 1 & 0 & 0 & 0 & 0 & 0 \\
                          0 & 0 & 1 & 0 & 0 & 0 & 0 \\
                          0 & 0 & 0 & 1 & 0 & 0 & 0 \\
                          0 & 0 & 0 & 0 & 1 & 0 & 0 \\
                          0 & 0 & 0 & 0 & 0 & -5 & 0 \\
                          0 & 0 & 0 & 0 & 0 & 0 & 0 \end{pmatrix},
  \end{equation}
 and $\chi_{2}$ condenses in the component corresponding to 
 \begin{equation}
  \label{eq:su6u1vsu6u1secondleftover}
  T'_{1} = \begin{pmatrix} 1 & 0 & 0 & 0 & 0 & 0 & 0 \\
                          0 & 1 & 0 & 0 & 0 & 0 & 0 \\
                          0 & 0 & 1 & 0 & 0 & 0 & 0 \\
                          0 & 0 & 0 & 1 & 0 & 0 & 0 \\
                          0 & 0 & 0 & 0 & 1 & 0 & 0 \\
                          0 & 0 & 0 & 0 & 0 & 0 & 0 \\
                          0 & 0 & 0 & 0 & 0 & 0 & -5 \end{pmatrix}.
  \end{equation}
  There are therefore a couple of semi-delocalized $U(1)$ generators which may be taken to be
  \begin{equation}
  \label{eq:su6u1vsu6u1firstcharge}
  q_{1} = 5/6Q_{1}+7/6T_{1} = 5/6Q'_{1}+7/6T'_{1} = \begin{pmatrix} 2 & 0 & 0 & 0 & 0 & 0 & 0 \\
                                                                    0 & 2 & 0 & 0 & 0 & 0 & 0 \\
                                                                    0 & 0 & 2 & 0 & 0 & 0 & 0 \\
                                                                    0 & 0 & 0 & 2 & 0 & 0 & 0 \\
                                                                    0 & 0 & 0 & 0 & 2 & 0 & 0 \\
                                                                    0 & 0 & 0 & 0 & 0 & -5 & 0 \\
                                                                    0 & 0 & 0 & 0 & 0 & 0 & -5 \end{pmatrix},
  \end{equation}
 and
 \begin{equation}
  \label{eq:su6u1vsu6u1secondcharge}
 q_{2} = 1/6(Q_{1}-T_{1}) = 1/6(T'_{1}-Q'_{1}) = \begin{pmatrix} 0 & 0 & 0 & 0 & 0 & 0 & 0 \\
                                                                 0 & 0 & 0 & 0 & 0 & 0 & 0 \\
                                                                 0 & 0 & 0 & 0 & 0 & 0 & 0 \\
                                                                 0 & 0 & 0 & 0 & 0 & 0 & 0 \\
                                                                 0 & 0 & 0 & 0 & 0 & 0 & 0 \\
                                                                 0 & 0 & 0 & 0 & 0 & 1 & 0 \\
                                                                 0 & 0 & 0 & 0 & 0 & 0 & -1 \end{pmatrix}.
  \end{equation}
 Thus the full symmetry respected on the wall is $SU(5)\times{}U(1)_{q_{1}}\times{}U(1)_{q_{2}}$ but only the $SU(5)$ subgroup is fully localized to the junction and just as before in the other cases with adjoint scalars we will have to introduce additional Higgs fields to break the residual Abelian groups. 
 
  To work out if this arrangement is the most stable one, again we just analyze the effective quartic coupling constants coming from the interactions $\rm{Tr}[\chi_{1}^2]\rm{Tr}[\chi_{2}^2]$, $(\rm{Tr}[\chi_{1}\chi_{2}])^2$, $\rm{Tr}[\chi_{1}^2\chi_{2}^2]$ and $\rm{Tr}[\chi_{1}\chi_{2}\chi_{1}\chi_{2}]$. Firstly, note that a pattern generating a clash between identical $SU(6)\times{}U(1)$ subgroups is simply one where both $\chi_{1}$ and $\chi_{2}$ condense in the component proportional to $Q_{1}$ in Eq.~\ref{eq:su6u1vsu6u1firstvevsmclash}. In calculating the relevant traces of the generators involved, we find that the effective quartic coupling $\lambda_{\chi_{1}\chi_{2}eff} = 1/2\lambda^1_{\chi_{1}\chi_{2}}+1/72\lambda^{2}_{\chi_{1}\chi_{2}}+11/504(\lambda^{3}_{\chi_{1}}\chi_{2}+\lambda^{4}_{\chi_{1}\chi_{2}})$ for the $SU(5)\times{}U(1)\times{}U(1)$ breaking pattern and it is $\lambda_{\chi_{1}\chi_{2}eff} = 1/2(\lambda^1_{\chi_{1}\chi_{2}}+\lambda^{2}_{\chi_{1}\chi_{2}})+31/84(\lambda^{3}_{\chi_{1}}\chi_{2}+\lambda^{4}_{\chi_{1}\chi_{2}})$ for the $SU(6)\times{}U(1)$ pattern. Thus there is a very large parameter space where the $SU(5)\times{}U(1)\times{}U(1)$ has the lowest effective quartic coupling given that the coefficients coming from the traces of the $(\rm{Tr}[\chi_{1}\chi_{2}])^2$, $\rm{Tr}[\chi_{1}^2\chi_{2}^2]$ and $\rm{Tr}[\chi_{1}\chi_{2}\chi_{1}\chi_{2}]$ terms are much lower than those for the $SU(6)\times{}U(1)$ pattern. Indeed, one can ensure that $SU(5)\times{}U(1)\times{}U(1)$ has the lowest effective $\chi_{1}-\chi_{2}$ coupling by choosing all of  $\lambda^{2}_{\chi_{1}\chi_{2}}$, $\lambda^{3}_{\chi_{1}\chi_{2}}$ and $\lambda^{4}_{\chi_{1}\chi_{2}}$ to be positive. 
  
   Having now ensured that the desired Clash-of-Symmetries breaking pattern where we have a localized $SU(5)$ subgroup on the domain-wall intersection can be the most stable one, let us comment briefly on how to construct a realistic scenario. We obviously have to break $SU(5)$ on the domain-wall intersection. We do this by introducing another adjoint scalar since under $SU(5)\times{}U(1)_{A}\times{}U(1)_{B}$ the $48$ breaks down as
   \begin{equation}
   \label{eq:adjointsu5u1u1}
   48 = (24, 0, 0)+(5, +7, -1)+(\overline{5}, -7, +1)+(5, +7, +1)+(\overline{5}, -7, -1)+(1, 0, -2)+(1, 0, +2)+(1, 0, 0)+(1, 0, 0),
   \end{equation}
   and subsequently we perform dynamical localization on this additional adjoint scalar field. As usual, each of the different $SU(5)\times{}U(1)_{A}\times{}U(1)_{B}$ components of the $48$ will have their own set of discrete localized modes and continuum modes. To break to the SM, we need the $(24, 0, 0)$ component to have at least one localized mode and we need its lowest energy localized mode to attain a tachyonic mass on the domain-wall intersection. Although doing the exact full analysis is beyond the scope of this paper, it would be interesting to see if we can make the lowest energy localized mode of one of the $(1, 0, -2)$ and $(1, 0, +2)$ components tachyonic simultaneously with that of the $(24, 0, 0)$ component in order to efficiently break one of the semi-delocalized subgroups. 
   
    We need to break both the semi-delocalized $U(1)$ subgroups to produce a phenomenologically acceptable model. As noted above we can break one of them by utilizing some of the components inside the additional adjoint scalar. Under $SU(5)\times{}U(1)_{A}\times{}U(1)_{B}$ symmetry, the $7$, $21$ and $35$ reduce respectively to 
    \begin{equation}
   \label{eq:adjointsu5u1u1}
   7 = (5, +2, 0)+(1, -5, +1)+(1, -5, -1),
   \end{equation}
   \begin{equation}
   \label{eq:adjointsu5u1u1}
   21 = (10, +4, 0)+(5, -3, +1)+(5, -3, -1)+(1, -10, 0),
   \end{equation}
   \begin{equation}
   \label{eq:adjointsu5u1u1}
   35 = (\overline{10}, +6, 0)+(10, -1, +1)+(10, -1, -1)+(5, -8, 0).
   \end{equation}
 Thus, we can utilize the $(1, -5, \pm{}1)$ components inside the $7$ or the $(1, -10, 0)$ components in conjunction with one of the $(1, 0, \pm{}2)$ components inside the $48$ to break both the semi-delocalized Abelian groups. Alternatively, we could use any two $SU(5)$ singlet components which have different non-trivial charges under the Abelian symmetries in any combination of 7's and 21's. 
 
 From the above equations for the representations, we can easily see how to make the exotic and unwanted fermionic states much more massive than the $SU(5)$ states yielding the SM quark and lepton field content. If we choose the standard anomaly-free combination $\overline{7}+\overline{7}+\overline{7}+21$ for each generation, we can see that if we use the combination of a $7$ and a $21$ to break the semi-delocalized $U(1)$ gauge symmetries by giving the respective $(1, -5, +1)$ and $(1, -5, -1)$ components tachyonic masses, the quintets $(\overline{5}, -2, 0)$ from the extra two anti-fundamentals can form singlets with the $(5, -3, +1)$ and $(5, -3, -1)$ components inside the $21$ and thus decouple as heavy fermions.
 
  Finally, one needs to break electroweak symmetry. In principle one could do this with any of the quintets embedded in the $7$, $21$, or $35$ representations. If we embed the usual fermionic quintet in a $7$, we can form the electron and down quark mass matrices with two conjugates of the $(5, +2, 0)$ component and the $(10, +4, 0)$ component in the $21$. On the other hand, we can't use the same quintet to yield the up quark mass matrix: we instead require the $(5, -8, 0)$ component to give the SM fermions inside the $(10, +4, 0)$ component of the $21$ masses. Thus, it seems we require a two-Higgs doublet model in this scenario, along with more singlet Higgs fields than is necessary to break the semi-delocalized $U(1)$'s in order to give the exotic states masses.

 \subsection{An Alternative Path to the Standard Model with $\chi_{1}\sim{}21$ and $\chi_{2}\sim{}35$}
 \label{subsec:2135model}

  Finally, we give an example yielding a Standard Model gauge group where the scalar fields responsible for the breakings on each wall are not in the adjoint representation. Instead, the field $\chi_{1}$ will be chosen to transform under the $21$ representation and $\chi_{2}$ will be chosen to transform under the $35$ representation. With these representations, we can end up with exactly the Standard Model gauge group \emph{without} any semi-delocalized $U(1)$ gauge groups. 
  
  The full scalar potential is 
\begin{equation}
 \label{eq:2135scalarpot}
 V = V_{\eta_{1}\chi_{1}}+V_{\eta_{2}\chi_{2}}+V_{\eta_{1}\chi_{1}\eta_{2}\chi_{2}},
 \end{equation}
 where in this case the self-interaction potentials for each kink-lump generating pair are 
 \begin{equation}
\begin{aligned}
\label{eq:21kinklumpaction}
V_{\eta_{1}\chi_{1}} &= \frac{1}{4}\lambda_{\eta_{1}}(\eta_{1}^2-v_{1}^2)^2+\lambda_{\eta_{1}\chi_{1}}(\eta_{1}^2-v_{1}^2)\chi^{ab}_{1}\chi_{1ba}+\mu^2_{\chi_{1}}\chi^{ab}_{1}\chi_{1ba} \\
              &+\lambda^{1}_{\chi_{1}}[\chi^{ab}_{1}\chi_{1ab}]^{2}+\lambda^{2}_{\chi_{1}}\chi^{ab}_{1}\chi_{1bc}\chi^{cd}_{1}\chi_{1da},
\end{aligned}
\end{equation}
and 
\begin{equation}
\begin{aligned}
\label{eq:35kinklumpaction}
V_{\eta_{2}\chi_{2}} &= \frac{1}{4}\lambda_{\eta_{2}}(\eta_{2}^2-v_{2}^2)^2+\lambda_{\eta_{2}\chi_{2}}(\eta_{2}^2-v_{2}^2)\chi^{abc}_{2}\chi_{2abc}+\mu^2_{\chi_{2}}\chi^{abc}_{2}\chi_{2abc} \\
              &+\lambda^{1}_{\chi_{2}}[\chi^{abc}_{2}\chi_{2abc}]^{2}+\lambda^{2}_{\chi_{2}}\chi^{abc}_{1}\chi_{1bcd}\chi^{def}_{1}\chi_{1efa},
\end{aligned}
\end{equation}
and the interaction potential between the two sectors is
\begin{equation}
\begin{aligned}
\label{eq:eta1chi1eta2chi2intpotadjointcase}
V_{\eta_{1}\chi_{1}\eta_{2}\chi_{2}} &= \frac{1}{2}\lambda_{\eta_{1}\eta_{2}}(\eta_{1}^2-v_{1}^2)(\eta_{2}^2-v_{2}^2)+\lambda_{\eta_{1}\chi_{2}}(\eta_{1}^2-v_{1}^2)\chi^{abc}_{2}\chi_{2abc}+\lambda_{\chi_{1}\eta_{2}}(\eta_{2}^2-v_{2}^2)\chi^{ab}_{1}\chi_{1ab} \\
 &+2\lambda^{1}_{\chi_{1}\chi_{2}}\chi^{ab}_{1}\chi_{1ab}\chi^{cde}_{2}\chi_{2cde}+2\lambda^{2}_{\chi_{1}\chi_{2}}\chi^{ab}_{1}\chi_{1bc}\chi^{cde}_{2}\chi_{2dea}+2\lambda^{3}_{\chi_{1}\chi_{2}}\chi^{ab}_{1}\chi_{2abc}\chi^{cde}_{2}\chi_{1de}\\
 &+\lambda_{\chi_{1}\eta_{2}\chi_{2}}\epsilon^{abcdefg}\chi_{1ab}\chi_{1cd}\chi_{2efg}\eta_{2}+\lambda^{*}_{\chi_{1}\eta_{2}\chi_{2}}\epsilon_{abcdefg}\chi^{ab}_{1}\chi^{cd}_{1}\chi^{efg}_{2}\eta_{2}.
\end{aligned}
\end{equation}

 There are some clear advantages with regards to the energetics by choosing $\chi_{1}\sim{}21$ and $\chi_{2}\sim{}35$. Firstly, the $21$ representation corresponds to a rank 2 antisymmetric tensor. It was shown in Ref.~\cite{lingfongli74} that for a potential just involving a rank 2 antisymmetric $SU(N)$ tensor that for $\lambda^{2}_{\chi_{1}}>0$ the lowest energy breaking pattern was one where a single $2\times{}2$ block of the tensor is non-zero and proportional to the rank 2 alternating tensor while all other components vanish, yielding $SU(N-2)\times{}SU(2)$ as the unbroken subgroup. Thus, if we choose $\lambda^{2}_{\chi_{1}}>0$ then in the region where $\chi_{1}$ is tachyonic it should condense with this pattern and therefore the lowest energy 1D kink-lump solution should have $SU(7)$ broken to $SU(5)\times{}SU(2)$! Thus we have done what we had trouble doing in a simple way with an adjoint scalar in Sec.~\ref{subsec:su5su2u1versussu4su3u1yieldingsm} and ensured that one wall generates the same $SU(5)\times{}SU(2)$ subgroup. Furthermore, as the $21$ is an antisymmetric tensor rather than an adjoint, the $U(1)$ subgroup of $SU(5)\times{}SU(2)\times{}U(1)$ that we got with an adjoint scalar is already broken. 
 
  In a similar way to how the $21$ attains a VEV pattern with one block proportional to the rank 2 alternating tensor $\epsilon_{ij}$, one might think that for a certain region of parameter space that a rank 3 totally antisymmetric tensor such as the $35$ of $SU(7)$ might attain a VEV pattern in which just three indices trace over the elements of the rank 3 alternating tensor $\epsilon_{ijk}$ with all other components zero. If this were the case, since $\epsilon_{ijk}$ is an invariant tensor under $SU(3)$ and the VEV patttern of the $35$ would vanish for the remaining four indices, one would expect the unbroken subgroup would to be $SU(4)\times{}SU(3)$. Although obtaining the canonical form for a rank 3 alternating tensor is a much more non-trivial problem than that for a rank 2 antisymmetric tensor, this was indeed shown to be the case \cite{cumminskingrank3tensor, cumminsrank3tensorcanonicalforms}. Choosing $7\lambda^{1}_{\chi_{2}}+\lambda^{2}_{\chi_{2}}>0$ to ensure boundedness from below, if we choose $\lambda^{2}_{\chi_{1}}>0$ then the $35$ will indeed condense with the aforementioned pattern. In choosing the $35$ we also automatically break the $U(1)$ that usually comes with the $SU(4)\times{}SU(3)$ subgroup if we perform the breaking with an adjoint, which is analogous to how the $21$ breaks the $U(1)$ associated with $SU(5)\times{}SU(2)$. Hence, in choosing $\chi_{1}\sim{}21$ and $\chi_{2}\sim{}35$ we have already broken the semi-delocalized $U(1)$ subgroups that we get when we utilize adjoint scalars. 
  
  The last thing to check is whether we can guarantee that the pattern generating a Standard Model gauge group localized to the intersection is the most stable one. It is obvious that we can generate the same Standard Model gauge group given Sec.~\ref{subsec:su5su2u1versussu4su3u1yieldingsm}. It is this precise SM group since if we choose our VEV's such that $SU(5)\cap{}SU(4)\supset{}SU(3)$ and $SU(5)\cap{}SU(3)\supset{}SU(2)$, we still obtain the same leftover generators from each group given in Eqs.~\ref{eq:su5clashingu1generator}, \ref{eq:su2clashingu1generator}, \ref{eq:su4clashingu1generator} and \ref{eq:su3clashingu1generator}, yielding the same hypercharge generator as in Eq.~\ref{eq:localizedhyperchargegenerator}. To show this outcome can be achieved obviously requires looking at the possible VEV patterns. 
  
   As $\chi_{1}$ is a rank 2 antisymmetric tensor, it will attain a VEV of the form 
   \begin{equation}
   \label{eq:rank2tensorVEV}
   V^{21}_{ab} = A_{1}(\epsilon_{12}\delta^{m}_{a}\delta^{n}_{b}+\epsilon_{21}\delta^{n}_{a}\delta^{m}_{b}),
   \end{equation}
  where here $1\leq{}m<n\leq{}7$ denote some fixed, distinct integers. In a similar manner, $\chi_{2}\sim{}35$ will attain a VEV of the form
  \begin{equation}
   \label{eq:rank3tensorVEV}
   V^{35}_{abc} = A_{2}\big(\epsilon_{123}(\delta^{q}_{a}\delta^{r}_{b}\delta^{s}_{c}+\delta^{s}_{a}\delta^{q}_{b}\delta^{r}_{c}+\delta^{r}_{a}\delta^{s}_{b}\delta^{q}_{c})+\epsilon_{132}(\delta^{q}_{a}\delta^{s}_{b}\delta^{r}_{c}+\delta^{s}_{a}\delta^{r}_{b}\delta^{q}_{c}+\delta^{r}_{a}\delta^{q}_{b}\delta^{s}_{c})\big),
   \end{equation}
  where again $1\leq{}q<r<s\leq{}7$ are fixed, distinct integers. 
  
   Up to rearrangement of the indices and gauge transformations, there are three distinct clashing patterns. The first is where neither of the integers $m$ or $n$ of Eq.~\ref{eq:rank2tensorVEV} are equal to any of the integers $q$, $r$ or $s$ of Eq.~\ref{eq:rank3tensorVEV}. For this first pattern, the $SU(2)$ subgroup preserving the rank 2 alternating tensor of $V^{21}_{ab}$ is outside the $SU(3)$ alternating tensor preserving the rank 3 alternating tensor of the pattern $V^{35}_{abc}$, and thus the unbroken symmetry in the intersection region is $SU(3)\times{}SU(2)\times{}SU(2)$, with only one of the $SU(2)$ subgroups localized and the other $SU(2)$ and the $SU(3)$ semi-delocalized. 
   
   The second pattern is where, without loss of generality, $n=q$ with $m$ not equal to neither of $r$ or $s$. Here, since the two indices $r$ and $s$ overlap with the remaining five indices for which any element of $V^{21}_{ab}$ is zero, the $SU(2)$ subgroup of the $SU(3)$ preserving $V^{35}_{abc}$ is also contained in the $SU(5)$ subgroup preserved by $V^{21}_{ab}$. Also, three of the indices transformed by the $SU(4)$ subgroup left unbroken by $V^{35}_{abc}$ also transform under the $SU(5)$ subgroup left unbroken by $V^{21}_{ab}$. Hence this is the pattern we want, with $SU(3)_{c}\times{}SU(2)_{I}\times{}U(1)_{Y}$ localized to the domain-wall intersection.
   
   The last possible pattern is where, without loss of generality, $m=q$ and $n=r$. Here, the $SU(2)$ subgroup preserving $V^{21}_{ab}$ is also a subgroup of the $SU(3)$ subgroup preserving $V^{35}_{abc}$. Also, the $SU(4)$ subgroup left unbroken by $V^{35}_{abc}$ is also a subgroup of the $SU(5)$ subgroup preserved by $V^{21}_{ab}$. Thus, the group respected on the wall with this pattern is $SU(4)\times{}SU(2)$, with both non-Abelian factor groups semi-delocalized. 
   
   Having outlined the possible groups resulting from the clash-of-symmetries mechanism, we now need to look at the effective couplings between the relevant components of $\chi_{1}$ and $\chi_{2}$ involved in each clash. For simplicity of analysis, we will ignore the $\epsilon^{abcdefg}\chi_{1ab}\chi_{1cd}\chi_{2efg}\eta_{2}$ term and set $\lambda_{\chi_{1}\eta_{2}\chi_{2}}=0$. This leaves as the relevant terms $\chi^{ab}_{1}\chi_{1ba}\chi^{cde}_{2}\chi_{2cde}$, $\chi^{ab}_{1}\chi_{1bc}\chi^{cde}_{2}\chi_{2dea}$, and $\chi^{ab}_{1}\chi_{2abc}\chi^{cde}_{2}\chi_{1de}$. To determine the effective quartic coupling constants, we need to calculate the contractions of the various epsilon tensors involved in the products, which can be though of as products between $V^{21}_{ab}/A_{1}$ and $V^{35}_{abc}/A_{2}$. For $\chi^{ab}_{1}\chi_{1ba}\chi^{cde}_{2}\chi_{2cde}$, the resulting coefficient is always the same, namely we have $\chi^{ab}_{1}\chi_{1ba}\chi^{cde}_{2}\chi_{2cde}\propto{}\epsilon^{ij}\epsilon_{ij}\epsilon^{uvw}\epsilon_{uvw} = 2\times{}6=12$. Hence, the $\chi^{ab}_{1}\chi_{1bc}\chi^{cde}_{2}\chi_{2dea}$ and $\chi^{ab}_{1}\chi_{2abc}\chi^{cde}_{2}\chi_{1de}$ terms are ultimately the ones which determine which clash-of-symmetries group is favored. 
   
   For the $SU(3)\times{}SU(2)\times{}SU(2)$ pattern, $\chi^{ab}_{1}\chi_{1bc}\chi^{cde}_{2}\chi_{2dea}$ and $\chi^{ab}_{1}\chi_{2abc}\chi^{cde}_{2}\chi_{1de}$ both vanish since the rank 2 and rank 3 tensors contained in $V^{21}_{ab}$ and $V^{35}_{abc}$ do not have any indices in common. Therefore, the effective quartic coupling in this situation is simply $\lambda_{\chi_{1}\chi_{2}eff} = \lambda^{1}_{\chi_{1}\chi_{2}}$. 
   
   For the pattern generating a localized $SU(3)_{c}\times{}SU(2)_{I}\times{}U(1)_{Y}$ to the intersection, there is one index in common between the rank 2 alternating tensor from $V^{21}_{ab}$ and the rank 3 alternating tensor from $V^{35}_{abc}$. This means that $\chi^{ab}_{1}\chi_{2abc}\chi^{cde}_{2}\chi_{1de}$ must vanish because it involves a contraction between $V^{21}_{ab}$ and $V^{35}_{abc}$ over two indices rather than just one. On the other hand, since $\epsilon^{ij}\epsilon_{jk} = \delta^{i}_{k}$ and $V^{21ab}V^{21}_{bc}\propto{}\delta^{a}_{m}\delta^{m}_{c}+\delta^{a}_{n}\delta^{n}_{c}$, we have
\begin{equation}
  \begin{aligned}
  \label{eq:SMpatterncoefficient1}
  \chi^{ab}_{1}\chi_{1bc}\chi^{cde}_{2}\chi_{2dea} &\propto{} (\delta^{a}_{m}\delta^{m}_{c}+\delta^{a}_{n}\delta^{n}_{c})V^{35cde}V^{35}_{dea} \\
                                                   &= V^{35mde}V^{35}_{dem}+V^{35nde}V^{35}_{den} \\
                                                   &= 0+2V^{35qrs}V^{35}_{rsq} \\
                                                   &= 2V^{35qrs}V^{35}_{qrs} \\
                                                   &\propto{} 2\epsilon^{ijk}\epsilon_{ijk} \\
                                                   &= 12.
  \end{aligned}
  \end{equation}
  Thus for the pattern we want, $\lambda_{\chi_{1}\chi_{2}eff} = \lambda^{1}_{\chi_{1}\chi_{2}}+\lambda^{2}_{\chi_{1}\chi_{2}}$.
 For the $SU(4)\times{}SU(2)$ pattern, both indices of the rank 2 alternating tensor in $V^{21}_{ab}$ coincide with indices of the rank 3 alternating tensor in $V^{35}_{abc}$. Thus, in this case, both the non-trivial quartic coupling terms are non-vanishing. For $\chi^{ab}_{1}\chi_{1bc}\chi^{cde}_{2}\chi_{2dea}$ we have 
  \begin{equation}
  \begin{aligned}
  \label{eq:su4su2patterncoefficient1}
  \chi^{ab}_{1}\chi_{1bc}\chi^{cde}_{2}\chi_{2dea} &\propto{} (\delta^{a}_{m}\delta^{m}_{c}+\delta^{a}_{n}\delta^{n}_{c})V^{35cde}V^{35}_{dea} \\
                                                   &= V^{35mde}V^{35}_{dem}+V^{35nde}V^{35}_{den} \\
                                                   &= 2V^{35qrs}V^{35}_{rsq}+2V^{35rsq}V^{35}_{sqr} \\
                                                   &= 4V^{35qrs}V^{35}_{qrs} \\
                                                   &\propto{} 4\epsilon^{ijk}\epsilon_{ijk} \\
                                                   &= 24.
  \end{aligned}
  \end{equation}
  Given $V^{21}_{ab}V^{35cab}\propto{}\delta^{m}_{a}\delta^{n}_{b}V^{35cab}-\delta^{n}_{a}\delta^{m}_{b}V^{35cab}=2V^{35cmn}$ and $V^{21ab}V^{35}_{abc}\propto{}-\delta^{a}_{m}\delta^{b}_{n}V^{35}_{abc}+\delta^{a}_{n}\delta^{b}_{m}V^{35abc}=-2V^{35}_{mnc}$, we have
  \begin{equation}
  \begin{aligned}
  \label{eq:su4su2patterncoefficient2}
  \chi^{ab}_{1}\chi_{2abc}\chi^{cde}_{2}\chi_{1de} &\propto{} -4V^{35}_{mnc}V^{35cmn} \\
                                                   &= -4V^{35}_{mnc}V^{35mnc} \\
                                                   &\propto{} -4\epsilon_{ijk}\epsilon^{ijk} \\
                                                   &=-24.
  \end{aligned}
  \end{equation}
  Thus for the $SU(4)\times{}SU(2)$ pattern, $\lambda_{\chi_{1}\chi_{2}eff} = \lambda^{1}_{\chi_{1}\chi_{2}}+2\lambda^{2}_{\chi_{1}\chi_{2}}-2\lambda^{3}_{\chi_{1}\chi_{2}}$.
  
  We can easily choose parameters such that the pattern yielding the localized Standard Model has the lowest effective $\chi_{1}-\chi_{2}$ coupling and is thus the most stable solution. One can easily see by inspection that choosing $\lambda^{1}_{\chi_{1}\chi_{2}}>0$, $\lambda^{2}_{\chi_{1}\chi_{2}}<0$, $\lambda^{1}_{\chi_{1}\chi_{2}}+\lambda^{2}_{\chi_{1}\chi_{2}}>0$ and $\lambda^{2}_{\chi_{1}\chi_{2}}-\lambda^{3}_{\chi_{1}\chi_{2}}>0$ that $\lambda_{\chi_{1}\chi_{2}eff}$ will be positive for all three patterns and will always be lowest for the $SU(3)_{c}\times{}SU(2)_{I}\times{}U(1)_{Y}$ pattern and highest for the $SU(4)\times{}SU(2)$ pattern.
  
  We have successfully shown that an intersecting kink-lump solution with $\chi_{1}\sim{}21$ and $\chi_{2}\sim{}35$ yields a subgroup localized to the domain-wall intersection which is precisely the Standard Model gauge group with no other localized or semi-delocalized gauge symmetries respected there. Furthermore, we have shown that this solution can be the most stable one possible. From here, aside from the semi-delocalized $U(1)$'s which are already broken in this case, we face many of the same challenges as with the models produced from adjoint scalars. We need to localize the requisite Higgs fields to the intersection with tachyonic masses and we need to ensure that other unwanted components have positive definite squared masses. As the Standard Model produced here is equivalent to the one produced with two adjoint scalars in Sec.~\ref{subsec:su5su2u1versussu4su3u1yieldingsm}, we will have to embed the electroweak Higgs doublet inside another scalar field charged under the $35$ representation if we embed the Standard Model fermions inside a $\overline{7}$ and a $21$ with a couple of $\overline{7}$'s in addition to ensure anomaly cancellation. 
  
   Since the $21$ and $35$ representations are complex, the fermion couplings to $\chi_{1}$ and $\chi_{2}$ are not exactly vector-like as they are in the case in which they are adjoint scalars. They involve Dirac scalar products between spinor fields $\Psi_{7}\sim{}\overline{7}$ and $\Psi_{21}\sim{}21$ and their charge conjugates. Note that if a 5+1D spinor $\Psi$ transforms under the two discrete $\mathbb{Z}_{2}$ symmetries as $\Psi\rightarrow{}i\Gamma^{4}\Gamma^{7}\Psi$ and $\Psi\rightarrow{}i\Gamma^{5}\Psi$ respectively, then its charge conjugate $\Psi^{C}$ also transforms as $\Psi^{C}\rightarrow{}i\Gamma^{4}\Gamma^{7}\Psi^{C}$ and $\Psi\rightarrow{}i\Gamma^{5}\Psi^{C}$. This implies that it is also the case that $\overline{\Psi^{C}}\Psi{}\rightarrow{}\overline{\Psi^{C}}\Psi{}$ and $\overline{\Psi^{C}}\Gamma^{7}\Psi{}\rightarrow{}-\overline{\Psi^{C}}\Gamma^{7}\Psi{}$ under the first $\mathbb{Z}_{2}$ symmetry and $\overline{\Psi^{C}}\Psi{}\rightarrow{}-\overline{\Psi^{C}}\Psi{}$ and $\overline{\Psi^{C}}\Gamma^{7}\Psi{}\rightarrow{}\overline{\Psi^{C}}\Gamma^{7}\Psi{}$ under the second. Hence, in this scenario, the background Yukawa Lagrangian for one generation, with the SM fermions embedded in $\Psi_{7}$ and $\Psi_{21}$ and with the fermionic fields $K^1\sim{}\overline{7}$ and $K^2\sim{}\overline{7}$ added for anomaly cancellation, is
   \begin{equation}
   \begin{aligned}
   \label{eq:2135yukawalagrangian}
   \mathcal{L}_{Yuk} &= -ih_{7\eta_{1}}\overline{\Psi_{7}}\Gamma^{7}\Psi_{7}\eta_{1}+h_{7\eta_{2}}\overline{\Psi_{7}}\Psi_{7}\eta_{2}-ih_{7K^{i}\eta_{1}}\overline{\Psi_{7}}\Gamma^{7}K^{i}\eta_{1}+h_{7K^{i}\eta_{2}}\overline{\Psi_{7}}K^{i}\eta_{2}-ih^{*}_{7K^{i}\eta_{1}}\overline{K^{i}}\Gamma^{7}\Psi_{7}\eta_{1}+h^{*}_{7K^{i}\eta_{2}}\overline{K^{i}}\Psi_{7}\eta_{2} \\
   &-ih_{K^{i}K^{j}\eta_{1}}\overline{K^{i}}\Gamma^{7}K^{j}\eta_{1}+h_{K^{i}K^{j}\eta_{2}}\overline{K^{i}}K^{j}\eta_{2}+\overline{\Psi^{C}_{7}}-2ih_{21\eta_{1}}\rm{Tr}[\overline{\Psi_{21}}\Gamma^{7}\Psi_{21}]\eta_{1}+2h_{21\eta_{2}}\rm{Tr}[\overline{\Psi_{21}}\Psi_{21}]\eta_{2}-ih_{7\chi_{1}}\overline{\Psi_{7}}\chi^{\dagger}_{1}\Gamma^{7}\Psi^{C}_{7}  \\
   &-ih^{*}_{7\chi_{1}}\overline{\Psi^{C}_{7}}\chi_{1}\Gamma^{7}\Psi_{7}-ih_{7K^{i}\chi_{1}}\overline{\Psi_{7}}\chi^{\dagger}_{1}\Gamma^{7}K^{iC}-ih^{*}_{7K^{i}\chi_{1}}\overline{K^{iC}_{7}}\chi_{1}\Gamma^{7}\Psi_{7}-ih_{K^{i}K^{j}\chi_{1}}\overline{K^{i}}\chi^{\dagger}_{1}\Gamma^{7}K^{jC}-ih^{*}_{K^{i}K^{j}\chi_{1}}\overline{K^{jC}}\chi^{\dagger}_{1}\Gamma^{7}K^{i} \\
   &+h_{7\chi_{2}21}\overline{\Psi_{21}}^{ab}\Psi^{c}_{7}\chi_{2abc}
   \end{aligned}
   \end{equation}
  It would be interesting to see what effect some of these non-standard background couplings have on the profiles. There should still be chiral zero modes localized on the intersection since their existence is mainly due to the couplings to the fields generating the kinks $\eta_{1}$ and $\eta_{2}$. If the couplings are vector-like, the interactions with $\chi_{1}$ and $\chi_{2}$ tend to affect the localization centers although in this case we also have interactions mixing the fermionic fields so one would expect some mixing induced in the profiles. The analysis for fermion localization is beyond the scope of the paper. 
  
  In showing that there is an interesting solution in a case where the fields inducing the symmetry breaking on each wall are not adjoint scalar fields, we have demonstrated that the scope for application of this new realization of the clash-of-symmetries mechanism is broad. One of the advantages of using complex representations to induce the breakings on the walls is that the residual $U(1)$'s are automatically broken. Indeed, one can imagine using different representations from the ones chosen in this section to reproduce other interesting scenarios. For example, it is obvious that the SM-like gauge group produced and described in Sec.~\ref{subsec:su4su3vsu4su3smmodel} could alternatively be produced by utilizing two scalars in the $35$ representation since they both induce breakings to $SU(4)\times{}SU(3)$ subgroups. Likewise, an $SU(5)$ theory equivalent to the one produced in Sec.~\ref{subsec:su6vsu6su5model} could also be reproduced by replacing the adjoint scalars with fundamental scalars.

\section{Conclusion}
\label{sec:conclusion}

  In this paper, we have proposed a new version of the Clash-of-Symmetries mechanism, which is an extension of the Dvali-Shifman mechanism, in the context of intersecting domain walls in 5+1D spacetime. Here, a large gauge group $G$ was assumed to be in confinement phase in the 6D bulk away from both domain-wall branes and on the branes $G$ was broken to subgroups $H_{1}$ and $H_{2}$ on each wall by the fields which attain lump-like VEV patterns on the wall. $H_{1}$ and $H_{2}$ are taken to be localized via the Dvali-Shifman mechanism. In turn, there is a clash-of-symmetries on the intersection of these walls where the symmetry respected is $H_{1}\cap{}H_{2}$ and further that subgroups of $H_{1}\cap{}H_{2}$ that are proper subgroups of larger, non-Abelian subgroups of $H_{1}$ and $H_{2}$ are then taken to be localized on the intersection  by confinement dynamics of these smaller non-Abelian groups. Assuming that both 5D and 6D non-Abelian Yang-Mills gauge theories exhibit confinement, this is a plausible mechanism to localize subgroups of a larger group on the intersection of two domain walls.
 
  We then dealt with a toy $SU(7)$ model which yielded some interesting results. In a model in which both $\chi_{1}$ and $\chi_{2}$ were charged under the adjoint representation, we showed that two choices for the VEV patterns for these fields yielded SM-like gauge groups fully localized to the domain-wall intersection, and another yielded a localized $SU(5)$ gauge theory. We found that in these cases, there are always left-over photons that are semi-delocalized and thus must be broken. We then gave the most elegant example in the paper in which $\chi_{1}$ is charged under the $21$ representation and $\chi_{2}$ is charged under the $35$ representation, yielding exactly an SM-like gauge group localized to the intersection with no leftover semi-delocalized photons. This case also has another advantage over the case with adjoint scalars generating the same SM, namely that it is possible to ensure that the desired configuration is the most stable in a quartic scalar field theory. 
  
  In all the examples that we have given, we only briefly touched on some of the basics of how to construct realistic fermionic and scalar sectors localized to the domain-wall intersection. We did not, for example, go into the specifics of scalar and fermion localization and show that realistic masses for the Standard Model fermions could be generated and that all the extra exotic fermions and scalars could be made massive enough. In some of the examples we have used, this seems to be quite a formidable task and one that is truly beyond the scope of this paper. Nevertheless, we have achieved something quite non-trivial in showing that in principle it is possible to localize and break straight down to a Standard Model gauge group by using the Clash-of-Symmetries mechanism. We showed this could be done both by using adjoint scalars and scalars in complex representations, and we have thus shown that the scope for use of this particular version of the Clash-of-Symmetries mechanism is very broad. It may not turn out that the particular models we have described in this paper are of phenomenological relevance after a more thorough analysis of the fermionic and scalar sectors, but we have laid the foundations for building a successful intersecting domain-wall braneworld model with gauge bosons localized to the intersection. 
  
  There is still further work that needs to be done in the intersecting domain-wall braneworld framework. We also need to successfully localize gravity and we also need to analyze the local stability properties of these intersecting domain-wall solutions.

\newpage

\appendix{}

\section{Some $SU(7)$ Representations, Products and Embeddings}
\label{appendix:su7reptheory}

\subsection{Basic $SU(7)$ Representations}

\begin{equation}
\begin{aligned}
\label{eq:su7representations}
7 &= (1, 0, 0, 0, 0, 0)                \qquad{}\qquad{}\qquad{}   196 &&= (0, 2, 0, 0, 0, 0) \\
21 &= (0, 1, 0, 0, 0, 0)               \qquad{}\qquad{}\qquad{}   210' &&= (1, 0, 1, 0, 0, 0) \\
35 &= (0, 0, 1, 0, 0, 0)               \qquad{}\qquad{}\qquad{}   224 &&= (1, 0, 0, 1, 0, 0) \\
\overline{35} &= (0, 0, 0, 1, 0, 0)    \qquad{}\qquad{}\qquad{}   392 &&= (0, 1, 0, 0, 1, 0) \\
\overline{21} &= (0, 0, 0, 0, 1, 0)    \qquad{}\qquad{}\qquad{}   490' &&= (0, 1, 1, 0, 0, 0) \\
\overline{7} &= (0, 0, 0, 0, 0, 1)     \qquad{}\qquad{}\qquad{}   540 &&= (2, 0, 0, 0, 1, 0) \\
28 &= (2, 0, 0, 0, 0, 0)               \qquad{}\qquad{}\qquad{}   588 &&= (0, 1, 0, 1, 0, 0) \\
48 &= (1, 0, 0, 0, 0, 1)               \qquad{}\qquad{}\qquad{}   735 &&= (2, 0, 0, 0, 0, 2) \\
84 &= (3, 0, 0, 0, 0, 0)               \qquad{}\qquad{}\qquad{}   735' &&= (1, 1, 0, 0, 0, 1) \\
112 &= (1, 1, 0, 0, 0, 0)              \qquad{}\qquad{}\qquad{}   784 &&= (0, 0, 1, 1, 0, 0) \\
140 &= (1, 0, 0, 0, 1, 0)              \qquad{}\qquad{}\qquad{}   1323 &&= (1, 0, 1, 0, 0, 1) \\
189 &= (2, 0, 0, 0, 0, 1)                 
\end{aligned}
\end{equation}

\subsection{Some Tensor Products of $SU(7)$ Representations}

\begin{equation}
\begin{aligned}
\label{eq:su7products}
&7\times{}\overline{7} = 1+48               \qquad{}\qquad{}\qquad{}      &&21\times{}\overline{21} = 1+48+392 \\
&7\times{}7 = 21+28                         \qquad{}\qquad{}\qquad{}      &&21\times{}35 = \overline{21}+224+490' \\ 
&7\times{}21 = 35+112                       \qquad{}\qquad{}\qquad{}      &&21\times{}\overline{35} = \overline{7}+140+588 \\
&7\times{}\overline{21} = \overline{7}+140  \qquad{}\qquad{}\qquad{}      &&21\times{}48 = 21+28+\overline{224}+735' \\
&7\times{}35 = \overline{35}+210'           \qquad{}\qquad{}\qquad{}      &&35\times{}35 = \overline{7}+140+490'+588 \\
&7\times{}\overline{35} = \overline{21}+224 \qquad{}\qquad{}\qquad{}      &&35\times{}\overline{35} = 1+48+392+784 \\
&7\times{}48 = 7+\overline{140}+189         \qquad{}\qquad{}\qquad{}      &&35\times{}48 = 35+112+\overline{210'}+1323 \\
&21\times{}21 = \overline{35}+196+210'      \qquad{}\qquad{}\qquad{}      &&48\times{}48 = 1+48+48+392+540+\overline{540}+735 \\
\end{aligned}
\end{equation}

\subsection{Embeddings of Subgroups of $SU(7)$}

\subsubsection{$SU(7)\supset{}SU(6)\times{}U(1)$}

\begin{equation}
\begin{aligned}
\label{eq:su6u1repembeddings}
7 &= (6, +1)+(1, -6) \\
21 &= (15, +2)+(6, -5) \\
28 &= (21, +2)+(6, -5)+(1, -12) \\
35 &= (20, +3)+(15, -4) \\
48 &= (35, 0)+(6, +7)+(\overline{6}, -7)+(1, 0) \\ 
112 &= (70, +3)+(21, -4)+(15, -4)+(6, -11)
\end{aligned}
\end{equation}

\subsubsection{$SU(7)\supset{}SU(5)\times{}SU(2)\times{}U(1)$}

\begin{equation}
\begin{aligned}
\label{eq:su6u1repembeddings}
7 &= (5, 1, +2)+(1, 2, -5) \\
21 &= (10, 1, +4)+(5, 2, -3)+(1, 1, -10) \\
28 &= (15, 1, +4)+(5, 2, -3)+(1, 3, -10) \\
35 &= (\overline{10}, 1, +6)+(10, 2, -1)+(5, 1, -8) \\
48 &= (24, 1, 0)+(5, \overline{2}, +7)+(\overline{5}, 2, -7)+(1, 3, 0)+(1, 1, 0) \\ 
112 &= (40, 1, +6)+(15, 2, -1)+(10, 2, -1)+(5, 3, -8)+(5, 1, -8)+(1, 2, -15)
\end{aligned}
\end{equation}

\subsubsection{$SU(7)\supset{}SU(4)\times{}SU(3)\times{}U(1)$}

\begin{equation}
\begin{aligned}
\label{eq:su6u1repembeddings}
7 &= (4, 1, +3)+(1, 3, -4) \\
21 &= (6, 1, +6)+(4, 3, -1)+(1, \overline{3}, -8) \\
28 &= (10, 1, +6)+(4, 3, -1)+(1, 6, -8) \\
35 &= (\overline{4}, 1, +9)+(6, 3, +2)+(4, \overline{3}, -5)+(1, 1, -12) \\
48 &= (15, 1, 0)+(4, \overline{3}, +7)+(\overline{4}, 3, -7)+(1, 8, 0)+(1, 1, 0) \\ 
112 &= (20, 1, +9)+(10, 3, +2)+(6, 3, +2)+(4, 6, -5)+(4, \overline{3}, -5)+(1, 8, -12)
\end{aligned}
\end{equation}

\section{All Possible Clash-of-Symmetries Groups from $SU(7)$ With Two Adjoint Scalars}
\label{appendix:allpossiblecosgroups}

 In this appendix, we list all the possible Clash-of-Symmetries breaking patterns with both of $\chi_{1}$ and $\chi_{2}$ transforming under the adjoint representation. For each possibility, we give example VEV patterns for $\chi_{1}$ and $\chi_{2}$ which generate them. We also state which resultant gauge groups are localized to the domain-wall intersection under the Dvali-Shifman formalism and which are semi-delocalized. We start by detailing the possibilities when $H_{1}\simeq{}H_{2}\simeq{}SU(6)\times{}U(1)$.

\subsection{$H_{1} = SU(6)\times{}U(1)$ and $H_{2} = SU(6)'\times{}U(1)'$}

\subsubsection{Case 1: $H_{1}\cap{}H_{2}=H_{1}=H_{2}=SU(6)\times{}U(1)$}
 
\begin{itemize}
 
 \item{Example VEV pattern: both $\chi_{1}$ and $\chi_{2}$ condense in the component proportional to the generator $Q_{1} = \rm{diag}(1, 1, 1, 1, 1, 1, -6)$.}
\item{Here, $SU(6)\cap{}SU(6)'=SU(6)$}
\item{There are no leftover diagonal generators.}
\item{Hence, the only Abelian symmetry preserved on the wall is $Q_{1}$.}
\item{The full symmetry respected on the intersection is $H_{1}\cap{}H_{2} = SU(6)\times{}U(1)_{Q_{1}}$. Both the gauge groups are semi-delocalized and able to propagate along both walls.}

\end{itemize}

 \subsubsection{Case 2: $H_{1}\cap{}H_{2}=SU(5)\times{}U(1)\times{}U(1)$}
 
 \begin{itemize}
 
\item{Example VEV pattern: $\chi_{1}$ condenses in the component proportional to the generator $Q_{1} = \rm{diag}(1, 1, 1, 1, 1, 1, -6)$  and $\chi_{2}$ condenses in the component proportional to the generator $Q'_{1} = \rm{diag}(1, 1, 1, 1, 1, -6, 1)$.}
\item{Here, $SU(6)\cap{}SU(6)=SU(5)$}
\item{The leftover diagonal generators are $T_{1} = \rm{diag}(1 ,1 ,1 ,1 ,1, -5, 0)$ and $T'_{1} = \rm{diag}(1, 1, 1, 1, 1, 0, -5)$.}
\item{Hence, the Abelian symmetries preserved on the wall are $q_{1} = 5/6Q_{1}+7/6T_{1} = 5/6Q'_{1}+7/6T'_{1} = \rm{diag}(2, 2, 2, 2, 2, -5, -5)$ and $q_{2} = 1/6(Q_{1}-T_{1}) = 1/6(T'_{1}-Q'_{1}) = \rm{diag}(0, 0, 0, 0, 0, 1, -1)$.}
\item{The full symmetry respected on the intersection is $H_{1}\cap{}H_{2} = SU(5)\times{}U(1)_{q_{1}}\times{}U(1)_{q_{2}}$. The $SU(5)$ subgroup is fully localized, the Abelian subgroups are not localized to the intersection and are free to propagate along both walls}

\end{itemize}

\subsection{$H_{1}=SU(6)\times{}U(1)$ and $H_{2} = SU(5)\times{}SU(2)\times{}U(1)$}

\subsubsection{Case 1: $H_{1}\cap{}H_{2} = SU(5)\times{}U(1)\times{}U(1)$}

\begin{itemize}

\item{Example VEV pattern: $\chi_{1}$ condenses in the component proportional to the generator $Q_{1} = \rm{diag}(1, 1, 1, 1, 1, 1, -6)$  and $\chi_{2}$ condenses in the component proportional to the generator $Q'_{1} = \rm{diag}(2, 2, 2, 2, 2, -5, -5)$.}
\item{Here, $SU(6)\cap{}SU(5)=SU(5)$}
\item{The leftover diagonal generators are $T_{1} = \rm{diag}(1 ,1 ,1 ,1 ,1, -5, 0)$ and $T'_{1} = \rm{diag}(0, 0, 0, 0, 0, 1, -1)$}
\item{Hence, the Abelian symmetries preserved on the wall are $q_{1} = Q_{1}+T_{1} = Q'_{1}+T'_{1} = \rm{diag}(2, 2, 2, 2, 2, -4, -6)$ and $q_{2} = 1/6(Q_{1}-T_{1}) = T'_{1} = \rm{diag}(0, 0, 0, 0, 0, 1, -1)$.}
\item{The full symmetry respected on the intersection is $H_{1}\cap{}H_{2} = SU(5)\times{}U(1)_{q_{1}}\times{}U(1)_{q_{2}}$. None of the gauge groups are localized; the $SU(5)$ gauge bosons are free to propagate along the $H_{2}$-respecting wall, the $U(1)_{q_{2}}$ photon can propagate along the $H_{1}$-respecting wall and the $U(1)_{q_{1}}$ photon can propagate along both walls.}

\end{itemize}

\subsubsection{Case 2: $H_{1}\cap{}H_{2} = SU(4)\times{}SU(2)\times{}U(1)\times{}U(1)$}

\begin{itemize}

\item{Example VEV pattern: $\langle{}\chi_{1}\rangle{}\propto{}Q_{1} = \rm{diag}(1, 1, 1, 1, 1, 1, -6)$ and $\langle{}\chi_{2}\rangle{}\propto{}Q'_{1} = \rm{diag}(-5, -5, 2, 2, 2, 2, 2)$.}
\item{Hence, $SU(6)\cap{}SU(5) = SU(4)$ and $SU(6)\cap{}SU(2) = SU(2)$.}
\item{Leftover diagonal generators: $T_{1} = \rm{diag}(-2, -2, 1, 1, 1, 1, 0)$ from $H_{1}$ and $T'_{1} = \rm{diag}(0, 0, 1, 1, 1, 1, -4)$ from $H_{2}$.}
\item{Preserved Abelian generators: $q_{1} = Q_{1}-2T_{1} = T'_{1}-Q'_{1} = \rm{diag}(5, 5, -1, -1, -1, -1, -6)$ and $q_{2} = 2Q_{1}+T_{1} = 3T'_{1} = \rm{diag}(0, 0, 1, 1, 1, 1, -4)$.}
\item{Preserved symmetry on intersection: $H_{1}\cap{}H_{2} = SU(4)\times{}SU(2)\times{}U(1)_{q_{1}}\times{}U(1)_{q_{2}}$. The $SU(4)$ subgroup is fully localized, the $SU(2)$ and $U(1)_{q_{2}}$ subgroup is semi-delocalized and able to propagate along the $H_{2}$-respecting wall, $U(1)_{q_{1}}$ is semi-delocalized and able to propagate along both walls.}

\end{itemize}

\subsection{$H_{1}=SU(6)\times{}U(1)$ and $H_{2} = SU(4)\times{}SU(3)\times{}U(1)$}

\subsubsection{Case 1: $H_{1}\cap{}H_{2} = SU(4)\times{}SU(2)\times{}U(1)\times{}U(1)$}

\begin{itemize}

\item{Example VEV pattern: $\langle{}\chi_{1}\rangle{}\propto{}Q_{1} = \rm{diag}(1, 1, 1, 1, 1, 1, -6)$ and $\langle{}\chi_{2}\rangle{}\propto{}Q'_{1} = \rm{diag}(3, 3, 3, 3, -4, -4, -4)$.}
\item{Hence, $SU(6)\cap{}SU(4) = SU(4)$ and $SU(6)\cap{}SU(3) = SU(2)$.}
\item{Leftover diagonal generators: $T_{1} = \rm{diag}(1, 1, 1, 1, -2, -2, 0)$ from $H_{1}$ and $T'_{1} = \rm{diag}(0, 0, 0, 0, 1, 1, -2)$ from $H_{2}$.}
\item{Preserved Abelian generators: $q_{1} = Q_{1}+2T_{1} = Q'_{1}+T'_{1} = \rm{diag}(3, 3, 3, 3, -3, -3, -6)$ and $q_{2} = 2Q_{1}-T_{1} = 1/3(Q'_{1}+16T'_{1}) = \rm{diag}(1, 1, 1, 1, 4, 4, -12)$.}
\item{Preserved symmetry on intersection: $H_{1}\cap{}H_{2} = SU(4)\times{}SU(2)\times{}U(1)_{q_{1}}\times{}U(1)_{q_{2}}$. The $SU(2)$ subgroup is fully localized, the $SU(4)$ subgroup is semi-delocalized and able to propagate along the $H_{2}$-respecting wall, and the $U(1)_{q_{1}}$ and $U(1)_{q_{2}}$ subgroups are semi-delocalized and able to propagate along both walls.}

\end{itemize}

\subsubsection{Case 2: $H_{1}\cap{}H_{2} = SU(3)\times{}SU(3)\times{}U(1)\times{}U(1)$}

\begin{itemize}

\item{Example VEV pattern: $\langle{}\chi_{1}\rangle{}\propto{}Q_{1} = \rm{diag}(1, 1, 1, 1, 1, 1, -6)$ and $\langle{}\chi_{2}\rangle{}\propto{}Q'_{1} = \rm{diag}(3, 3, 3, -4, -4, -4, 3)$.}
\item{Hence, $SU(6)\cap{}SU(4) = SU(3)_{1}$ and $SU(6)\cap{}SU(3) = SU(3)_{2}$.}
\item{Leftover diagonal generators: $T_{1} = \rm{diag}(1, 1, 1, -1, -1, -1, 0)$ from $H_{1}$ and $T'_{1} = \rm{diag}(1, 1, 1, 0, 0, 0, -3)$ from $H_{2}$.}
\item{Preserved Abelian generators: $q_{1} = 3T_{1}-Q_{1} = Q'_{1}-T'_{1} = \rm{diag}(2, 2, 2, -4, -4, -4, 6)$ and $q_{2} = 3Q_{1}+T_{1} = 1/2(11T'_{1}-Q'_{1}) = \rm{diag}(4, 4, 4, 2, 2, 2, -18)$.}
\item{Preserved symmetry on intersection: $H_{1}\cap{}H_{2} = SU(3)_{1}\times{}SU(3)_{2}\times{}U(1)_{q_{1}}\times{}U(1)_{q_{2}}$. The $SU(3)_{1}$ subgroup is fully localized while the $SU(3)_{2}$ subgroup is semi-delocalized and able to propagate along the $H_{2}$-respecting wall, and the $U(1)_{q_{1}}$ and $U(1)_{q_{2}}$ subgroups are semi-delocalized and able to propagate along both walls.}

\end{itemize}

\subsection{$H_{1} = SU(5)\times{}SU(2)\times{}U(1)$ and $H_{2} = SU(5)'\times{}SU(2)'\times{}U(1)'$}

\subsubsection{Case 1: $H_{1}\cap{}H_{2} = H_{1}=H_{2}=SU(5)\times{}SU(2)\times{}U(1)$}

\begin{itemize}

\item{Example VEV pattern: $\langle{}\chi_{1}\rangle{}\propto{}Q_{1} = \rm{diag}(2, 2, 2, 2, 2, -5, -5)$ and $\langle{}\chi_{2}\rangle{}\propto{}Q'_{1} = \rm{diag}(2, 2, 2, 2, 2, -5, -5)$.}
\item{Hence, $SU(5)\cap{}SU(5)' = SU(5)$ and $SU(2)\cap{}SU(2)' = SU(2)$.}
\item{Leftover diagonal generators: None}
\item{Preserved Abelian generators: $q_{1} = Q_{1} = Q'_{1} = \rm{diag}(2, 2, 2, 2, 2, -5, -5)$}
\item{Preserved symmetry on intersection: $H_{1}\cap{}H_{2} = SU(5)\times{}SU(2)\times{}U(1)_{q_{1}}$. All the factor gauge groups are semi-delocalized and free to propagate along both walls.}
\end{itemize}

\subsubsection{Case 2: $H_{1}\cap{}H_{2} = SU(4)\times{}U(1)\times{}U(1)\times{}U(1)$}

\begin{itemize}

\item{Example VEV pattern: $\langle{}\chi_{1}\rangle{}\propto{}Q_{1} = \rm{diag}(2, 2, 2, 2, 2, -5, -5)$ and $\langle{}\chi_{2}\rangle{}\propto{}Q'_{1} = \rm{diag}(2, 2, 2, 2, -5, -5, 2)$.}
\item{Hence, $SU(5)\cap{}SU(5)' = SU(4)$.}
\item{Leftover diagonal generators: $T_{1} = \rm{diag}(1, 1, 1, 1, -4, 0, 0)$, $T_{2} = \rm{diag}(0, 0, 0, 0, 0, 1, -1)$ from $H_{1}$ and $T'_{1} = \rm{diag}(1, 1, 1, 1, 0, 0, -4)$, $T'_{2} = \rm{diag}(0, 0, 0, 0, 1, -1, 0)$ from $H_{2}$.}
\item{Preserved Abelian generators: $q_{1} = T_{1}+4T_{2} = T'_{1}-4T'_{2} = \rm{diag}(1, 1, 1, 1, -4, 4, -4)$, $q_{2} = 1/2(Q_{1}-7T_{2}) = 1/2(Q'_{1}+7T'_{2}) = \rm{diag}(1, 1, 1, 1, 1, -6, 1)$, and $q_{3} = 1/5(2Q_{1}+T_{1}) = 1/5(Q'_{1}+3T'_{1}+5T'_{2} = \rm{diag}(1, 1, 1, 1, 0, -2, -2))$.}
\item{Preserved symmetry on intersection: $H_{1}\cap{}H_{2} = SU(4)\times{}U(1)_{q_{1}}\times{}U(1)_{q_{2}}\times{}U(1)_{q_{3}}$. The $SU(4)$ and $U(1)_{q_{1}}$ subgroups are fully localized to the intersection while the $U(1)_{q_{2}}$ and $U(1)_{q_{}}$ subgroups are semi-delocalized and able to propagate along both walls.}

\end{itemize}

\subsubsection{Case 3: $H_{1}\cap{}H_{2} = SU(3)\times{}SU(2)\times{}SU(2)\times{}U(1)\times{}U(1)$}

\begin{itemize}

\item{Example VEV pattern: $\langle{}\chi_{1}\rangle{}\propto{}Q_{1} = \rm{diag}(2, 2, 2, 2, 2, -5, -5)$ and $\langle{}\chi_{2}\rangle{}\propto{}Q'_{1} = \rm{diag}(2, 2, 2, -5, -5, 2, 2)$.}
\item{Hence, $SU(5)\cap{}SU(5)' = SU(3)$, $SU(2)\cap{}SU(5)' = SU(2)_{1}$ and $SU(5)\cap{}SU(2)' = SU(2)_{2}$.}
\item{Leftover diagonal generators: $T_{1} = \rm{diag}(2/3, 2/3, 2/3, -1, -1, 0, 0)$ from $H_{1}$ and $T'_{1} = \rm{diag}(2/3, 2/3, 2/3, 0, 0, -1, -1)$ from $H_{2}$.}
\item{Preserved Abelian generators: $q_{1} = 9/5Q_{1}+3/5T_{1} = 3/5Q'_{1}-9/5T'_{1} = \rm{diag}(4, 4, 4, 3, 3, -9, -9)$ and $q_{2} = Q_{1}-3T_{1} = -Q'_{1}+3T'_{1} = \rm{diag}(0, 0, 0, 5, 5, -5, -5)$.}
\item{Preserved symmetry on intersection: $H_{1}\cap{}H_{2} = SU(3)\times{}SU(2)_{1}\times{}SU(2)_{2}\times{}U(1)_{q_{1}}\times{}U(1)_{q_{2}}$. The $SU(3)$ subgroup is fully localized to the intersection, the $SU(2)_{1}$ gauge bosons are semi-delocalized and free to propagate along the $H_{1}$-respecting wall, similarly the $SU(2)_{2}$ gauge bosons are semi-delocalized and free to propagate along the $H_{2}$-respecting wall and the Abelian groups $U(1)_{q_{1}}$ and $U(1)_{q_{2}}$ are semi-delocalized and free to propagate along both walls.}

\end{itemize}

\subsection{$H_{1} = SU(5)\times{}SU(2)\times{}U(1)$ and $H_{2} = SU(4)\times{}SU(3)\times{}U(1)$}

\subsubsection{Case 1: $H_{1}\cap{}H_{2} = SU(4)\times{}SU(2)\times{}U(1)\times{}U(1)$}

\begin{itemize}

\item{Example VEV pattern: $\langle{}\chi_{1}\rangle{}\propto{}Q_{1} = \rm{diag}(2, 2, 2, 2, 2, -5, -5)$ and $\langle{}\chi_{2}\rangle{}\propto{}Q'_{1} = \rm{diag}(3, 3, 3, 3, -4, -4, -4)$.}
\item{Hence, $SU(5)\cap{}SU(4) = SU(4)$ and $SU(2)\cap{}SU(3) = SU(2)$.}
\item{Leftover diagonal generators: $T_{1} = \rm{diag}(1, 1, 1, 1, -4, 0, 0)$ from $H_{1}$ and $T'_{1} = \rm{diag}(0, 0, 0, 0, -2, 1, 1)$ from $H_{2}$.}
\item{Preserved Abelian generators: $q_{1} = Q_{1}+T_{1} = Q'_{1}-T'_{1} = \rm{diag}(3, 3, 3, 3, -2, -5, -5)$ and $q_{2} = -Q_{1}+5T_{1} = Q'_{1}+9T'_{1} = \rm{diag}(3, 3, 3, 3, -22, 5, 5)$.}
\item{Preserved symmetry on intersection: $H_{1}\cap{}H_{2} = SU(4)\times{}SU(2)\times{}U(1)_{q_{1}}\times{}U(1)_{q_{2}}$. None of the subgroups are fully localized. The $SU(4)$ gauge bosons are semi-delocalized and free to propagate along the $H_{2}$-respecting wall, similarly the $SU(2)$ gauge bosons are semi-delocalized and free to propagate along the $H_{1}$-respecting wall and the Abelian groups $U(1)_{q_{1}}$ and $U(1)_{q_{2}}$ are semi-delocalized and free to propagate along both walls.}

\end{itemize}

\subsubsection{Case 2: $H_{1}\cap{}H_{2} = SU(3)\times{}SU(2)\times{}U(1)\times{}U(1)\times{}U(1)$}

\begin{itemize}

\item{Example VEV pattern: $\langle{}\chi_{1}\rangle{}\propto{}Q_{1} = \rm{diag}(2, 2, 2, 2, 2, -5, -5)$ and $\langle{}\chi_{2}\rangle{}\propto{}Q'_{1} = \rm{diag}(3, 3, 3, -4, -4, -4, 3)$.}
\item{Hence, $SU(5)\cap{}SU(4) = SU(3)$ and $SU(5)\cap{}SU(3) = SU(2)$.}
\item{Leftover diagonal generators: $T_{1} = \rm{diag}(2/3, 2/3, 2/3, -1, -1, 0, 0)$, $T_{2} = \rm{diag}(0, 0, 0, 0, 0, 1, -1)$ from $H_{1}$ and $T'_{1} = \rm{diag}(2/3, 2/3, 2/3, 0, 0, 0, -2)$, $T'_{2} = \rm{diag}(0, 0, 0, 1, 1, -2, 0)$ from $H_{2}$.}
\item{Preserved Abelian generators: $q_{1} = -T_{1}-2T_{2} = -T'_{1}+T'_{2} = \rm{diag}(-2/3, -2/3, -2/3, 1, 1, -2, 2)$, $q_{2} = 4Q_{1}+7T_{1}-6T_{2} = 2Q'_{1}+10T'_{1}+9T'_{2} = \rm{diag}(38/3, 38/3, 38/3, 1, 1, -26, -14)$ and $q_{3} = -3Q_{1}+12T_{1}+12T_{2} = 3/2Q'_{1}-3/8T'_{1}-15'T_{2} = \rm{diag}(2, 2, 2, -18, -18, 27, 3)$.}
\item{Preserved symmetry on intersection: $H_{1}\cap{}H_{2} = SU(3)\times{}SU(2)\times{}U(1)_{q_{1}}\times{}U(1)_{q_{2}}\times{}U(1)_{q_{3}}$. The $SU(3)$, $SU(2)$ and $U(1)_{q_{1}}$ subgroups are fully localized to the domain-wall intersection. The $U(1)_{q_{2}}$ and $U(1)_{q_{3}}$ subgroups are semi-delocalized and their photons can propagate along both walls.}

\end{itemize}

\subsubsection{Case 3: $H_{1}\cap{}H_{2} = SU(3)\times{}SU(2)\times{}SU(2)\times{}U(1)\times{}U(1)$}

\begin{itemize}

\item{Example VEV pattern: $\langle{}\chi_{1}\rangle{}\propto{}Q_{1} = \rm{diag}(2, 2, 2, 2, 2, -5, -5)$ and $\langle{}\chi_{2}\rangle{}\propto{}Q'_{1} = \rm{diag}(-4, -4, -4, 3, 3, 3, 3)$.}
\item{Hence, $SU(5)\cap{}SU(3) = SU(3)$, $SU(5)\cap{}SU(4) = SU(2)_{1}$ and $SU(2)\cap{}SU(4) = SU(2)_{2}$.}
\item{Leftover diagonal generators: $T_{1} = \rm{diag}(2/3, 2/3, 2/3, -1, -1, 0, 0)$ from $H_{1}$ and $T'_{1} = \rm{diag}(0, 0, 0, 1, 1, -1, -1)$ from $H_{2}$.}
\item{Preserved Abelian generators: $q_{1} = TQ_{1}+3T_{1} = -Q'_{1}+2T'_{1} = \rm{diag}(4, 4, 4, -1, -1, -5, -5)$ and $q_{2} = 3Q_{1}-T_{1}= -4/3Q'_{1}+11T'_{1} = \rm{diag}(16/3, 16/3, 16/3, 7, 7, -15, -15)$.}
\item{Preserved symmetry on intersection: $H_{1}\cap{}H_{2} = SU(3)\times{}SU(2)_{1}\times{}SU(2)_{2}\times{}U(1)_{q_{1}}\times{}U(1)_{q_{2}}$. Only the $SU(2)_{1}$ subgroup is fully localized to the domain-wall intersection. The $SU(3)$ subgroup is semi-delocalized and its gauge bosons can propagate along the $H_{2}$-respecting wall. The $SU(2)_{2}$ subgroup is semi-delocalized and its gauge bosons can propagate along the $H_{1}$-respecting wall. The $U(1)_{q_{2}}$ and $U(1)_{q_{3}}$ subgroups are semi-delocalized and their photons can propagate along both walls.}

\end{itemize}

\subsection{$H_{1} = SU(4)\times{}SU(3)\times{}U(1)$ and $H_{2} = SU(4)'\times{}SU(3)'\times{}U(1)'$}

\subsubsection{Case 1: $H_{1}\cap{}H_{2} = SU(4)\times{}SU(3)\times{}U(1)$}

\begin{itemize}

\item{Example VEV pattern: $\langle{}\chi_{1}\rangle{}\propto{}Q_{1} = \rm{diag}(3, 3, 3, 3, -4, -4, -4)$ and $\langle{}\chi_{2}\rangle{}\propto{}Q'_{1} = \rm{diag}(3, 3, 3, 3, -4, -4, -4)$.}
\item{Hence, $SU(4)\cap{}SU(4)' = SU(4)$ and $SU(3)\cap{}SU(3)' = SU(3)$.}
\item{Leftover diagonal generators: None}
\item{Preserved Abelian generators: $q_{1} = Q_{1} = Q'_{1} = \rm{diag}(3, 3, 3, 3, -4, -4, -4)$}
\item{Preserved symmetry on intersection: $H_{1}\cap{}H_{2} = SU(4)\times{}SU(3)\times{}U(1)_{q_{1}}$. All the factor gauge groups are semi-delocalized and free to propagate along both walls.}
\end{itemize}

\subsubsection{Case 2: $H_{1}\cap{}H_{2} = SU(3)\times{}SU(2)\times{}U(1)\times{}U(1)\times{}U(1)$}

\begin{itemize}

\item{Example VEV pattern: $\langle{}\chi_{1}\rangle{}\propto{}Q_{1} = \rm{diag}(3, 3, 3, -4, -4, -4, 3)$ and $\langle{}\chi_{2}\rangle{}\propto{}Q'_{1} = \rm{diag}(3, 3, 3, -4, -4, 3, -4)$.}
\item{Hence, $SU(4)\cap{}SU(4)' = SU(3)$ and $SU(3)\cap{}SU(3)' = SU(2)$.}
\item{Leftover diagonal generators: $T_{1} = \rm{diag}(2/3, 2/3, 2/3, 0, 0, 0, -2)$, $T_{2} = \rm{diag}(0, 0, 0, 1, 1, -2, 0)$ from $H_{1}$ and $T'_{1} = \rm{diag}(2/3, 2/3, 2/3, 0, 0, -2, 0)$, $T'_{2} = \rm{diag}(0, 0, 0, 1, 1, 0, -2)$ from $H_{2}$.}
\item{Preserved Abelian generators: $q_{1} = -T_{1}-T_{2} = -T'_{1}-T'_{2} = \rm{diag}(-2/3, -2/3, -2/3, -1, -1, 2, 2)$, $q_{2} = 4Q_{1}+T_{1}-T_{2} = 2Q'_{1}+10T'_{1}-9T'_{2} = \rm{diag}(38/3, 38/3, 38/3, -17, -17, -14, 10)$, and $q_{3} = Q_{1}-2T_{1}+2T_{2} = -1/2Q'_{1}+29/4T'_{1}-6T'_{2} = \rm{diag}(5/3, 5/3, 5/3, -2, -2, -8, 7))$.}
\item{Preserved symmetry on intersection: $H_{1}\cap{}H_{2} = SU(3)\times{}SU(2)\times{}U(1)_{q_{1}}\times{}U(1)_{q_{2}}\times{}U(1)_{q_{3}}$. The $SU(3)$, $SU(2)$ and $U(1)_{q_{1}}$ subgroups are fully localized to the domain-wall intersection. The $U(1)_{q_{2}}$ and $U(1)_{q_{3}}$ subgroups are semi-delocalized and their photons can propagate along both walls.}

\end{itemize}

\subsubsection{Case 3: $H_{1}\cap{}H_{2} = SU(2)\times{}SU(2)\times{}SU(2)\times{}U(1)\times{}U(1)\times{}U(1)$}

\begin{itemize}

\item{Example VEV pattern: $\langle{}\chi_{1}\rangle{}\propto{}Q_{1} = \rm{diag}(3, 3, 3, 3, -4, -4, -4)$ and $\langle{}\chi_{2}\rangle{}\propto{}Q'_{1} = \rm{diag}(3, 3, -4, -4, -4, 3, 3)$.}
\item{Hence, $SU(4)\cap{}SU(4)' = SU(2)_{1}$, $SU(4)\cap{}SU(3)' = SU(2)_{2}$ and $SU(3)\cap{}SU(4)' = SU(2)_{3}$.}
\item{Leftover diagonal generators: $T_{1} = \rm{diag}(1, 1, -1, -1, 0, 0, 0)$, $T_{2} = \rm{diag}(0, 0, 0, 0, -2, 1, 1)$ from $H_{1}$ and $T'_{1} = \rm{diag}(1, 1, 0, 0, 0, -1, -1)$, $T'_{2} = \rm{diag}(0, 0, 1, 1, -2, 0, 0)$ from $H_{2}$.}
\item{Preserved Abelian generators: $q_{1} = T_{1}-T_{2} = T'_{1}-T'_{2} = \rm{diag}(1, 1, -1, -1, 2, -1, -1)$, $q_{2} = Q_{1}+2T_{1}+2T_{2} = 1/2(Q'_{1}+7T'_{1}+6T'_{2}) = \rm{diag}(5, 5, 1, 1, -8, -2, -2)$, and $q_{3} = 4Q_{1}-T_{1}-T_{2} = -Q'_{1}+14T'_{1}+9T'_{2} = \rm{diag}(11, 11, 13, 13, -14, -17, -17))$.}
\item{Preserved symmetry on intersection: $H_{1}\cap{}H_{2} = SU(2)_{1}\times{}SU(2)_{2}\times{}SU(2)_{3}\times{}U(1)_{q_{1}}\times{}U(1)_{q_{2}}\times{}U(1)_{q_{3}}$. The $SU(2)_{1}$, $SU(2)_{2}$, $SU(2)_{3}$ and $U(1)_{q_{1}}$ subgroups are fully localized to the domain-wall intersection. The $U(1)_{q_{2}}$ and $U(1)_{q_{3}}$ subgroups are semi-delocalized and their photons can propagate along both walls.}

\end{itemize}

\subsubsection{Case 4: $H_{1}\cap{}H_{2} = SU(3)\times{}SU(3)\times{}U(1)\times{}U(1)$}

\begin{itemize}

\item{Example VEV pattern: $\langle{}\chi_{1}\rangle{}\propto{}Q_{1} = \rm{diag}(3, 3, 3, 3, -4, -4, -4)$ and $\langle{}\chi_{2}\rangle{}\propto{}Q'_{1} = \rm{diag}(-4, -4, -4, 3, 3, 3, 3)$.}
\item{Hence, $SU(3)\cap{}SU(4)' = SU(3)_{1}$ and $SU(4)\cap{}SU(3)' = SU(3)_{2}$.}
\item{Leftover diagonal generators: $T_{1} = \rm{diag}(1, 1, 1, -3, 0, 0, 0)$ from $H_{1}$ and $T'_{1} = \rm{diag}(0, 0, 0, -3, 1, 1, 1)$ from $H_{2}$.}
\item{Preserved Abelian generators: $q_{1} = 1/4(Q_{1}+T_{1}) = -1/4(Q'_{1}+T'_{1}) = \rm{diag}(1, 1, 1, 0, -1, -1, -1)$ and $q_{2} = 1/2(Q_{1}-T_{1}) = -1/4(Q'_{1}+5T'_{1}) = \rm{diag}(1, 1, 1, 3, -2, -2, -2)$.}
\item{Preserved symmetry on intersection: $H_{1}\cap{}H_{2} = SU(3)_{1}\times{}SU(3)_{2}\times{}U(1)_{q_{1}}\times{}U(1)_{q_{2}}$. The $SU(3)_{1}$ subgroup is semi-delocalized and its gauge bosons are able to propagate along the $H_{1}$-respecting wall. The $SU(3)_{2}$ subgroup is semi-delocalized and its gauge bosons are able to propagate along the $H_{2}$-respecting wall. The $U(1)_{q_{1}}$ and $U(1)_{q_{2}}$ subgroups are semi-delocalized and their photons can propagate along both walls.}

\end{itemize}

\bibliographystyle{ieeetr}
\bibliography{bibliography2.bib}

\end{document}